\documentclass[AMA,STIX1COL]{WileyNJD-v2}

\articletype{}%

\usepackage{rotating}
\usepackage{dsfont}        
\usepackage{bbold}         

\received{XX.XX.XXXX}
\revised{YY.YY.YYYY}
\accepted{ZZ.ZZ.ZZZZ}

\providecommand{\prob}{\mathrm{P}}
\providecommand{\differential}{\mathrm{d}}
\providecommand{\expect}{\mathrm{E}}
\providecommand{\var}{\mathrm{Var}}
\providecommand{\cov}{\mathrm{Cov}}
\providecommand{\normaldistn}{\mathrm{N}}
\providecommand{\unifdistn}{\mathrm{Uniform}}
\providecommand{\halfnormaldistn}{\mathrm{HN}}
\providecommand{\halftdistn}{\mbox{H$t$}}

\providecommand{\invgamma}{\mbox{inv-$\Gamma$}}
\providecommand{\invchi}{\mbox{inv-$\chi$}}
\providecommand{\realline}{\mathds{R}}

\providecommand{\sigmau}{\sigma_\mathbb{1}} 
\providecommand{\su}{s_\mathbb{1}}          
\providecommand{\scale}{s}                
\providecommand{\varcoef}{c_{\mathrm{v}}}    
\providecommand{\sigmaw}{\sigma_\mathrm{w}} 

\begin{document}

\title{On weakly informative prior distributions for the heterogeneity parameter in Bayesian random-effects meta-analysis}

\author[1]{Christian R\"{o}ver}
\author[2]{Ralf Bender}
\author[3]{Sofia Dias}
\author[4]{Christopher H. Schmid}
\author[5]{Heinz Schmidli}
\author[2]{Sibylle Sturtz}
\author[6]{Sebastian Weber}
\author[1]{Tim Friede}

\authormark{C.~R\"{O}VER, \textsc{et al.}}

\address[1]{\orgdiv{Department of Medical Statistics}, \orgname{University Medical Center G\"{o}ttingen}, \orgaddress{\state{G\"{o}ttingen}, \country{Germany}}}
\address[2]{\orgdiv{Department of Medical Biometry}, \orgname{Institute for Quality and Efficiency in Health Care (IQWiG)}, \orgaddress{\state{K\"{o}ln}, \country{Germany}}}
\address[3]{\orgdiv{Centre for Reviews and Dissemination}, \orgname{University of York}, \orgaddress{\state{York}, \country{UK}}}
\address[4]{\orgdiv{Department of Biostatistics and Center for Evidence Synthesis in Health}, \orgname{Brown University School of Public Health}, \orgaddress{\state{Providence, RI}, \country{USA}}}
\address[5]{\orgdiv{Statistical Methodology, Development}, \orgname{Novartis Pharma AG}, \orgaddress{\state{Basel}, \country{Switzerland}}}
\address[6]{\orgdiv{Advanced Exploratory Analytics}, \orgname{Novartis Pharma AG}, \orgaddress{\state{Basel}, \country{Switzerland}}}

\corres{Christian R\"{o}ver, \email{christian.roever@med.uni-goettingen.de}}


\abstract[Abstract]{The normal-normal hierarchical model (NNHM)
  constitutes a simple and widely used framework for meta-analysis.
  In the common case of only few studies contributing to the
  meta-analysis, standard approaches to inference tend to perform
  poorly, and Bayesian meta-analysis has been suggested as a potential
  solution.  The Bayesian approach, however, requires the sensible
  specification of prior distributions.  While noninformative priors
  are commonly used for the overall mean effect, the use of
  \emph{weakly informative} priors has been suggested for the
  heterogeneity parameter, in particular in the setting of (very) few
  studies.  To date, however, a consensus on how to generally specify
  a weakly informative heterogeneity prior is lacking. Here we
  investigate the problem more closely and provide some guidance on
  prior specification.}

\keywords{marginal likelihood, Bayes factor, hierarchical model, 
          variance component, GLMM}

\jnlcitation{\cname{%
\author{C.~R\"{o}ver}, 
\author{R.~Bender}, 
\author{S.~Dias}, 
\author{C.~Schmid}, 
\author{H.~Schmidli}, 
\author{S.~Sturtz}, 
\author{S.~Weber}, and 
\author{T.~Friede}} (\cyear{2021}), 
\ctitle{On weakly informative prior distributions for the heterogeneity parameter in Bayesian random-effects meta-analysis}, \cjournal{(submitted for publication)}, \cvol{2021}.}

\maketitle


\section{Introduction}
  In meta-analysis, researchers commonly encounter a certain amount of
  variability between experiments, to a degree going beyond what could
  be attributed to measurement error alone. Hierarchical models are
  commonly used in order to account for such (``between-study'')
  heterogeneity.\citep{BDA3rd,GelmanHill} In the present paper, we
  focus on the special simple case of meta-analysis within the
  framework of the \emph{normal-normal hierarchical model (NNHM)}\@.
  The NNHM approximates estimates from separate sources and their
  standard errors via normal distributions, and implements
  heterogeneity at a second level using another normal variance
  component.  In meta-analysis applications, the NNHM provides a good
  approximation for many types of endpoints or effect
  measures.\citep{HedgesOlkin,HartungKnappSinha} The normal
  approximation has its limitations,\citep{JacksonWhite2018} some of
  which are less of a problem in a Bayesian
  context.\citep{RoeverFriede2018} A small number of studies tends to
  pose a problem especially for frequentist methods, in particular
  regarding the construction of confidence intervals (CIs) with good
  coverage
  properties.\citep{FriedeRoeverWandelNeuenschwander2017a,FriedeRoeverWandelNeuenschwander2017b,BenderEtAl2018,GonnnermannEtAl2015}
  A common convention is to exercise extra caution when the number of
  studies is small.\citep{BenderEtAl2018}

  Bayesian approaches to meta-analysis have been advocated for quite a
  while,\citep{SmithSpiegelhalterThomas1995,SuttonAbrams2001,Schmid2001,Spiegelhalter2004,SpiegelhalterEtAl,HigginsThompsonSpiegelhalter2009,LunnEtAl2013}
  and analyses may technically be performed using MCMC
  methods\citep{BDA3rd} or semi-analytical
  integration.\citep{RoeverFriede2017} Within the \textsf{R}~software,
  for example the \texttt{bayesmeta}\citep{bayesmeta,Roever2020} or
  \texttt{bmeta}\citep{R:bmeta} packages are available.  Performing a
  Bayesian analysis is not technically challenging; computations are
  straightforward and valid for any number of studies, although less
  data will mean that results are more sensitive to prior
  specifications (especially when it comes to variance parameters).  A
  crucial condition is that the explicitly implemented normal
  approximation needs to hold, which may break down e.g.\ for
  meta-analyses of \emph{small}
  studies.\citep{JacksonWhite2018,RoeverFriede2018} While for large
  numbers of studies, the choice of prior distributions usually has
  little impact, for few studies the exact form of the prior
  distributions chosen may become crucial, as one cannot rely on the
  prior information being overruled by the data in that case.  At
  least part of this problem may be considered ``shared'' for
  frequentist and Bayesian methods {as long as one tries to get by
    without using a proper, informative prior}.\citep{Senn2007} Some
  supposedly \emph{noninformative} prior distributions can probably be
  argued to be less influential than others, but ultimately these are
  unlikely to be the best choice in few-study problems.  Beyond
  meta-analysis, the use of informative priors for regularisation in
  the estimation of certain parameters is also
  common.\citep{vanDongen2006} Especially for few studies, this may be
  a promising approach.\citep{WilliamsEtAl2018} The case of ``few''
  studies is hard to define; there is no obvious threshold, and in
  fact there may actually be no need to distinguish: use of an
  informative prior will not be harmful for analyses of ``many''
  studies. Indeed, a proper prior is necessary irrespective of the
  number of studies in case the analysis requires the calculation of
  marginal likelihoods. In the present manuscript, we will investigate
  examples ranging in size between~2 and 5~studies. These are the
  cases where the use of an informative prior will make the greatest
  difference, and such situations have been discussed in the context
  of up to~4,\citep{BenderEtAl2018},
  3--10,\citep{FriedeRoeverWandelNeuenschwander2017a} or only
  2~studies.\citep{FriedeRoeverWandelNeuenschwander2017b}

  Heterogeneity priors have been investigated previously from
  different angles; some discussed general considerations for variance
  parameters\citep{SpiegelhalterEtAl,Gelman2006,NeuenschwanderSchmidli2020}
  while others motivated particular settings for specific example
  cases\citep{PrevostEtAl2000,DiasEtAl2013} or investigated commonly
  used settings in a systematic literature
  review.\citep{DebrayEtAl2015}
  The aim of the present investigation is to provide general guidance
  for judging and deriving \emph{weakly informative} heterogeneity
  priors, and to suggest consensus examples for some common types of
  effect measures.  This may also aid in the design and justification
  of prior settings, or the prospective pre-specification of Bayesian
  meta-analyses\citep{StewartMoherShekelle2012} and it may help avoid
  (suspicion of) post-hoc tweaking of prior assumptions.

  The remainder of this article is structured as follows. In the next
  section, the normal-normal hierarchical model (NNHM) along with its
  parameters and prior distributions are formally introduced.
  Section~\ref{sec:heteroPriors} discusses prior distributions for the
  heterogeneity parameter and some general motivating considerations
  and implications.  Section~\ref{sec:specialcases} motivates
  heterogeneity priors for a selection of common types of endpoints
  and effect measures based on the previously discussed ideas.  In
  Section~\ref{sec:examples}, examples of meta-analyses with different
  endpoints are introduced, and analyses are performed using the
  suggested prior settings.  Section~\ref{sec:discussion} closes with
  conclusions and recommendations.

\section{The statistical model}\label{sec:model}
  \subsection{The normal-normal hierarchical model (NNHM)}\label{sec:nnhm}
    The \emph{normal-normal hierarchical model (NNHM)} represents
    measurements~$y_i$ from $k$~different sources using two hierarchy
    levels. Along with the estimates, their associated standard
    errors~$\sigma_i$ need to be available. The~$\sigma_i$ are assumed
    to be fixed and known (which commonly is only an
    approximation.\citep{JacksonWhite2018,Stevens2011}) Each
    estimate~$y_i$ is assumed to measure an underlying true
    value~$\theta_i$, which is not necessarily identical across all
    $k$~measurements; (``between-study'') variability among
    the~$\theta_i$ is accounted for by an additional variance
    component whose magnitude is given by the 
    heterogeneity~$\tau \geq 0$:
    \begin{eqnarray}
      y_i|\theta_i & \sim & \normaldistn(\theta_i, \sigma_i^2)\mbox{,}
      \label{eqn:NNHM1} \\
      \theta_i|\mu,\tau    & \sim & \normaldistn(\mu,\tau^2)
      \quad \mbox{for $i=1,\ldots,k$}\mbox{,}
      \label{eqn:NNHM2}
    \end{eqnarray}
    where the estimates~$y_i$ (as well as the~$\theta_i$) are modelled
    as exchangeable.  The overall mean effect $\mu$ is often the
    figure of primary interest.  By marginalizing over the
    $\theta_i$~values, the model may be written in simplified form:
    \begin{eqnarray}\label{eqn:NNHM3}
      y_i|\mu,\tau & \sim & \normaldistn(\mu,\,\sigma_i^2+\tau^2)\mbox{.} \label{eqn:NNHM}
    \end{eqnarray}
    This is a random-effects model, which in the special case
    of~$\tau=0$ simplifies to the common-effect model (also known as
    the fixed-effect
    model).\citep{HedgesOlkin,HartungKnappSinha,Roever2020,BorensteinEtAl2010}
    The NNHM provides a good approximation for many types of effect
    measures where the estimates as well as between-study variability
    may be assumed to be (approximately) normally
    distributed.\citep{JacksonWhite2018}

    While often the aim of a meta-analysis is estimation of the
    overall mean~$\mu$, it is sometimes useful to also infer the
    study-specific means~$\theta_i$ or a
    prediction~$\theta_{k+1}$. The amount of information gained on
    $\theta_i$ or $\theta_{k+1}$ through the joint meta-analysis
    depends very much on the amount of heterogeneity~$\tau$. If there
    was no heterogeneity ($\tau=0$), then we would have
    $\theta_1=\theta_2=\ldots=\theta_{k+1}=\mu$, and all data would
    essentially contribute to the estimation of a single common
    parameter. If, on the other hand, $\tau$~was very large, then
    different parameters~$\theta_i$ would only be very loosely
    connected \eqref{eqn:NNHM2}, and consideration of additional data
    would only add very little to the estimation of any
    particular~$\theta_i$ or to a prediction~$\theta_{k+1}$. In
    between, for moderate $\tau$~values, estimates of~$\theta_i$ are
    somewhat ``shrunk'' towards the overall mean~$\mu$, and the
    prediction~$\theta_{k+1}$ is also more tightly
    constrained. Estimation of the heterogeneity~$\tau$ hence also has
    distinct effects on the so-called ``shrinkage
    estimates''~$\theta_i$ as well as
    predictions~$\theta_{k+1}$.\citep{Roever2020,WandelNeuenschwanderRoeverFriede2017}

  \subsection{Prior distributions}
    \subsubsection{Effect and heterogeneity priors}
      In the NNHM, there are two unkowns requiring prior
      specification, namely the overall mean effect~$\mu$ and the
      heterogeneity~$\tau$.  In the following, we will assume that the
      prior may be factored into $p(\mu,\tau)=p(\mu)\times p(\tau)$,
      implying prior independence of~$\mu$ and~$\tau$; note though
      that one may also argue in favour of a dependent
      prior.\citep{Senn2007,Pullenayegum2011} In a sense, dependence
      is often implicitly implemented e.g. in the case of
      log-transformed effect scales: on the back-transformed
      (exponentiated) scale, the amount of heterogeneity then scales
      with the value of the effect.

      The effect prior~$p(\mu)$ may often, also for technical
      convenience, be taken to be (improper) uniform or
      normal.\citep{Roever2020} In case a proper, informative effect
      prior is used, this may also have implications for the
      heterogeneity prior; in particular the prior variance of~$\mu$
      may be relevant when considering reasonable $\tau$~values (see
      also Section~\ref{sec:effectMagnitude} below).

      Here we are first of all concerned with the prior distribution
      for the heterogeneity, $p(\tau)$.  A number of priors have been
      proposed that may be considered ``noninformative'' in
      particular senses (e.g., improper uniform or Jeffreys priors,
      which may be motivated using invariance or information-theoretic
      arguments),\citep[][Sec.~2.2]{Roever2020} but these usually
      cause problems especially when the number of studies~($k$) is
      sufficiently small, or when the computation of marginal
      likelihoods (or Bayes factors) is desired. In the following, we
      will hence be concerned with proper, (weakly) informative
      priors.

    \subsubsection{Different views of prior specification}\label{sec:priorViews}
      There may be different perspectives on the role or purpose of
      prior specification within a Bayesian analysis; we sketch three
      aspects here:
      \begin{description}
        \item[(i) Epistemic point-of-view:] The posterior distribution
          depends on the prior via Bayes' theorem; the prior
          inevitably needs to enter inference, reflecting the state of
          information beyond the data at hand.\citep{BDA3rd,Jaynes}
          Prior assumptions simply add to the line of other
          assumptions being made, like a normal likelihood,
          independence, known standard errors, etc.
        \item[(ii) Regularisation point-of-view:] The aim is to
          introduce ``\emph{weakly informative priors, which attempt
            to let the data speak while being strong enough to exclude
            various `unphysical' possibilities which, if not blocked,
            can take over a posterior distribution in settings with
            sparse data}'' (Gelman; 2009).\citep{Gelman2009} This
          perspective is closely connected to regularisation or
          penalization approaches in
          general.\citep{ColeChuGreenland2013} While in the likelihood
          framework it may sometimes be perceived as a rather
          \emph{ad~hoc} fix, it constitutes a transparent, readily
          interpretable model component in the Bayesian case.
        \item[(iii) Pragmatic point-of-view:] The resulting estimates
          may be judged solely based on their operating
          characteristics (which may be frequentist or
          Bayesian,\citep[Sec.~4.4]{BDA3rd}) without worrying about
          their exact theoretic underpinning.
      \end{description}
      The first viewpoint is probably the most ``constructive'' one
      here, in the sense of providing guidance on sensible prior
      choices.  An example of a regularisation approach in the NNHM
      context is given by the procedure proposed by Chung
      \emph{et~al.} (2013),\citep{ChungEtAl2013a} where regularisation
      is used to implement preference for positive
      $\tau$~values. Alternatively, one may also give preference to
      small $\tau$~values, as these imply a less complex model, which
      is the idea behind penalized complexity
      priors\citep{KleinKneib2016} (and which here would lead to an
      exponential prior).  Comparisons of operating characteristics
      (also including frequentist approaches) were done e.g. by Friede
      \emph{et~al.}
      (2017).\citep{FriedeRoeverWandelNeuenschwander2017a} There are
      probably more perspectives beyond or between these three (e.g.,
      \citep{BayarriBerger2004,Kass2011}).  For example, meta-analyses
      may be thought of as constituting draws from a ``population''
      whose associated heterogeneities are reflected in the prior
      distribution --- an ``\emph{aleatory}'' interpretation of
      (prior) probability, which may lend a somewhat frequentist
      flavour to the analysis.  An important point to stress is that
      there is not necessarily a single ``correct'' prior: the use of
      different priors may be seen as basing inferences on different
      preconditions, and the choice of prior depends on which
      information one is willing to incorporate into the analysis;
      different analysts may hence draw different conclusions from the
      same data, when these are founded on differing prior
      beliefs.\citep{SpiegelhalterEtAl1994} In a sense, the posterior
      inherits its meaning from the prior to some
      extent.\citep{Jaynes1976} Other common shortcuts taken or
      approximations and asymptotics relied upon may in fact often be
      potentially more influential and relevant than the choice among
      the (usually limited) set of reasonable prior distributions
      (see, e.g., Jackson and White (2018)\citep{JacksonWhite2018}).

    \subsubsection{Implications for interval estimation} 
      While (frequentist) confidence intervals aim to provide coverage
      of the true parameter \emph{uniformly}, independent of the
      actual current parameter value, this is generally not the case
      for (Bayesian) credible intervals.  In some cases, it is
      possible to specify (often improper) priors leading to posterior
      distributions that also provide proper frequentist coverage, but
      usually such a prior is not available.\citep{DattaSweeting2005b}
      Credible intervals are calibrated and yield proper coverage
      \emph{on average} across the prior distribution; for the
      point-wise coverage this means that there may be overcoverage in
      certain regions of parameter space and undercoverage in
      others.\citep{Roever2020,Dawid1982,Mandelkern2002,GneitingEtAl2007}
      For example, in the present case this may mean that long-run
      coverage may be above the nominal level if data were repeatedly
      generated based on heterogeneity values from the lower end of
      the prior range, and below the nominal level otherwise.

\section{Heterogeneity priors}\label{sec:heteroPriors}
  \subsection{Aim}\label{sec:aim}
    For meta-analyses involving many studies (large~$k$), the choice
    of prior distribution often has little impact, and an (improper)
    uniform prior for~$\tau$ may be a good choice, not least due to
    its invariance property.\citep{Roever2020,Gelman2006} Here we are
    concerned first of all with the case of few studies (small~$k$); a
    uniform prior may not actually be an option here, as it requires
    $k\geq 3$ studies in order to yield a proper, integrable
    posterior,\citep{Gelman2006} and it may otherwise generally be
    considered overly
    conservative.\citep{FriedeRoeverWandelNeuenschwander2017a,FriedeRoeverWandelNeuenschwander2017b,Gelman2006}
    Similar problems arise also with the Jeffreys prior for the NNHM
    model;\citep[][Sec.~2.2]{Roever2020} this kind of issue is common
    in Bayesian analysis.\citep{vanDongen2006} Another case where a
    proper, weakly informative prior may be required (not only for few
    studies) is when marginal likelihoods or Bayes factors are of
    interest.

    While the availability of a ``noninformative'' prior comes with a
    certain convenience (one less issue to worry about), in the
    present case its failure to provide reasonable estimates in
    certain instances will often appear somewhat contradictory to
    common sense. The introduction of an informative prior then may
    entail a trade-off of the introduced regularisation versus
    simplicity and robustness.  On the other hand, the explicit
    consideration of relevant prior information may also be seen as an
    advantage.

    From a merely ``technical'' perspective, a heterogeneity prior
    must (in order to ensure integrability of the posterior) have a
    shorter-than-uniform upper tail (an eventually decreasing,
    integrable density function) and also an integrable density
    towards zero.  In that spirit, it may also make sense to consider
    near-origin- and upper-tail-behaviours separately.  While an
    (improper) uniform prior may be considered noninformative for
    several reasons (e.g., due to its scale-invariance
    property\citep[][Sec.~2.2]{Roever2020}), its overly heavy upper
    tail may also be considered
    ``anti-conservative''.\citep{StanHelpWebsite} On the other hand,
    it may be possible to ``rescue'' some of the desirable behaviour
    and robustness e.g. by the use of heavy-tailed
    priors.\citep{OHaganPericchi2012} Besides upper-tail
    considerations, priors may also behave quite differently near
    zero; for example, depending on whether the prior density
    approaches zero, a finite value, or infinity. A finite prior
    density may ensure a near-zero behaviour roughly like a uniform
    prior, while a zero density may be useful e.g. in bounding
    maximum-a-posteriori (MAP) point estimates away from
    zero;\citep{ChungEtAl2013a} in particular from the regularisation
    perspective, the prior density's derivative near zero may also be
    of interest (as it determines how small $\tau$~values may be
    pushed towards or away from zero).

    While the concept of ``weak informativeness'' remains somewhat
    elusive (just like that of a ``noninformative'' prior), the
    information content (or ``vagueness'') of a prior is commonly
    related to its variance, its entropy,\citep{Jaynes1968} or its
    associated effective sample size
    (ESS).\citep{MoritaEtAl2008,NeuenschwanderEtAl2020} In many cases
    it is also helpful to consider the informativeness of a prior
    relative to a reference,\citep{EvansJang2011} for example, a unit
    information
    prior.\citep{NeuenschwanderSchmidli2020,KassWasserman1995} Since
    the posterior draws its interpretation in part from the prior, it
    is important to make the prior specification plausible and
    transparent.  Following the parsimony principle (\emph{Ockham's
      razor}), it may be contructive to seek the (in some sense)
    \emph{simplest} prior distribution within any relevant
    constraints.\citep{Lazar2010} Possible approaches to implement
    such a notion in practice may work, e.g., via maximization of the
    entropy,\citep{Jaynes1968} pre-specification of an effective
    sample size,\citep{MoritaEtAl2008,NeuenschwanderEtAl2020} or
    matching of moments.

    Despite the aim of a weakly informative formulation, one
    should also anticipate the case where the data have little
    information to add, so that the posterior closely resembles the
    prior and hence the analysis results are largely determined by the
    prior settings. This may happen especially in cases of few studies
    and is also suggested in some of the examples that will be
    discussed below (see Figure~\ref{fig:tauposteriors}); such cases
    highlight the importance of a transparent and convincing prior
    specification.

    In the remainder of this section, we aim to facilitate a
    structured approach to interpreting heterogeneity and specifying
    heterogeneity prior distributions by pointing out relevant
    perspectives and highlighting consequences of certain
    heterogeneity settings.  Similar ideas are to some degree also
    utilized in prior elicitation in
    general.\citep{RenOakleyStevens2018,HampsonEtAl2014} A set of
    guiding questions is eventually suggested in
    Table~\ref{tab:questions}.

  \subsection{General properties of the NNHM}\label{sec:nnhmGeneral}
    When considering prior distributions for the heterogeneity~$\tau$,
    it is useful to recall that~$\tau \geq 0$ is a scale parameter,
    and that its square~$\tau^2$ denotes a variance component within
    the NNHM\@.  Immediate associations of variance priors useful in a
    simple normal model however may be misleading: inverse-gamma (or
    inverse-$\chi^2$) distributions are usually not recommended, as
    these arise as conjugate distributions only in related, yet
    distinctly different circumstances.  An inverse-gamma distribution
    is conjugate in the simple case of estimating the variance of a
    normal distribution with known mean.\citep{BDA3rd} In such a case,
    an unequal pair of two data points for example implies that the
    variance must be positive (a zero variance would have a zero
    likelihood); in the present NNHM context, however, unequal
    $y_i$~values may be consistent with zero heterogeneity ($\tau=0$),
    so that such priors are not a natural choice here, and their use
    is generally
    discouraged.\citep{GelmanHill,Gelman2006,PolsonScott2012,CunananEtAl2018}
    Supposedly noninformative settings based on inverse-gamma
    distributions commonly tend to result in sensitivity to
    specification details,\citep{Gelman2006} and often too much
    probability is allocated to very large heterogeneity
    values.\citep{SchmidCarlinWelton2021}

    For uniform or normal effect prior distributions, the resulting
    \emph{conditional} effect posterior $p(\mu|\tau,y)$ again is
    normal.  While for increasing~$\tau$ the (conditional) posterior
    mean of~$\mu$ shifts from the inverse-variance weighted mean
    towards the unweighted average of the estimates~$y_i$, the
    (conditional) posterior variance of~$\mu$ is proportional
    to~$\tau$.\citep{Roever2020} At the same time, larger
    heterogeneity values also imply wider prediction intervals and
    less shrinkage
    \citep{HigginsThompsonSpiegelhalter2009,Roever2020,AdesLuHiggins2005,RoeverFriede2020,RoeverFriede2020b}
    (see also Section~\ref{sec:nnhm}).  Varying $\tau$ between zero
    and infinity essentially also means varying between the extremes
    of \emph{pooled} and \emph{separate} analyses of individual
    studies.  In a sense, overestimation of~$\tau$ may hence often be
    considered a ``conservative'' or ``less harmful'' form of bias.
    In that spirit, one might argue that ---within reasonable
    limits--- a prior that is \emph{stochastically larger} than
    another is also \emph{more
      conservative}.\citep{ShakedShanthikumar} A simple way to
    implement stochastically ordered distribution families is by using
    parametisations that include a scale
    parameter.\citep[][Sec.~VII.6.2]{MGB} Use of a scale parameter
    does not actually impose a restriction; if not already included in
    the parametrisation, it may easily be introduced.  Note that
    simple re-scaling of a prior distribution~$p(\tau)$ then also
    implies a (re)scaling of the corresponding marginal prior
    predictive distributions~$p(\theta_i|\mu)$ by the same factor.  In
    general, stochastically ordered priors also imply the same
    ordering for the resulting
    posteriors.\citep{RoeverFriede2020b,Meczarski2015,BartoszewiczSkolimowska2006}
    Consideration of stochastically ordered alternative priors may
    hence also offer a framework for sensitivity analyses 
    (see also Appendix~\ref{sec:SensitivityAppendix}).

  \subsection{Reasonable (proper) distributional families}\label{sec:priorFamilies}
\begin{figure}[b]
      \centering
      \makebox{\includegraphics[width=0.60\linewidth]{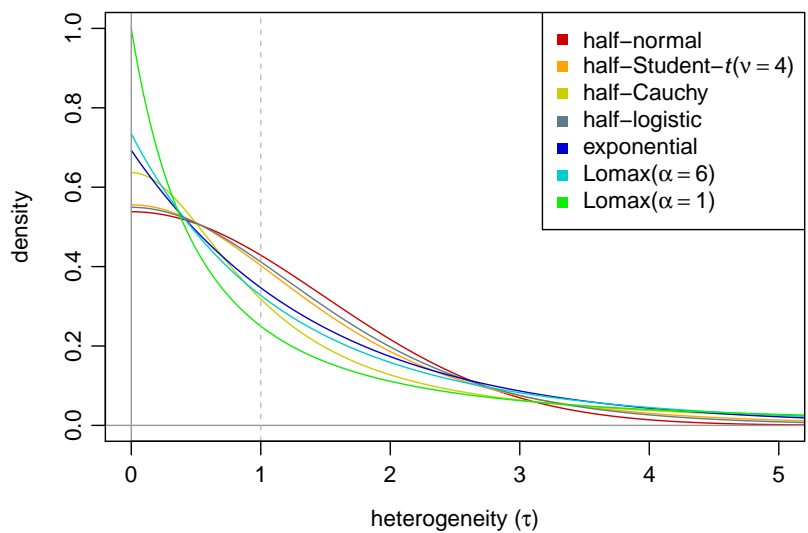}}
      \caption{\label{fig:densities}A selection of potential probability densities for the
        heterogeneity. All distributions are scaled so that their prior
        median is at unity ($\tau\!=\!1$, dashed line; see also Table~\ref{tab:predictive-2}).}
    \end{figure}

    A simple way to implement the ``technical'' requirements (as
    suggested in Section~\ref{sec:aim}) may be to require roughly
    uniform behaviour near zero (implying indifference among small
    heterogeneity values on the $\tau$~scale and ensuring
    integrability in the lower tail), and a monotonically decaying tail
    with increasing heterogeneity values (implying decreasing
    probability for increasing $\tau$ values and ensuring
    integrability in the upper tail).  This may be achieved e.g. by
    using half-normal, half-Student\mbox{-}$t$, half-Cauchy,
    half-logistic, exponential or Lomax distributions. A sample of
    such distributions is sketched in Figure~\ref{fig:densities}.
    Note that for comparability, the distributions in the figure are
    all scaled such that they have a common median of~$1$; their
    corresponding parameters are also listed in
    Table~\ref{tab:predictive-2} below.  In particular, half-normal,
    half-Student\mbox{-}$t$, or half-Cauchy distributions have been
    recommended as appropriate families within the NNHM, also due to
    favourable frequentist
    properties.\citep{GelmanHill,Gelman2006,PolsonScott2012} The
    half-Student\mbox{-}$t$ distribution (including the half-Cauchy as
    a special case, and the half-normal as a limiting case) may be
    derived as conditionally conjugate distributions in an extended
    parametrisation of the NNHM\@.\citep[][Sec.~19.6]{GelmanHill} The
    exponential distribution might be motivated as the \emph{maximum
      entropy} distribution for a pre-specified prior
    expectation,\citep{Jaynes1968} or as the \emph{penalised
      complexity} prior.\citep{KleinKneib2016} The half-logistic
    distribution combines a zero derivative (implying near-uniform
    behaviour) at the origin with an upper tail behaviour close to
    that of an exponential distribution.

    Half-Student\mbox{-}$t$ and Lomax distributions here may be
    considered as heavy-tailed variants of the half-normal and
    exponential distributions, respectively.  In the spirit of a
    \emph{contaminated} prior, encompassing priors ``close to an
    elicited
    one'',\citep{BergerBerliner1986}\citep[][Sec.~3.5.3]{Berger} these
    may also be motivated as scale mixtures, where the (exponential or
    half-normal) scale parameter is associated with some variability
    or uncertainty.  The scale mixture connection is also derived in
    detail in Appendix~\ref{sec:MixtureAppendix} below.  The special
    case of a Lomax($\alpha\!=\!1$) distribution also coincides with
    the form of prior distribution suggested by DuMouchel (a
    log-logistic prior
    for~$\tau$).\citep{DuMouchel1996,DuMouchelNormand2000} Similarly,
    the exponential distribution may also be motivated as a scale
    mixture of a half-normal distribution with Rayleigh-distributed
    scale.  The use of heavy-tailed prior distributions has the
    advantage of ensuring some degree of robustness against prior
    misspecification (or prior/data
    conflict)\citep{OHaganPericchi2012} at the cost of sacrificing
    some of its ``regularisation'' power.  Another simple way of
    implementing some degree of robustness is by combining
    ``informative'' and ``heavy-tailed'' elements in a two-component
    mixture
    distribution.\citep{SchmidliEtAl2014,RoeverWandelFriede2018}

    Another simple and common prior distribution is the (proper)
    bounded uniform distribution defined on an interval~$[0,a]$. It
    inherits certain qualities from the (improper) uniform
    distribution, but it introduces a sharp cutoff at the upper
    bound~$a$, which may be hard to motivate or justify.  Although, if
    the bound is large enough, then it may be very reasonable (e.g. for
    log-ORs).

    Among the above examples, the Student\mbox{-}$t$ and Lomax
    distributions possess ``shape'' parameters in addition to scale
    parameters, which here essentially regulate the degree of
    heavy-tailedness. If considered desirable, more complex prior
    assumptions may be implemented using more complex distributions,
    e.g., using folded non-central Student\mbox{-}$t$ distributions
    with a non-zero mode,\citep{GelmanHill,Gelman2006,PolsonScott2012}
    however, additional degrees of complexity would probably require
    solid justification to be convincing.  In the context of a
    \emph{penalisation} interpretation of the prior, a mode at zero
    also implies a corresponding ``penalty term'' that is
    monotonically increasing in $\tau$; this applies e.g. for a
    penalized-complexity prior\citep{KleinKneib2016} that aims to give
    preference to sparse models.  In empirical investigations based on
    meta-analyses archived in the \emph{Cochrane Database of
      Systematic Reviews}, log-Normal and log-Student\mbox{-}$t_5$
    distributions have been fitted to empirical
    data.\citep{RhodesEtAl2015,TurnerEtAl2015} The log-normal and
    log\mbox{-}$t$ distributions here were found to fit the predictive
    distributions best, however, only few alternatives (log-normal,
    log\mbox{-}$t_5$ and inverse-gamma,\citep{RhodesEtAl2015} or
    log-normal, inverse-gamma and gamma
    distributions\citep{TurnerEtAl2015} for $\tau^2$) were considered
    as candidates in these comparisons.  Some properties of the
    distributions discussed here are also listed in
    Appendix~\ref{sec:PriorTableAppendix}.

    In practice, the half-normal distribution is quite commonly used;
    the reasons for its popularity are probably its simple and
    familiar form, its near-uniform behaviour at the origin along with
    a reasonably quickly decaying upper tail, as well as
    considerations of numerical stability.  In the following, we will
    focus mostly on half-normal distributions. In our experience,
    minor differences between similar prior densities are of rather
    minor practical relevance, while it is most important what
    heterogeneity ranges the bulk of prior probability is assigned to.

    When eventually formulating prior assumptions in terms of a
    parametric prior probability distribution, it is first of all
    necessary to be able to judge the meaning and implications of
    certain heterogeneity settings; these issues will be discussed in
    the following section.

  \subsection{Interpreting heterogeneity values}\label{sec:interpretingTau}
    \subsubsection[Units of tau]{Units of~$\tau$}
      Informative priors naturally \emph{always} need to be considered
      in the context of the endpoint under consideration.  In order to
      specify a sensible prior for~$\tau$, it is important to
      recapitulate its role in the NNHM (see
      Section~\ref{sec:nnhm}). The heterogeneity $\tau$~is a
      \emph{scale parameter} that relates to the probable size of
      \emph{differences} (between-study differences) in effects
      ($\theta_i$ and $\mu$; see equation~\eqref{eqn:NNHM2}).  With
      that, the units of measurements ($y_i$), effects ($\theta_i$,
      $\mu$) and heterogeneity ($\tau$) are the same; if the effect is
      measured, say, in metres, then so is the heterogeneity.  Or both
      may be dimensionless, as e.g. in the case of log-transformed
      ratios (like log-odds-ratios (log-ORs), log-incidence-rate-ratios
      (log-IRRs), log-hazard-ratios (log-HRs),\ldots) or standardized
      mean differences (SMDs).  One may in fact argue that the nature
      of the effect scale is the most important aspect to consider for
      prior specification.\citep{WilliamsEtAl2018} In case the
      effects~$y_i$ have been transformed prior to analysis, then it
      is often useful to consider implications on the back-transformed
      scale.  Transformations are usually introduced to achieve a
      better fit to the normality assumptions within the NNHM; for
      example, using logarithmic or arcsine
      transforms.\citep{HedgesOlkin,HartungKnappSinha,RueckerSchwarzerCarpenterOlkin2009}
      In such cases, also considering the back-transformed
      (exponential or sine) effect scales is often instructive.

      In case the effect scale has definite upper and lower bounds
      (which is often the case e.g. for endpoints measured as scores),
      this also provides information on the plausible (and possible)
      between-study variability. In case of bounded scales, it may for
      example be useful to consider the extreme cases of a continuous
      uniform distribution across the considered range (which would
      have standard
      deviation~$\frac{b-a}{\sqrt{12}}=\frac{b-a}{3.46}$, where~$a$
      and~$b$ are the lower and upper bounds, respectively), or a
      discrete distribution with probabilities of~$\frac{1}{2}$
      concentrated at both margins~$a$ and~$b$ (which would have
      standard deviation~$\frac{b-a}{2}$).  Such considerations may
      define absolute ``worst-case'' settings for the
      heterogeneity. Any normal approximation employed on a bounded
      parameter space with a standard deviation of, say,
      $>\frac{b-a}{4}$ would inevitably have substantial overlap with
      out-of-domain values; any heterogeneity value that is not $\ll
      \frac{b-a}{4}$ should raise suspicion and might actually call
      for a different approach (e.g., transformation to a different
      parameter space).

    \subsubsection{Magnitudes of other effects}\label{sec:effectMagnitude}
      Relevant hints may originate from considering the magnitude of
      other (known or plausible) effects of interventions or
      covariates.  The reasonable range for the overall mean
      effect~$\mu$ may also have implications for the expected range
      of study-specific means~$\theta_i$; in case an informative prior
      for~$\mu$ is used (or is at least plausible), its variance may
      help constraining also the between-trial variability.  
      Heterogeneity may often be attributed to differences in the
      composition of the populations underlying each estimate, and the
      distribution of relevant covariates within (which may be
      observed or unobserved).
      If the observed heterogeneity is assumed to be due to different
      constitutions of populations, then the heterogeneity relates to
      accumulated effects of associated covariates.
      With that, within- and between-study variability in effects are
      related to within- and between-study differences among subjects
      and the plausible magnitude of covariates' effects.
      For example, if a treatment effect is known to differ between
      males and females by a certain amount, this difference between
      genders may help judging or motivating plausible magnitudes of
      effect differences between studies.
      In case the variability between centers within the same study
      has been investigated, this may also provide a hint on
      between-study variability (which will then most likely be
      larger).

    \subsubsection{Implications of a fixed heterogeneity value}\label{sec:fixedTau}
      Specific values of the heterogeneity~$\tau$ may be judged and
      compared based on the implied distribution of true
      effects~$\theta_i$, which is given by the \emph{(conditional)
        prior predictive distribution} $p(\theta_i|\mu,\tau)$ (see
      equation~\eqref{eqn:NNHM2}), where $\tau$ defines the
      distribution's standard deviation.  The effects~$\theta_i$
      (conditional on~$\mu$) then vary within a range of $\mu \pm
      1.96\tau$ with 95\% probability. For a randomly picked pair of
      effects ($\theta_i$ and $\theta_j$), their difference
      ($\theta_i-\theta_j$) follows a
      $\normaldistn(0,2\tau^2)$-distribution \eqref{eqn:NNHM2}, and
      their absolute difference $|\theta_i-\theta_j|$ then has a
      median of $0.95\tau$.  Quite commonly, the effects~$\theta_i$
      are transformed prior to analysis, so that it may be helpful to
      consider the implications on the back-transformed scale. A very
      common example is the logarithmic transformation, which is often
      used for analyses involving e.g. odds ratios (ORs), relative
      risks (RRs) or hazard ratios (HRs), and where the inverse
      transform is the exponential function. 95\%~predictive intervals
      and median differences are shown for a range of $\tau$~values in
      Table~\ref{tab:conditional} along with the corresponding
      exponentiated figures.

      \begin{table}[b] 
        \caption{\label{tab:conditional}Implications of certain
          \emph{fixed} heterogeneity values~$\tau$ on the probable
          ranges of true effects~$\theta_i$ (\emph{conditional} prior
          predictive distributions) and the corresponding
          exponentiated ranges (the latter are relevant for
          log-transformed effect scales).}  
        \centering
        \begin{tabular}{ccccc}
          \toprule
                 & \multicolumn{2}{c}{95\% predictive interval}
                 & \multicolumn{2}{c}{random pair $|\theta_i-\theta_j|$} \\
          \cmidrule(lr){2-3}
          \cmidrule(lr){4-5}
          $\tau$ & $\theta_i-\mu$ & $\exp(\theta_i-\mu)$ 
                 & median & $\exp(\mbox{median})$ \\
          \midrule
          0.1 & [-0.20, 0.20] & [0.82, 1.22]  & 0.10 & 1.10 \\ 
          0.2 & [-0.39, 0.39] & [0.68, 1.48]  & 0.19 & 1.21 \\ 
          0.5 & [-0.98, 0.98] & [0.38, 2.66]  & 0.48 & 1.61 \\ 
          1.0 & [-1.96, 1.96] & [0.14, 7.10]  & 0.95 & 2.60 \\ 
          2.0 & [-3.92, 3.92] & [0.020, 50.4] & 1.91 & 6.74 \\ 
          \bottomrule
        \end{tabular}
      \end{table}

      An extensive discussion of these conditional distributions is
      given in Spiegelhalter \emph{et~al.}
      (2004).\citep[][Sec.~5.7]{SpiegelhalterEtAl} By working out what
      \emph{range} of $\theta_i$~values is expected, or what
      \emph{difference} between a randomly picked pair of
      $\theta_i$~values is expected, corresponding plausible ranges of
      $\tau$~values may be determined.  Based on such considerations,
      Spiegelhalter \emph{et~al.}  (2004)\citep{SpiegelhalterEtAl}
      categorized ranges of $\tau$ values in the context of log-ORs as
      ``reasonable'', ``fairly high'' or ``fairly extreme'' as shown
      in Table~\ref{tab:SpiegelhalterCategories}.
      \begin{table} 
        \caption{\label{tab:SpiegelhalterCategories}Categories of
          heterogeneity and corresponding $\tau$~ranges in the context
          of log-ORs, according to Spiegelhalter
          \emph{et~al.} (2004).\citep[][Sec.~5.7]{SpiegelhalterEtAl}}
        \centering
        \begin{tabular}{lc}
          \toprule
          \multicolumn{1}{c}{category} & \multicolumn{1}{c}{range}\\
          \midrule
          ``reasonable''     & $0.1 < \tau < 0.5$\\
          ``fairly high''    & $0.5 < \tau < 1.0$\\
          ``fairly extreme'' & $\tau > 1.0$\\
          \bottomrule
        \end{tabular}
      \end{table}
      Such investigations may help judging what $\tau$ values are
      reasonable or unrealistic and with that may help specifying
      e.g. the heterogeneity prior's tail quantiles.

      For example, Prevost \emph{et~al.}
      (2000)\citep[][Sec.~4]{PrevostEtAl2000} aimed to constrain the
      predictive interval ($\exp(\theta_i-\mu)$) to a range of
      $[0.5,\,2.0]$, which is achieved for $\tau=0.35$.  Considering
      this range as extreme and unlikely, a half-Normal prior with
      scale~$0.18$ (implying $\prob(\tau\leq0.35)=0.95$) was
      eventually suggested for a log-RR\@.  \textsf{R}~code to
      illustrate these arguments using Monte Carlo sampling and exact
      calculations is provided in Appendix~\ref{sec:PrevostRCode}.

    \subsubsection{Implications of a heterogeneity distribution}\label{sec:randomTau}
      Besides considering the \emph{conditional} distribution for
      fixed $\tau$~values ($p(\theta_i|\mu,\tau)$, see previous
      subsection), one may also investigate the \emph{marginal} prior
      predictive distribution $p(\theta_i|\mu)$, marginalized over a
      particular heterogeneity prior, which technically results as the
      integral
      $p(\theta_i|\mu) = \int_0^\infty p(\theta_i|\mu,\tau)\,p(\tau)\,\differential\tau$.
      Since $p(\theta_i|\mu,\tau)$ is normal \eqref{eqn:NNHM2}, the
      marginal $p(\theta_i|\mu)$ is a \emph{normal (scale) mixture}
      distribution. Its form may usually either be derived
      numerically,\citep{RoeverFriede2017,bayesmeta,Roever2020} or it
      may easily be explored using \emph{collapsed Gibbs sampling},
      that is, generating a Monte Carlo sample by repeatedly sampling
      from the heterogeneity prior ($p(\tau)$), and subseqently from
      the conditional predictive distribution ($p(\theta_i|\tau)$).
      Investigating the marginal prior predictive distribution may
      help judging the prior scale or distributional family.

      \begin{table}[b] 
        \caption{\label{tab:predictive-1}Implications of a range of
          half-normal heterogeneity priors~$p(\tau)$ on probable
          values of heterogeneity~$\tau$ and predicted
          effects~$\theta_i$ (\emph{marginal} prior predictive
          distributions). The three rightmost columns show the
          corresponding probabilities for the three categories from
          Table~\ref{tab:SpiegelhalterCategories}.}  
        \centering
        \begin{tabular}{cccccccccc}
          \toprule
                     & \multicolumn{3}{c}{heterogeneity~$\tau$} & \multicolumn{2}{c}{95\% predictive interval} & \multicolumn{3}{c}{category probability (\%)}\\ 
           \cmidrule(lr){2-4} \cmidrule(lr){5-6} \cmidrule(lr){7-9}
           $p(\tau)$ & median & mean & 95\% quant. & $\theta_i-\mu$ & $\exp(\theta_i-\mu)$ & \scalebox{0.8}{\shortstack{reason-\\able}} & \scalebox{0.8}{\shortstack{fairly\\high}} & \scalebox{0.8}{\shortstack{fairly\\extreme}} \\ 
          \midrule
          half-normal(0.1) & 0.07 & 0.08 & 0.20 & [-0.22, 0.22] & [0.80, 1.24] & 32 & \phantom{0}0 & \phantom{0}0 \\ 
          half-normal(0.2) & 0.13 & 0.16 & 0.39 & [-0.44, 0.44] & [0.65, 1.55] & 60 & \phantom{0}1 & \phantom{0}0 \\ 
          half-normal(0.5) & 0.34 & 0.40 & 0.98 & [-1.09, 1.09] & [0.34, 2.98] & 52 & 27 & \phantom{0}5 \\ 
          half-normal(1.0) & 0.67 & 0.80 & 1.96 & [-2.18, 2.18] & [0.11,  8.89] & 30 & 30 & 32 \\ 
          half-normal(2.0) & 1.35 & 1.60 & 3.92 & [-4.37, 4.37] & [0.013, 79.0] & 16 & 19 & 62 \\ 
          \bottomrule
        \end{tabular}
      \end{table}

      Table~\ref{tab:predictive-1} illustrates a range of prior
      predictive distributions for a set of half-normal priors that
      differ in their scale.  The implied probabilities for the
      (log-OR) categories shown in
      Table~\ref{tab:SpiegelhalterCategories} are also given.  Note
      that a simple re-scaling of the heterogeneity prior implies
      proportional scaling of mean and quantiles for $\tau$ as well as
      $\theta_i$ (as can be seen in Table~\ref{tab:predictive-1}).  In
      this spirit, Dias \emph{et~al.} (2013)\citep{DiasEtAl2013} for
      example proposed a half-normal(0.32)-prior for a log-OR based on
      the implied prediction interval for $\exp(\theta_i-\mu)$ of
      $[0.5,\,2.0]$.  \textsf{R}~code to illustrate these arguments
      using Monte Carlo sampling and exact calculations is provided in
      Appendix~\ref{sec:DiasRCode}

      \begin{table}
        \caption{\label{tab:predictive-2}Implications of a range of
          heterogeneity priors~$p(\tau)$ from different families on
          probable values of of heterogeneity~$\tau$ and predicted
          effects~$\theta_i$ (\emph{marginal} prior predictive
          distributions). For comparability, the different priors are
          all scaled to a common median of~$1.0$\@. Except for the
          exponential distribution, which is commonly parameterized by
          its \emph{rate} (or inverse scale), all distributions have a
          scale parameter.}  
        \centering
        \begin{tabular}{lccccccc}
          \toprule
                    && \multicolumn{3}{c}{heterogeneity~$\tau$} & \multicolumn{2}{c}{95\% predictive interval} \\ 
          \cmidrule(lr){3-5} \cmidrule(lr){6-7}
          \multicolumn{1}{c}{$p(\tau)$} & scale & median & mean & 95\% quant. & $\theta_i-\mu$ & $\exp(\theta_i-\mu)$ \\ 
          \midrule
          half-normal(1.48)      & 1.48 & 1.00 & 1.18 & \phantom{0}2.91 & [-3.24, 3.24] & [0.039, 25.5] \\ 
          half-Student-$t_{\nu=4}$(1.35) & 1.35 & 1.00 & 1.28 & \phantom{0}3.75 & [-3.85, 3.85] & [0.021, 46.8] \\ 
          half-Cauchy(1.00)      & 1.00 & 1.00 &      & 12.7\phantom{1} & [-10.10, 10.10] & [0.000$\,$041, 24$\,$371] \\ 
          half-logistic(0.91)    & 0.91 & 1.00 & 1.26 & \phantom{0}3.33 & [-3.55, 3.55] & [0.029, 34.7] \\ 
          exponential(0.69)      & 1.44 & 1.00 & 1.44 & \phantom{0}4.32 & [-4.33, 4.33] & [0.013, 75.9] \\ 
          Lomax$_{\alpha=6}$(8.17) & 8.17 & 1.00 & 1.63 & \phantom{0}5.29 & [-5.04, 5.04] & [0.0065, 155] \\
          Lomax$_{\alpha=1}$(1.00) & 1.00 & 1.00 &      & 19.0\phantom{0} & [-14.74, 14.74] & [0.000$\,$000$\,$40, 2$\,$520$\,$157] \\
          \bottomrule
        \end{tabular}
      \end{table}

      Similarly, Table~\ref{tab:predictive-2} illustrates a range of
      prior predictive distributions for a set of heterogeneity priors
      from different distributional families; what they have in common
      is the prior median of~$1.0$ for~$\tau$. Quantiles or mean
      of~$\tau$ or $\theta_i$ for other scalings of $p(\tau)$ may be
      derived by proportional re-scaling (as in
      Table~\ref{tab:predictive-1}). For example, a half-Cauchy
      distribution that has its median heterogeneity matched to that
      of a half-normal distribution requires a scale parameter that is
      smaller by a factor of $\approx 2/3$. From the table, one can
      also read off the ratio of 95\%~quantile over the median, which
      may be a useful indicator of the heavy-tailedness of the
      different distribution families.  The distributions from
      Table~\ref{tab:predictive-2} are also illustrated in
      Figure~\ref{fig:densities}.  Some additional properties of these
      distributions are provided in
      Appendix~\ref{sec:PriorTableAppendix}.

      Different distributional families for the prior~$p(\tau)$ imply
      differing marginal prior predictive
      distributions~$p(\theta_i|\mu,\tau)$. Concrete prior information
      on $p(\theta_i|\mu,\tau)$ then may help constraining the shape
      of~$p(\tau)$, however, the prior family may also be selected
      based on considerations of heavy-tailedness, near-zero
      behaviour, or simplicity.

    \subsubsection{The role of the unit information standard deviation (UISD)}\label{sec:sigmau}
\begin{figure}[t]
        \centering
        \makebox[0.45\linewidth][c]{(a) $\;\tau \ll \sigmau$}
                 \hspace{0.05\linewidth}
        \makebox[0.45\linewidth][c]{(b) $\;\tau \gg \sigmau$} \\
        \makebox{\includegraphics[width=0.45\linewidth]{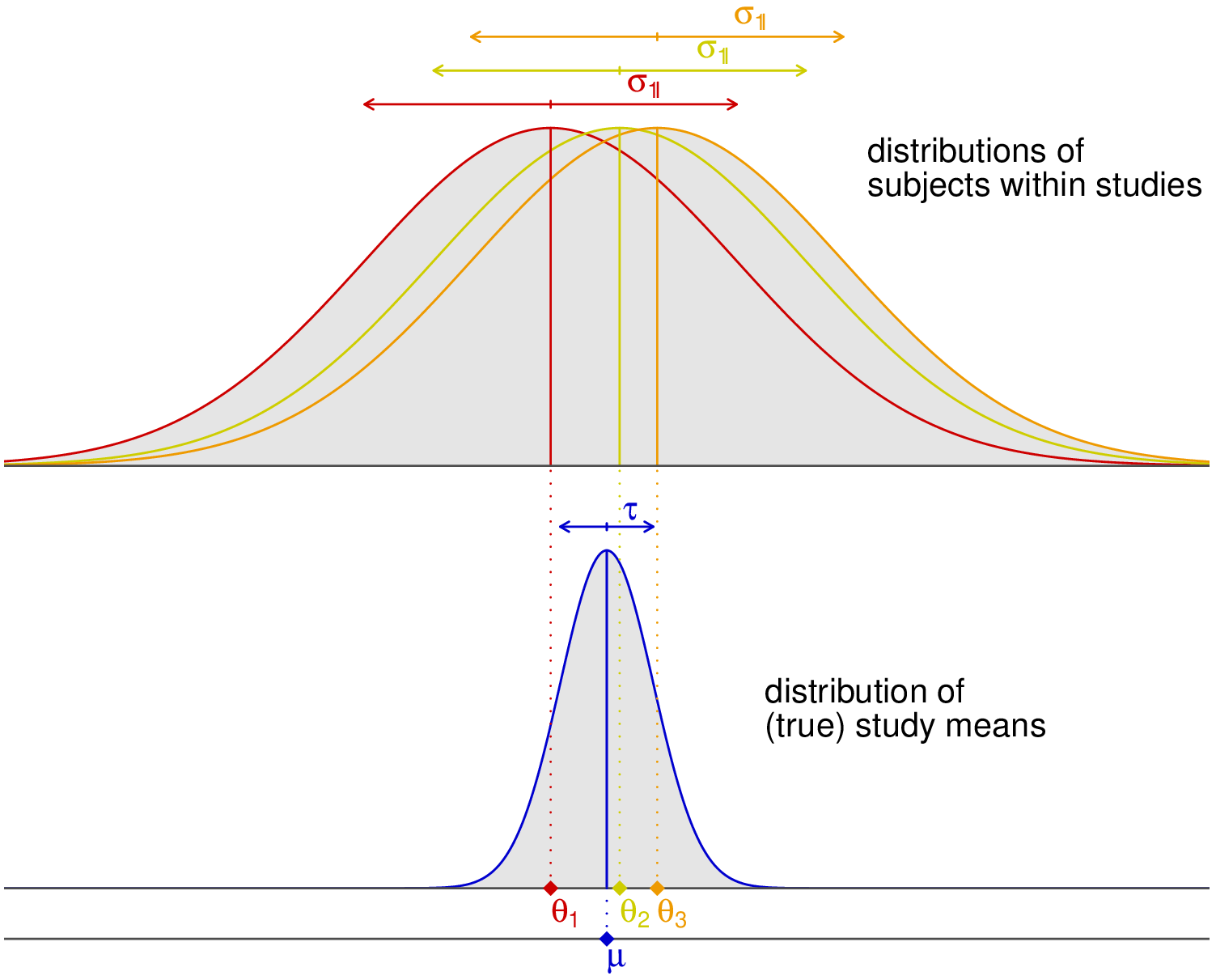}
                 \hspace{0.05\linewidth}
                 \includegraphics[width=0.45\linewidth]{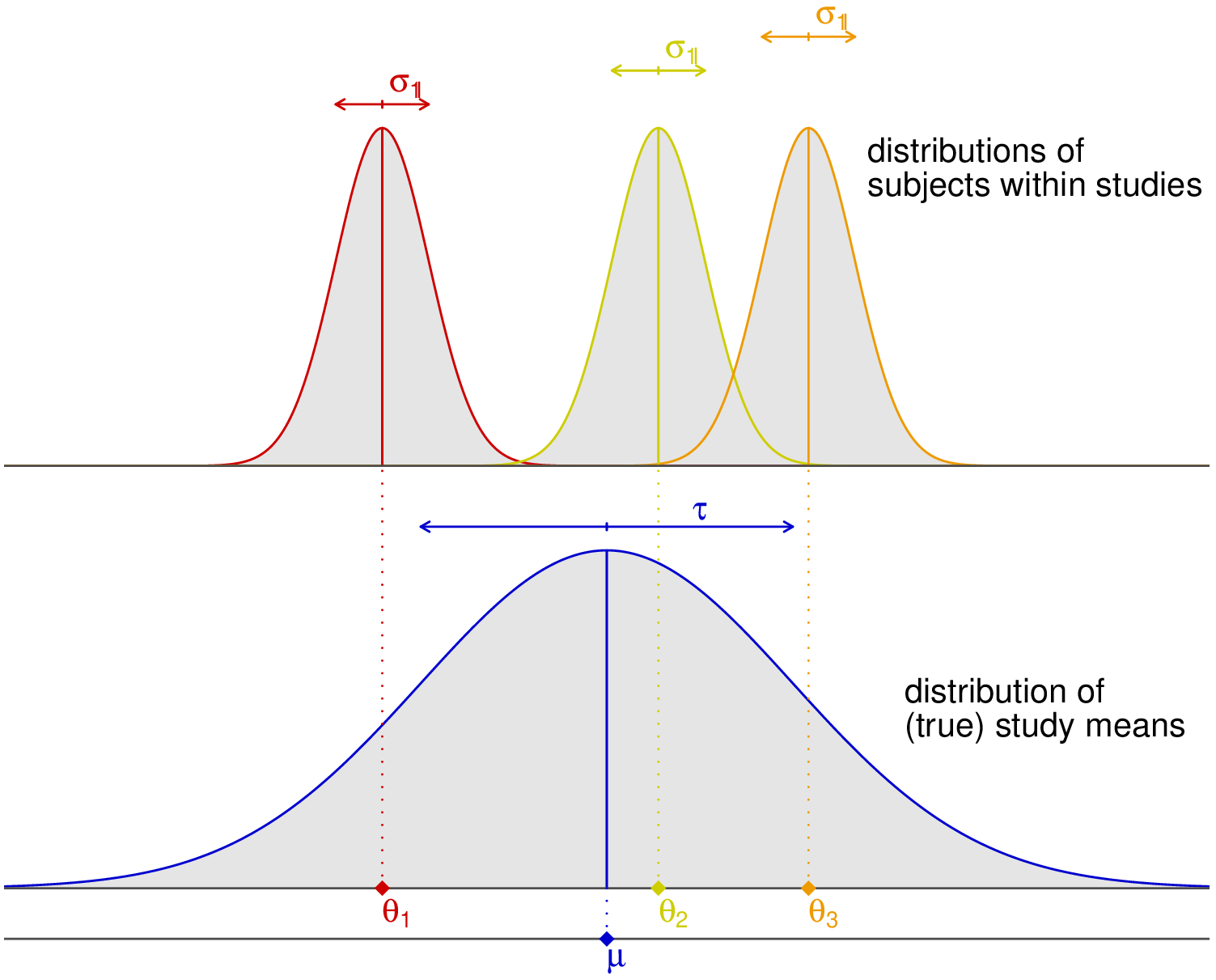}}
        \caption{\label{fig:unitInfo}Illustration of the relationship
          of between-study heterogeneity~$\tau$ and unit information
          standard deviation (UISD)~$\sigmau$.  The left panel~(a)
          shows the commonly expected setup, in which the
          heterogeneity~$\tau$ is relatively small compared to the
          within-study standard deviation ($\tau \!\ll\! \sigmau$).
          The right panel~(b) shows that a larger~$\tau$ would imply
          that the distributions of subjects from different studies
          were eventually barely overlapping.  Note that the eventual
          \emph{estimates} ($y_i$) resulting from the different
          studies then may have different standard
          errors~$\sigma_i=\frac{\sigmau}{\sqrt{n_i}}<\sigmau$
          associated, depending on the studies' sample sizes~$n_i$.  }
      \end{figure}

      Consider the simple case of an effect measure that for each
      study is determined as an average of independent identically
      distributed observations. In such a case, the associated
      standard error is simply of the form
      \begin{equation}\label{eqn:unitInfoStdev}
        \sigma_i \;=\; \frac{\sigmau}{\sqrt{n_i}}\mbox{,}
      \end{equation}
      where $n_i$~is the sample size, and $\sigmau$~is the common
      ``population'' standard deviation of each single observation
      that was averaged over.  This figure describes the
      \emph{population-}, or \emph{within-study-standard
        deviation},\citep{KassWasserman1995} which for the moment we
      take to be constant across studies.  This figure is also called
      the \emph{unit information standard deviation (UISD)}, as it
      relates to an observational unit's contribution to a study's
      likelihood.  One may now relate the heterogeneity~$\tau$ to
      $\sigmau$ and ask whether the between-study variability~($\tau$)
      is likely to exceed the within-study variability~($\sigmau$), or
      what ratios of these two are plausible.
      Figure~\ref{fig:unitInfo} illustrates the relationship of
      within-study and between-study standard deviations~$\sigmau$
      and~$\tau$. Usually, one would expect $\tau\!\ll\!\sigmau$,
      implying that while study means ($\theta_i$) may differ to some
      degree, the distributions of subjects within studies will still
      be largely overlapping (see Figure~\ref{fig:unitInfo}, left
      panel).  In that sense, the UISD~$\sigmau$ may constitute an
      important ``landmark'' on the heterogeneity continuum and thus
      may help constraining the range of plausible heterogeneity
      values.\citep{NeuenschwanderSchmidli2020}

      This concept of within-study standard deviation may be extended
      to other types of effect scales --- for example, the standard
      error of a log-OR derived from a $2 \times 2$-table is
      approximately given by $\sigma_i = \frac{4}{\sqrt{n_i}}$, so
      that, \emph{heuristically}, the UISD here equals~$\sigmau\!=\!4$
      per subject (at least).\citep[][Appendix~A.1]{Roever2020}
      Sometimes it may also make more sense to define UISDs not
      \emph{per subject} but rather \emph{per event} (see also
      Appendix~\ref{sec:UnitInfoIRR} for an example), but care also
      needs to be taken in order not to confuse these two figures.
      For a given set of log-OR estimates, the UISD may alternatively
      also be investigated by inverting
      equation~\ref{eqn:unitInfoStdev} (see also \eqref{eqn:su} and
      the examples in Section~\ref{sec:LOGexample} below).

      Another link may be drawn between~$\sigmau$ and~$\tau$ via
      shrinkage estimation (see Section~\ref{sec:nnhm}) and the
      consideration of \emph{prior effective sample
        sizes}.\citep{NeuenschwanderEtAl2020,NeuenschwanderEtAl2010}
      Consider the case where a meta-analysis of $k$~studies is
      available, and a new ($k\!+\!1$th) study is conducted. The
      previous meta-analysis of course provides (prior) information on
      the new study's estimate~$\theta_{k+1}$, the exact amount of
      which is determined by the number of studies~$k$, their sample
      sizes~$n_i$, the UISD~$\sigmau$, but also by the amount of
      heterogeneity.\citep{WandelNeuenschwanderRoeverFriede2017,SchmidliEtAl2014}
      If $\tau$~is large, then separate studies are only loosely
      related and the previous data add little information. If on the
      other hand $\tau$~is very small (i.e., studies are almost
      homogeneous), then they may contribute a lot of information.
      With that, the amount of heterogeneity is related to whether
      studies should rather be pooled or viewed as essentially
      independent pieces of information.  One may then consider the
      idealized limiting case of infinitely many
      ($k\rightarrow\infty$) infinitely large ($n_i\rightarrow\infty$)
      studies as the previous data source, so that the amount of
      contributed information solely depends on~$\tau$.  In that case,
      the historical data may be thought of as effectively
      contributing a number of $n^\star_\infty$ additional subjects to
      the $k\!+\!1$th study.  This \emph{prior maximum sample size}
      then relates to~$\sigmau$ and~$\tau$
      as\citep{NeuenschwanderEtAl2010}
      \begin{equation}\label{eqn:pmss}
        \frac{\tau}{\sigmau} \; = \; \frac{1}{\sqrt{n^\star_\infty}}\mbox{.}
      \end{equation}
      Table~\ref{tab:pmss} illustrates this
      relationship.
      \begin{table}[t]
        \caption{\label{tab:pmss}Correspondence between prior maximum
          sample sizes~($n^\star_\infty$) and the magnitude of the
          heterogeneity~($\tau$) relative to the unit information
          standard deviation (UISD)~($\sigmau$)
          (see~\eqref{eqn:pmss}).\citep{NeuenschwanderEtAl2010}}
        \centering
        \begin{tabular}{lccccccc}
          \toprule
          $\tau / \sigmau \quad$ & 0 & 1/16 & 1/8 & 1/4 & 1/2 & 1 & $\infty$ \\
          \midrule
          $n^\star_\infty$          & $\infty$ & 256 & 64 & 16 & 4 & 1 & 0 \\
          \bottomrule
        \end{tabular}
      \end{table}
      For example, if in the ideal case (i.e., $k=\infty$,
      $n_i=\infty$) the additional data should add information
      equivalent to \emph{at most} 16~subjects, then this would
      correspond to $\tau$~amounting to \emph{at most} a quarter
      of~$\sigmau$.
      If one has an idea of how much information a meta-analysis may
      (or should) contribute to a single study's shrinkage estimate
      (in the idealized case of very many very large studies), then
      such considerations may help constraining probable magnitudes
      of~$\tau$, or associating probabilities with ranges of
      $\tau$~values.

      Note that a number of priors have been proposed which are
      defined relative to the magnitude of the $\sigma_i$~values (or
      their harmonic mean), e.g., the \emph{Jeffreys},
      \emph{DuMouchel} or \emph{uniform shrinkage}
      priors.\citep[][Sec.~2.2]{Roever2020} In view of the above
      arguments, it might also make sense to define priors relative to
      the UISD, or its estimated
      value. Inverting~\eqref{eqn:unitInfoStdev} yields $\sigmau =
      \sqrt{n_i\,\sigma_i^2\,}$ for a single study, and based on a
      given data set we suggest the more general empirical estimate
      \begin{equation}\label{eqn:su}
        \su \;=\; \sqrt{\bar{n} \; \bar{s}_\mathrm{h}^2\,}
            \;=\; \sqrt{\frac{\sum_{i=1}^kn_i}{\sum_{i=1}^k\sigma_i^{-2}}}
      \end{equation}
      where $\bar{n}=\frac{1}{k}\sum_{i=1}^kn_i$ is the average
      (arithmetic mean) sample size, and $\bar{s}_\mathrm{h}^2 =
      \bigl(\frac{1}{k}\sum_{i=1}^k\sigma_i^{-2}\bigr)^{-1}$ is the
      harmonic mean of the squared standard errors (variances). This
      estimator is defined so that in the special case of a
      common-effect analysis (i.e., assuming $\tau=0$), the overall
      mean estimate's variance (which then is given by
      $\bigl(\sum_{i=1}^k\sigma_i^{-2}\bigr)^{-1}$) consistently also
      equals $\frac{\su^2}{\sum_i n_i}$.

    \subsubsection[Empirical information on tau]{Empirical information on~$\tau$}
      Empirical data, e.g. from earlier investigations in a related
      area,\citep{HigginsWhitehead1996} may also contribute to
      a-priori information.  Informative priors based on empirical
      information have been derived for standardized mean differences
      (SMDs) and log-ORs in medical applications by investigating
      large numbers of meta-analyses published in the \emph{Cochrane
        Database of Systematic Reviews} by Rhodes \emph{et~al.}
      (2015)\citep{RhodesEtAl2015} and Turner \emph{et~al.}
      (2015).\citep{TurnerEtAl2015} Additional evidence for certain
      types of effect scales may be found e.g.\ in the works by
      Pullenayegum (2011),\citep{Pullenayegum2011} Turner
      \emph{et~al.} (2012),\citep{TurnerEtAl2012} Kontopantelis
      \emph{et~al.} (2013),\citep{KontopantelisSpringateReeves2013}
      Steel \emph{et~al.} (2015),\citep{SteelEtAl2015}
      van~Erp \emph{et~al.} (2017),\citep{vanErpEtAl2017} Seide
      \emph{et~al.} (2019),\citep{SeideEtAl2018b,SeideEtAl2019} and
      G\"{u}nhan \emph{et~al.} (2020).\citep{GunhanRoeverFriede2020}
      Note that some references provide information directly on the
      heterogeneity \emph{parameter}, while others summarize
      \emph{estimates} of heterogeneity.

      Empirical information often entails the question of how
      representative the external information is for the study at
      hand, or what may be the relevant data subset, or what to do if
      no such sample may be available.  In terms of the
      \emph{epistemic} view discussed in Section~\ref{sec:priorViews},
      the inclusion of empirical evidence in the prior specification
      affects the interpretation of the prior, and with that, of the
      posterior.  Empirical data may then often be seen as a somewhat
      complementary source of evidence.  When there is doubt about the
      immediate applicability of empirical information for the problem
      at hand, this may also be reflected e.g. in a robustified
      two-component mixture
      prior.\citep{SchmidliEtAl2014,RoeverWandelFriede2018}

  \subsection{Guiding questions}
    In order to summarize the above arguments,
    Table~\ref{tab:questions} lists some guiding questions that may
    aid in structuring the specification of a prior for the
    heterogeneity. These are mostly based on the arguments laid out in
    Sections~\ref{sec:priorFamilies} and~\ref{sec:interpretingTau}.
    Firstly, plausible heterogeneity magnitudes (in terms of~$\tau$
    or~$\theta_i$ ranges) need to be be determined. These reflections
    may then also help choosing a parametric family for the prior, or
    the distributional family may also be selected based on
    considerations of near-zero behaviour, heavy-tailedness or
    simplicity.  Beyond the mere type of endpoint or effect measure,
    the context also may determine whether smaller or larger amounts
    of heterogeneity are to be expected, e.g., depending on whether
    studies' designs and populations were similar.  Special
    considerations in the context of specific common types of effect
    scales are discussed in detail in
    Section~\ref{sec:specialcases}. These are then illustrated using
    actual data examples in Section~\ref{sec:examples}.

    \begin{table}[h]
      \caption{\label{tab:questions}Some guiding questions for judging
        reasonable prior distributions for the heterogeneity
        parameter~$\tau$.}  
      \centering
      \begin{tabular}{ll}
        \toprule
        \multicolumn{2}{l}{\emph{Prior information:}}\\
        (i)    & \parbox[t]{0.8\linewidth}{What is the effect scale, what \emph{(between-study) differences} are expected or plausible?}\\
        (ii)   & \parbox[t]{0.8\linewidth}{What is the magnitude of other known (or plausible) effects? Do these provide guidance?\\[-0.2ex] Is an informative effect prior used? If so, what is its variance? Does it provide guidance?}\\
        (iii)  & \parbox[t]{0.8\linewidth}{Is a plausible ``unit information standard deviation (UISD)'' available? Does it provide guidance?}\\
        (iv)   & \parbox[t]{0.8\linewidth}{Is relevant external empirical information on heterogeneity available? Should it be considered\\ in the analysis?}\\
        \midrule
        \multicolumn{2}{l}{\emph{Translation into a prior probability distribution:}}\\
        (v)    & \parbox[t]{0.8\linewidth}{Does the prior information help pinpointing prior quantiles (of~$\tau$)?}\\
        (vi)   & \parbox[t]{0.8\linewidth}{Does the prior information help pinpointing prior predictive quantiles (of~$\theta_i$)?}\\
        (vii)  & \parbox[t]{0.8\linewidth}{Does the prior information suggest particular properties for the prior (-density)?\\(Monotonicity? A non-zero mode? A heavy tail? Certain near-zero behaviour? \ldots)}\\
        \bottomrule
      \end{tabular}
    \end{table}

\section{Motivating heterogeneity priors in various settings}\label{sec:specialcases}
  \subsection{Means and mean differences}\label{sec:specialcase:MD}
    This general case covers endpoints measured on \emph{absolute}
    scales, hence it is not possible to give universally applicable
    advice on a plausible prior scale.  For example, the same analysis
    may require different scalings of the prior depending on whether
    an endpoint is expressed, say, in terms of hours or minutes.  In
    particular, in case of effects that are defined as averages, the
    UISD (see also Section~\ref{sec:sigmau}) may provide some
    guidance; if standard errors~$\sigma_i$ scale with sample size
    ($\sigma_i\approx\frac{\sigmau}{\sqrt{n_i}}$, see also
    equation~\eqref{eqn:unitInfoStdev}), then $\sigmau$ (or an
    estimate~$\su$, \eqref{eqn:su}) may provide some orientation based
    on the considered (or other related) data.  Relating effects to
    ``within-population standard deviations'' is actually an approach
    that is also formalized in the case of \emph{standardized} mean
    differences (SMDs); see the following section.

    \emph{Mean differences} are another very common special
    case. These are often used in order to ``normalize'' outcomes; for
    example, in controlled clinical trials, each study's
    \emph{treatment} group is usually related to a \emph{control}
    group in order to express the treatment effect \emph{relative to
      the unexposed group}.  In the simplest case, the study's outcome
    then is defined as $y_i = \bar{x}_{2;i} - \bar{x}_{1;i}$,
    where~$\bar{x}_{1;i}$ and~$\bar{x}_{2;i}$ are the $i$th study's
    averages from control and treatment group, respectively. When
    considering UISDs, the relevant sample size will then result as
    the sum of the two treatment groups' sizes
    ($n_i=n_{1;i}+n_{2;i}$).  In the simple case of two equally-sized
    groups ($n_{1;i}=n_{2;i}=\frac{n_i}{2}$) and equal variances
    within groups (so that
    $\var(\bar{x}_{1;i})=\var(\bar{x}_{2;i})=\frac{\sigmaw^2}{n_i/2}$)
    the UISD simply results as $\sigmau = \sqrt{2 \sigmaw^2}$, where
    $\sigmaw^2$ is the within-group variance.

    Again a special case arises when considering \emph{paired
      differences}.\citep{HsuLachenbruch2005} In general, analogous
    considerations apply for un-paired as well as for paired
    differences; only for the latter case the UISD~$\sigmau$ may be
    expressed as $\sigmau^2 = \var(x_{1;ij}) + \var(x_{2;ij}) -
    2\cov(x_{1;ij},\, x_{2;ij})$ where $j$~is the index identifying
    the $j$th pair of observations in the $i$th study.  We can see how
    the individual (paired) observation's variance contribution
    results as a sum of the two observations' marginal variances and
    their covariance.  Now, since any pair of observations ($y_{1;ij}$
    and~$y_{2;ij}$) is usually positively correlated
    ($\cov(y_{1;ij},\,y_{2;ij})>0$), the sum of individual variances
    ($\var(x_{1;ij}) + \var(x_{2;ij})$), if known, may provide an
    upper bound on $\sigmau$.

    Finally, there are generic cases of parameter estimates that are
    reported along with a standard error, but which do not necessarily
    have a ``sample size'' ($n_i$) associated, as is sometimes the
    case, e.g., for laboratory experiments.\citep{BakerJackson2013}

  \subsection{Standardized mean differences}\label{sec:specialcase:SMD}
    Standardized mean differences (SMDs) aim to compare mean
    differences measured on different scales by normalizing them
    through their population standard deviation. Effectively, these
    measure \emph{by how many standard deviations} the two study
    groups differ; SMDs are always dimensionless. Their aim is to
    estimate $\delta_i = \frac{\mu_{2;i}-\mu_{1;i}}{\varsigma_i}$,
    where~$\mu_{2;i}$ and~$\mu_{1;i}$ are the two groups' true means
    and $\varsigma_i$~is the within-group standard deviation (which
    may be defined with respect to one or the other or both treatment
    groups, or which may also be externally informed).  Note that
    $\varsigma_i$ here bears some similarity to the UISD~$\sigmau$
    (when considering the latter with respect to the
    \emph{unstandardized} differences).  Slightly differing, but
    essentially similar approaches are given e.g. by the
    ``Cohen's~$d$'', ``Hedges'~$g$'' or ``Glass'~$\Delta$''
    estimators, which differ in details like bias correction or
    standardization terms.\citep{HedgesOlkin,HartungKnappSinha}
    Essentially, these aim to estimate the mean difference
    ($\mu_{2;i}-\mu_{1;i}$) by the difference of averages
    ($\bar{x}_{2;i} - \bar{x}_{1;i}$), and also the standard deviation
    by an empirical one.  SMDs (along with the correlations treated
    below) are somewhat different here from the ``general'' mean
    differences, in that they are explicitly designed and utilized in
    order to compare endpoints measured on different scales, which are
    not \emph{directly} comparable. A heterogeneity of $\tau=0$ may
    hence be considered particularly unlikely.  A value of
    $\tau\!=\!1$ would mean that the between-study heterogeneity
    (among $\delta_i$~values) was equal to the within-group
    variability~$\varsigma_i$.
    Closely related to SMDs are \emph{standardized regression
      coefficients}, which are re-scaled as if both the regressor's as
    well as the response's variance were normalized to
    unity.\citep{Menard2004} Similar arguments would apply for
    analyses involving standardized regression coefficients, and
    arguments applicable to correlation coefficients (see
    Section~\ref{sec:specialcase:correlation} below) may also be
    relevant.

    Effects on the SMD scale have been categorized as
    0.2=``small'',
    0.5=``medium'',
    0.8=``large'',\citep[][Sec.~2.2.3]{Cohen}
    where an extension has recently been proposed to include the grades of
    0.1=``very small'',
    1.2=``very large'', and
    2.0=``huge''.\citep{Sawilowsky2009}
    Consequently, such a ranking might be utilized in order to bound
    between-study effects to mostly non-extreme values, e.g. by
    anticipating mostly up to ``large'' heterogeneity and hence
    formulating a bound on $\prob(\tau \leq 1)$.  Neglecting
    estimation uncertainty for the denominator, and for simplicity
    assuming equal sample sizes for each of the $i$th study's groups,
    leads to a UISD of $\sigmau=2$ (see
    Appendix~\ref{sec:UnitInfoSMD}).

    Empirical evidence on heterogeneities between SMDs based on an
    analysis of studies archived in the \emph{Cochrane Database of
      Systematic Reviews} is given by Rhodes \emph{et~al.}
    (2015);\citep{RhodesEtAl2015} for a general healthcare setting
    (not restricted to a particular outcome type), a
    log-Student\mbox{-}$t$ distribution with
    parameters~$\mu\!=\!-1.72$, $\sigma\!=\!1.295$, and $5$~degrees of
    freedom was derived (implying a median and 95\%~quantile of~0.18
    and~2.43, respectively).  Heterogeneity \emph{estimates} reported
    in studies published in the \emph{Psychological Bulletin} are
    provided by van~Erp \emph{et~al.} (2017);\citep{vanErpEtAl2017}
    the 189 $\tau$-estimates for SMDs that were quoted in
    32~publications had a median and 95\%~quantile of~0.20 and~0.66,
    respectively.

  \subsection{Log-transformed odds, rates and effect scales}\label{sec:specialcase:logtrafo}
    Many outcomes are commonly analyzed on a logarithmic scale, which
    may be advantageous for several reasons; firstly, the domain of
    positive numbers is mapped to the complete real line, which makes
    strictly positive scales tractable for normal models like the
    NNHM, which is often convenient.  Secondly, additive effects on
    the log-scale translate to multiplicative effects on the original
    scale.  Symmetry of the normal distribution \eqref{eqn:NNHM2} on
    the log-scale then implies a ``symmetric'' treatment of
    multiplicative factors and their inverses (since
    $\exp(\mu+x)=\exp(\mu)\times\exp(x)$ while
    $\exp(\mu-x)=\exp(\mu)\times\frac{1}{\exp(x)}$).  This is useful,
    e.g, when dealing with outcomes like rates, odds, rate ratios,
    odds ratios, relative risks, hazard ratios or concentration
    measurements.  An offset of, say,~$0.1$ on the log-scale
    translates (approximately) to a change of~$10\%$ on the
    back-transformed (exponentiated) scale, regardless of the original
    value.  Thirdly, the normal approximation to the likelihood that
    is used in the NNHM \eqref{eqn:NNHM1} may provide a better fit on
    the logarithmic scale.

    When considering heterogeneity values on the logarithmic scale, a
    more intuitive approach is usually to examine the corresponding
    implications on the back-transformed scale.  Note that a normal
    model on the log-scale actually corresponds to a log-normal model
    on the original scale.  In a sense, an analysis on the logarithmic
    scale may also be viewed as an implementation of a dependent joint
    prior for effect and
    heterogeneity\citep{Senn2007,Pullenayegum2011} on the original
    (exponentiated) scale.  The consequences of certain heterogeneity
    values or heterogeneity distributions were already investigated in
    some detail in Sections~\ref{sec:fixedTau}
    and~\ref{sec:randomTau}; the important issue to judge is what
    \emph{relative} (multiplicative) difference between studies is
    deemed plausible; see also the extensive discussion by
    Spiegelhalter \emph{et~al.}
    (2004).\citep[][Sec.~5.7]{SpiegelhalterEtAl}

    A common type of effect are log-transformed odds (or
    \emph{logits}).\citep{Viechtbauer2010,TrikalinosEtAl2013} For
    example, in epidemiology or at the design stage of a clinical
    trial it may be of interest to infer the magnitude and variability
    of the prevalence of a certain condition, or historical
    information may be utilized to support the control group in a
    clinical trial.\citep{SchmidliEtAl2014} The prevalence may be
    expressed in terms of the probablity~$p \in [0,1]$ or the
    odds~$\frac{p}{1-p}\in[0,\infty]$, while for meta-analysis
    purposes it then makes sense to move to the log-odds
    scale~$\log\bigl(\frac{p}{1-p}\bigr)\in\realline$.  Rather than
    viewing this as a case of a logarithmic transformation of the
    odds, one might as well consider this as a \emph{logit}
    transformation of probabilities, mapping the interval [0,1] to the
    real line via the \emph{logit} function
    $f(p)=\log\bigl(\frac{p}{1-p}\bigr)$.  Besides considerations of
    what ratios the odds may plausibly be spanning, here it may be
    helpful to consider a uniform distribution in proportions as an
    extreme case; for the log-odds, this implies a logistic
    distribution that has a standard deviation
    of~$\frac{\pi}{\sqrt{3}}=1.81$. The UISD in this case amounts to
    (at least) $\sigmau=2$ (see Appendix~\ref{sec:UnitInfoLogit}).
    Similarly, event rates (based on a Poisson model) are commonly
    combined in meta-analyses based on a log-transformation.

    Similarly to the cases of means and mean differences discussed
    earlier, a log-transform is also commonly applied in the context
    of two-group comparisons, for example, for log-OR, log-IRR, log-RR
    or log-HR effect measures.  Logarithmic ORs are a natural
    extension of the log-odds case above, since the logarithmic
    \emph{ratio} of odds is simply a \emph{difference} of log-odds;
    other pairwise group comparisons generalize similarly from
    single-group estimates.  UISDs for log-ORs and log-RRs are derived
    in R\"{o}ver (2020),\citep{Roever2020} and for log-IRRs in
    Appendix~\ref{sec:UnitInfoIRR}; the corresponding figures for
    log-HRs are discussed by Spiegelhalter \emph{et~al.}
    (2004).\citep[][Sec.~2.4.2]{SpiegelhalterEtAl} When discussing
    UISDs for count outcomes, it is important to clearly indicate
    whether these relate to \emph{subjects} or \emph{events} (e.g.,
    for ORs the numbers are $4$~per subject\citep{Roever2020} and
    $2$~per event\citep{SpiegelhalterEtAl}).

    Empirical evidence on the magnitude of heterogeneities within
    meta-analyses published in the \emph{Cochrane Database of
      Systematic Reviews} is given by Turner \emph{et~al.}
    (2015).\citep{TurnerEtAl2015,TurnerEtAl2012} For example, for a
    log-OR effect in a general healthcare setting (without restricting
    to a specific type of outcome), a log-normal distribution with
    $\mu\!=\!-1.28$ and $\sigma\!=\!0.87$ was derived, implying a
    median and 95\%~quantile of~0.28 and~1.16, respectively (see also
    Table~\ref{tab:predictive-1}).  Similarly, G\"{u}nhan
    \emph{et~al.}  (2020)\citep{GunhanRoeverFriede2020} in a
    re-analysis of data from the Cochrane Database of Systematic
    Reviews determined a 95\% quantile of heterogeneity
    \emph{estimates} of~$1.05$ for analyses based on binary data and
    log-ORs.

\begin{figure}
      \centering
      \makebox{\includegraphics[width=0.80\linewidth]{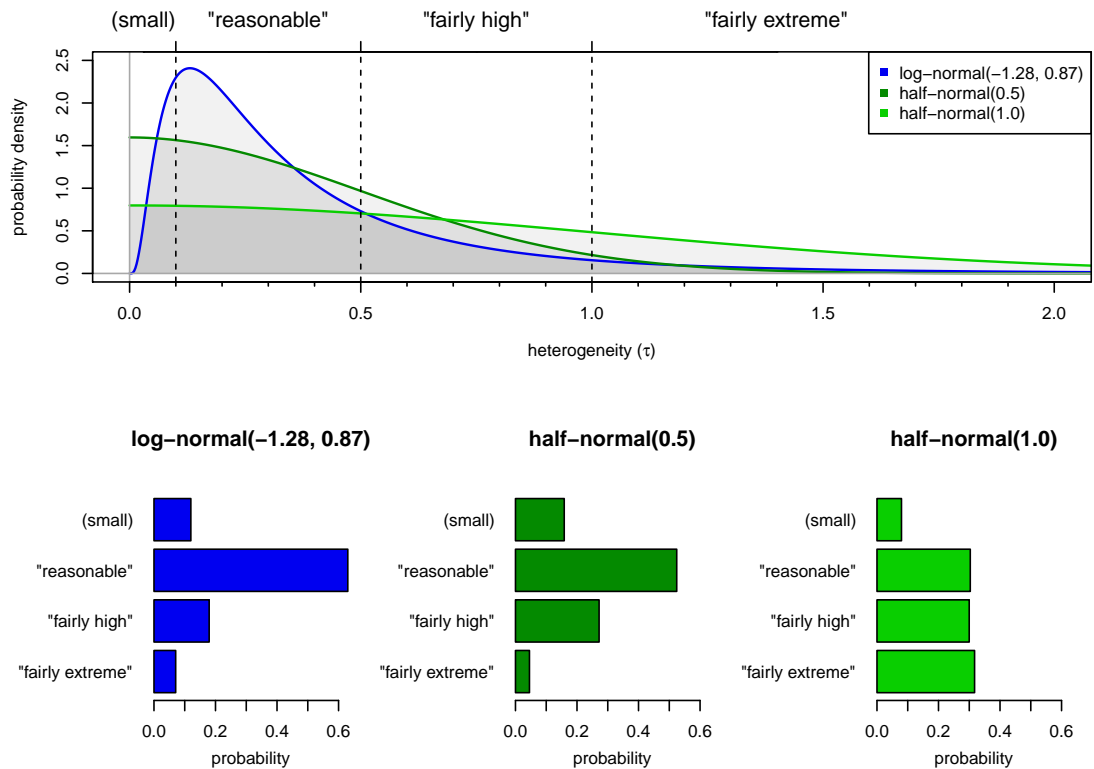}}
      \caption{\label{fig:TurnerPrior} Comparison of the heterogeneity
        prior proposed by Turner \emph{et~al.}
        (2015)\citep{TurnerEtAl2015} for log-ORs in a general setting
        (a log-normal distribution with $\mu\!=\!-1.28$ and
        $\sigma\!=\!0.87$, shown in blue) with half-normal priors
        (with scales~0.5 and~1.0). The bottom plots especially
        contrast the implied prior probabilities for the heterogeneity
        categories proposed by Spiegelhalter \emph{et~al.}
        (2004)\citep[][Sec.~5.7]{SpiegelhalterEtAl} (see also
        Tables~\ref{tab:SpiegelhalterCategories}
        and~\ref{tab:predictive-1}).}
    \end{figure}

    Consider for example the common case of a meta-analysis of log-OR
    estimates.  If we want to restrict prior probabilities mostly to
    ``reasonable'' to ``fairly high'' heterogeneity levels (according
    to Table~\ref{tab:SpiegelhalterCategories} in
    Section~\ref{sec:fixedTau}), one could use a half-normal prior
    with scale~$0.5$, implying $\prob(\tau> 1.0) = 4.6\%$ and
    assigning $52\%$ and $27\%$ probability to the ``reasonable'' and
    ``fairly high'' categories, respectively.
    Figure~\ref{fig:TurnerPrior} illustrates the half-normal(0.5)
    prior along a half-normal(1.0) prior, and the prior proposed by
    Turner \emph{et~al.} (2015)\citep{TurnerEtAl2015} (log-normal
    with $\mu\!=\!-1.28$ and $\sigma\!=\!0.87$).
    The heterogeneity categories from
    Table~\ref{tab:SpiegelhalterCategories} are marked, and at the
    bottom, the probabilities for the categories are shown. The
    probabilities assigned by the half-normal(0.5) prior and the
    ``empirical'' prior are roughly in agreement, while the
    half-normal(1.0) prior would assign more or less equal
    probabilities to the ``reasonable'', ``fairly high'' and ``fairly
    extreme'' categories, and leave only $8\%$ probability for smaller
    values.  Similar arguments hold also for other log-transformed
    effect scales.

  \subsection{Regression slopes}\label{sec:specialcase:slopes}
    Very closely related to mean differences is the more general case
    of meta-analysis of regression parameters (slopes or interactions)
    and their standard errors.\citep{BeckerWu2007} In the special case
    of a single binary covariate, the regression effectively reduces
    to a two-group comparison, and consideration of additional
    covariates then may allow for some ``adjustment''.  When the
    covariate is continuous, however, extra care needs to be taken,
    since not only the endpoint's scaling is relevant (the
    regression's ``$y$~variable''), but also the regressor's scaling
    (the regression's ``$x$~variable''). Whether the regressor is
    expressed in, say, days or weeks, affects the resulting slope
    parameter (and its standard error) by a corresponding re-scaling
    by a factor of seven. The regressor's scaling will then similarly
    also affect the scale of the anticipated heterogeneity: when
    combining estimated (linear) regression coefficients, which are to
    be interpreted as \emph{``the expected change in~$y$ for a
      one-unit change in~$x$''}, the heterogeneity between estimates
    depends on the units of~$x$. For example, the variability expected
    among temporal changes that are expressed on a \emph{per-week}
    scale rather than a \emph{per-day} scale should be seven times as
    large.

    The immediate question then is what increment in the regressor to
    base heterogeneity considerations on; what is eventually needed is
    a statement of the form \emph{``for a change in the regressor by a
      difference of~$\Delta_x$, the associated effects are anticipated
      to vary by a magnitude of~$\tau$''}, and that
    difference~$\Delta_x$ needs to be specified. Sometimes there may
    be obvious ``natural'' units to be used, for example in the common
    case of a binary (zero/one) coded covariate (e.g. for treatment
    vs. control or males vs. females); the obvious difference to
    consider here is an increment of~$\Delta_x=1$.  Otherwise the
    width of the regressor's distribution may be
    relevant.\citep{GelmanEtAl2008} Consider again the case of a
    binary covariate and a balanced setup; the standard deviation of
    the binary variable will then be~$\frac{1}{2}$, so that
    \emph{twice the standard deviation} might generally be a sensible
    scale to consider. Note though that this is by no means
    universally applicable, as such scales may be affected by many
    factors (e.g., inclusion criteria in clinical trials) and might
    also be very different between studies.
    Note that the $\Delta_x$~value needs to be the same across the
    considered studies.

    Once the ``reference'' increment~$\Delta_x$ has been determined, a
    prior for the associated heterogeneity may be formulated. In case
    the actual analysis then is done with respect to a differing
    scaling, the prior needs to be re-scaled accordingly. For example,
    if a prior with scale~$\scale$ was determined for a
    \emph{per-week} increment, but the actual analysis is based on the
    \emph{per-day} regression coefficients, then their prior should
    have scale~$\frac{\scale}{7}$. The UISD~$\sigmau$ then also scales
    proportionally.

    Note that the above arguments extend beyond simple linear
    regressions with continuous outcomes, for example, logistic
    regressions, Poisson regressions or survival analyses, in which
    regression parameters then relate to log-ORs, log-IRRs or log-HRs.
    Once a reference increment~$\Delta_x$ has been determined, the
    arguments regarding log-transformed endpoints discussed earlier in
    Section~\ref{sec:specialcase:logtrafo} apply, and potential
    re-scaling issues still need to be considered.  A way to
    circumvent considerations of regressor's or response's scales may
    be to move to \emph{standardized regression coefficients} instead,
    which are unitless and are somewhat similar to SMDs (see also
    Section~\ref{sec:specialcase:SMD}) or correlations (see
    Section~\ref{sec:specialcase:correlation}).\citep{Menard2004}. Depending
    on the exact type of regression analysis and the standardization
    technique (e.g., in case of a logistic regression, and when
    standardization is done based only on the regressor's
    scale),\citep{NewmanBrowner1991,GreenlandEtAl1991,Bring1994}
    arguments relevant for log-transformed endpoints might also apply.

  \subsection{Correlation coefficients}\label{sec:specialcase:correlation}
    Estimated correlation coefficients (Pearson's~$r$) are commonly
    quoted and summarized for studies dealing with paired
    observations.\citep{HedgesOlkin,HartungKnappSinha,Schulze}
    Correlation coefficients are restricted to the domain $[-1,\,1]$,
    with values of $|r| = 1$ indicating perfectly linear (positive or
    negative) correlation, and $r = 0$ indicating
    uncorrelatedness.\citep{Armitage2005}
    Due to the problems with bounded parameter spaces, correlation
    coefficients are commonly analyzed after an appropriate
    transformation using \emph{Fisher's~$z$ transform}, which is
    defined as
    $z_i=\frac{1}{2}\log\Bigl(\frac{1+r_i}{1-r_i}\Bigr)=\mathrm{arctanh}(r_i)$.
    This transformation maps the original domain to the real line, and
    in particular, it is also a \emph{variance stabilizing
      transformation}; the (approximate) standard error of the
    transformed $z_i$~value only depends on the $i$th study's sample
    size~$n_i$ and is given by~$\frac{1}{\sqrt{n_i-3}}$.  Correlation
    values within the range $-0.5 < r_i < 0.5$ are little affected by
    the transformation, which makes more of a difference for more
    extreme values.

    An upper limit to the expected heterogeneity may be specified by
    considering a uniform distribution of $\theta_i$ values across the
    range of correlation coefficients as a ``worst case''. For plain
    (correlation~$r$) values, this would imply a variance of
    $\frac{1}{3}=0.58^2$. On the scale of $z$-transformed values, this
    implies a distribution with probability density function
    $p(z)=\frac{2}{(\exp(-z)+exp(z))^2}$, that has a zero mean and a
    variance of $\frac{\pi^2}{12} \approx 0.91^2$ (these moments might
    actually motivate a prior for the overall effect~$\mu$, too).  The
    standard error of $z_i$~values after transformation (see above)
    implies a UISD of approximately~$\sigmau=1.0$.  With that, it
    should usually be safe to expect heterogeneity values well below
    $\tau=1.0$.

    If $\tau$ values near unity (or~$0.91$) already imply rather
    extreme heterogeneity, the question remains what constitutes
    ``large'', yet reasonable heterogeneity.  For that, we may
    consider the somewhat more moderate cases of
    $r\sim\unifdistn(-0.5,0.5)$ or $r\sim\unifdistn(0.0,0.8)$.  Both
    these cases happen to lead to similar variances of
    $\var(z)=0.30^2$ on the transformed scale, so that $\tau=0.30$ may
    already be considered ``large'' heterogeneity.

    While the use of ``plain'', un-transformed correlation values
    within the NNHM framework is a bit problematic due to the bounded
    parameter space that is not reflected in the model, it is not
    uncommon. We have already seen some hints of what amounts of
    between-study variance for plain correlations may be possible or
    plausible in the considerations above; a value of
    $\tau=\frac{1}{\sqrt{3}}=0.58$ (corresponding to a uniform
    distribution in~$r$) would already be extreme; one would most
    likely expect values way less than even half as much.

    Van~Erp \emph{et~al.} (2017)\citep{vanErpEtAl2017} collected
    heterogeneity \emph{estimates} reported in studies that were
    published in the \emph{Psychological Bulletin}. Although the
    figures were not identified as being based on Fisher-$z$
    transformation or not (apparently a mix of both was encountered),
    these numbers may provide some empirical motivation.  Among the
    observed heterogeneity estimates for correlation endpoints in
    539~analyses from 25~studies, a median and 95\%~quantile of 0.12
    and 0.29, respectively, were found.
    Similarly, Steel \emph{et~al.} (2015)\citep{SteelEtAl2015} quote
    heterogeneity estimates from 292 management-related meta-analyses
    in the range of~0.0 to~0.4, with a median of~0.16.

\section{Example applications}\label{sec:examples}
  \subsection{Mean differences}\label{sec:MDexample}
    Grande \emph{et~al.} (2015)\citep[Analysis~1.5]{GrandeEtAl2015}
    investigated the effect of physical exercise (vs.\ no~exercise as
    control) on the duration of acute respiratory infections (ARIs).
    Four studies were jointly considered in a meta-analysis, the
    endpoint of interest was the mean difference in the \emph{number
      of symptom days per episode}.  The relevant data are shown in
    Table~\ref{tab:example1}.

    \begin{table}[b] 
      \caption{\label{tab:example1}Mean difference (MD) example data
        due to Grande \emph{et~al.} (2015).\citep{GrandeEtAl2015}
        $\bar{x}$, $s$ and $n$ denote the treatment and control
        groups' empirical means, standard deviations and sample
        sizes. The $y_i$~are the derived MDs and $\sigma_i$ the
        associated standard errors that eventually go into the
        analysis (see Section~\ref{sec:nnhm}).  Here, mean differences
        are on the scale of days (change in disease
        duration). Negative estimates~$y_i$ indicate a beneficial
        effect.}  \centering
      \begin{tabular}{clcccccccc}
        \toprule
              && \multicolumn{3}{c}{treatment group}
               & \multicolumn{3}{c}{control group}
               & \multicolumn{2}{c}{MD} \\
        \cmidrule(lr){3-5}
        \cmidrule(lr){6-8}
        \cmidrule(lr){9-10}
        $i$ & study & $\bar{x}_{1;i}$ & $s_{1;i}$ & $n_{1;i}$ & $\bar{x}_{2;i}$ & $s_{2;i}$ & $n_{2;i}$
        & $y_i$ & $\sigma_i$\\
        \midrule
        1 & Nieman (1990)          & 3.60 & 2.97 & 18  &  \phantom{0}7.00 & 5.94 & 18  &  -3.40 & 1.57 \\
        2 & \c{C}ilo\u{g}lu (2005) & 5.15 & 1.56 & 60  &  \phantom{0}6.10 & 1.00 & 30  &  -0.95 & 0.27 \\
        3 & Barrett (2012)         & 9.30 & 5.13 & 47  &  11.40 & 5.75 & 51  &  -2.10 & 1.10 \\
        4 & Sloan (2013)           & 5.30 & 1.50 & 16  &  \phantom{0}6.30 & 2.20 & 16  &  -1.00 & 0.67 \\
        \bottomrule
      \end{tabular}
    \end{table}

    The outcome here is measured in units of \emph{days} (change in
    symptom duration for treated patients relative to the control
    group).  For the purpose of the present analysis, ARIs were
    defined as ``infections of the respiratory tract that last for
    less than 30~days'',\citep{GrandeEtAl2015} while ARI durations
    generally are substantially shorter, lasting of the order of a
    week.\citep{DelMar2000,ErsWhiteBookCh18} With that, the reduction
    in symptom days cannot be more than (roughly) a week. ARIs may be
    caused by bacterial or viral pathogens; the effect of antibiotic
    treatment is in a shortening of the order of one
    day.\citep{SandersDoustDelMar2008} From the data
    (Table~\ref{tab:example1}), we can derive estimates of the UISD,
    which here is at an average of $\su=3.9$.

    The treatment effect may be expected to be of the order of days
    (anything below 1~day would probably not be considered clinically
    meaningful), and a similar magnitude may be expected for the
    heterogeneity.  Values $\tau > 1$ would make the between-study
    heterogeneity larger than the effect of antibiotics, which seems
    implausible. Variations in treatment effects of the order of
    several days would probably imply that the effect was several
    times larger in some studies than in others.

    A $\tau$~value of $1.0$ would imply a median difference in true
    effects of $\approx 1$~day for a random pair of studies (see
    Table~\ref{tab:conditional}), which might be at the upper end of
    the plausible range.  A half-normal(0.5)~prior would imply
    $\prob(\tau \leq 1) \approx 95\%$, and considering the
    corresponding prior predictive distribution (see
    Table~\ref{tab:predictive-1}), we can see that this implies a 95\%
    prior predictive interval of roughly $\pm 1$~day around the
    overall mean effect.

    For the present example, we would hence suggest a half-normal(0.5)
    prior.  Note that this is a common, well-researched condition. For
    more uncertain cases, one might want to go for a heavier-tailed
    prior.  A meta-analysis based on the half-normal(0.5)~prior is
    illustrated in Figure~\ref{fig:forest01}.  Among the four studies
    considered, one suggests a stronger effect than the others,
    however, due to its relatively small size and correspondingly
    large associated standard error, it is still consistent with the
    remaining three.
    The estimated heterogeneity (the median and 95\% credible interval
    (CI) are shown in the bottom left of the forest plot) here has
    barely changed from the a~priori anticipated amount (see
    Table~\ref{tab:predictive-1}). The heterogeneity's posterior is
    also illustrated in Figure~\ref{fig:tauposteriors}; prior and
    posterior are very similar in this case.
    The resulting combined estimate then also suggests a more moderate
    effect, namely, a reduction of the order of one symptom day, with
    an uncertainty of about a factor of two.  The estimated
    heterogeneity is relatively low compared to the width of the
    overall mean's CI, and so the prediction interval is only slightly
    longer, and the shrinkage intervals show substantially greater
    precision than the original estimates.
    Sensitivity to other prior choices is also investigated for this example in Appendix~\ref{sec:SensitivityAppendix}.

  \subsection{Standardized mean differences}\label{sec:SMDexample}
    Aalbers \emph{et~al.} (2017)\citep[Analysis~1.1]{AalbersEtAl2017}
    investigated the short-term effect of music therapy on depression
    symptoms; four studies comparing music therapy plus
    treatment-as-usual (TAU) versus TAU alone were found. Within these
    four studies, differing clinician-rated symptom scores were
    utilized in order to quantify depression severity: the Hamilton
    rating scale for depression (\mbox{HAM-D}), considering
    potentially differing numbers of items between studies, as well as
    the Montgomery-{\AA}sberg depression rating scale (MADRS). In
    order to facilitate a joint analysis, the meta-analysis was based
    on SMDs (here: Hedges'~$g$); the relevant data are shown in
    Table~\ref{tab:example2}.

    \begin{table} 
      \caption{\label{tab:example2}Standardized mean difference (SMD)
        example data due to Aalbers \emph{et~al.}
        (2017).\citep{AalbersEtAl2017} $\bar{x}$, $s$ and $n$ denote
        the treatment and control groups' empirical means, standard
        deviations and sample sizes. The $y_i$~are the derived SMDs
        and $\sigma_i$ the associated standard errors that eventually
        go into the analysis (see Section~\ref{sec:nnhm}). The
        original data are based on different depression symptom scores
        that are measured on different scales. Negative
        estimates~$y_i$ indicate a reduction in symptom severity.}
      \centering
      \begin{tabular}{clcccccccc}
        \toprule
              && \multicolumn{3}{c}{treatment group}
               & \multicolumn{3}{c}{control group}
               & \multicolumn{2}{c}{SMD} \\
        \cmidrule(lr){3-5}
        \cmidrule(lr){6-8}
        \cmidrule(lr){9-10}
        $i$ & study & $\bar{x}_{1;i}$ & $s_{1;i}$ & $n_{1;i}$ & $\bar{x}_{2;i}$ & $s_{2;i}$ & $n_{2;i}$
        & $y_i$ & $\sigma_i$\\
        \midrule
        1 & Chen (1992)        & -98.23 & 15.19 & 34  &  -67.06 & 15.19 & 34  &  -2.03 & 0.30 \\
        2 & Radulovic (1996)   & -16.50 & 10.00 & 30  &  -10.60 & 10.00 & 30  &  -0.58 & 0.26 \\
        3 & Albornoz (2011)    & \phantom{0}-8.17 &  \phantom{0}5.89 & 12  &   \phantom{0}-3.83 &  \phantom{0}5.31 & 12  &  -0.75 & 0.42 \\
        4 & Erkkil\"{a} (2011) & -10.70 & \phantom{0}8.40 & 30  &   \phantom{0}-6.05 &  \phantom{0}8.06 & 37  &  -0.56 & 0.25 \\
        \bottomrule
      \end{tabular}
    \end{table}

    The outcome measured on the SMD scale means that a unit change in
    $y_i$ corresponds to a \emph{one standard deviation change} in the
    symptom severity score.  Considering e.g. the Albornoz (1992)
    study,\citep{Albornoz2011} which was measuring change in symptom
    severity using the \emph{17-item \mbox{HAM-D}} scale with a
    within-group standard deviation of about~5 (see
    Table~\ref{tab:example2}), a difference of~1 on the SMD scale here
    would roughly correspond to a 5-point change in \mbox{HAM-D}
    score.\citep{Hamilton1960,KristonWolff2011,FurukawaEtAl2007,MassonTejani2013}
    In terms of SMD, this would already be considered a ``large''
    effect.\citep{Cohen,Sawilowsky2009} The UISD for SMDs is predicted
    at $\sigmau=2$, while from the present data here we get a very
    similar empirical average of $\su=2.2$.

    For the between-study differences, we would assume that they would
    be mostly in the ``small'' to ``medium'' range ($\ll 1$) ---
    otherwise effects would be differing by a standard deviation or
    more between studies, and also the studies' confidence intervals
    (which are roughly of the size
    $\sigma_i\approx\frac{\sigmau}{\sqrt{n_i}}= \frac{2}{\sqrt{n_i}}$)
    would be unlikely to have any overlap.  Rhodes \emph{et~al.}
    (2015)\citep{RhodesEtAl2015} in their empirical investigation
    based on the \emph{Cochrane Database of Systematic Reviews}
    predicted a median and 95\%~quantile of~0.18 and~2.43 for the
    heterogeneity~$\tau$ (where the large upper quantile appears
    rather extreme, based on the above arguments).  Similarly, van~Erp
    \emph{et~al} (2017)\citep{vanErpEtAl2017} inferred a median and
    95\%~quantile of~0.20 and~0.66, respectively, based on
    heterogeneity \emph{estimates} within a smaller data base.

    A value of $\tau=1.0$ would imply a median difference of $\approx
    0.95$ (``large'') for a random pair of true study means~$\theta_i$
    (see Table~\ref{tab:conditional}), which already appears like a
    rather extreme amount; values of $\tau=0.5$ (implying mostly
    ``medium'' sized between-study differences) or below seem to be
    more plausible.  A half-normal(0.5) prior would cover this range
    and would imply a prior median (for~$\tau$) slightly above the
    magnitude suggested the empirical investigations (see also
    Table~\ref{tab:predictive-1}).
  
    \begin{figure}[t]
      \centering
      \makebox{\includegraphics[width=0.47\linewidth]{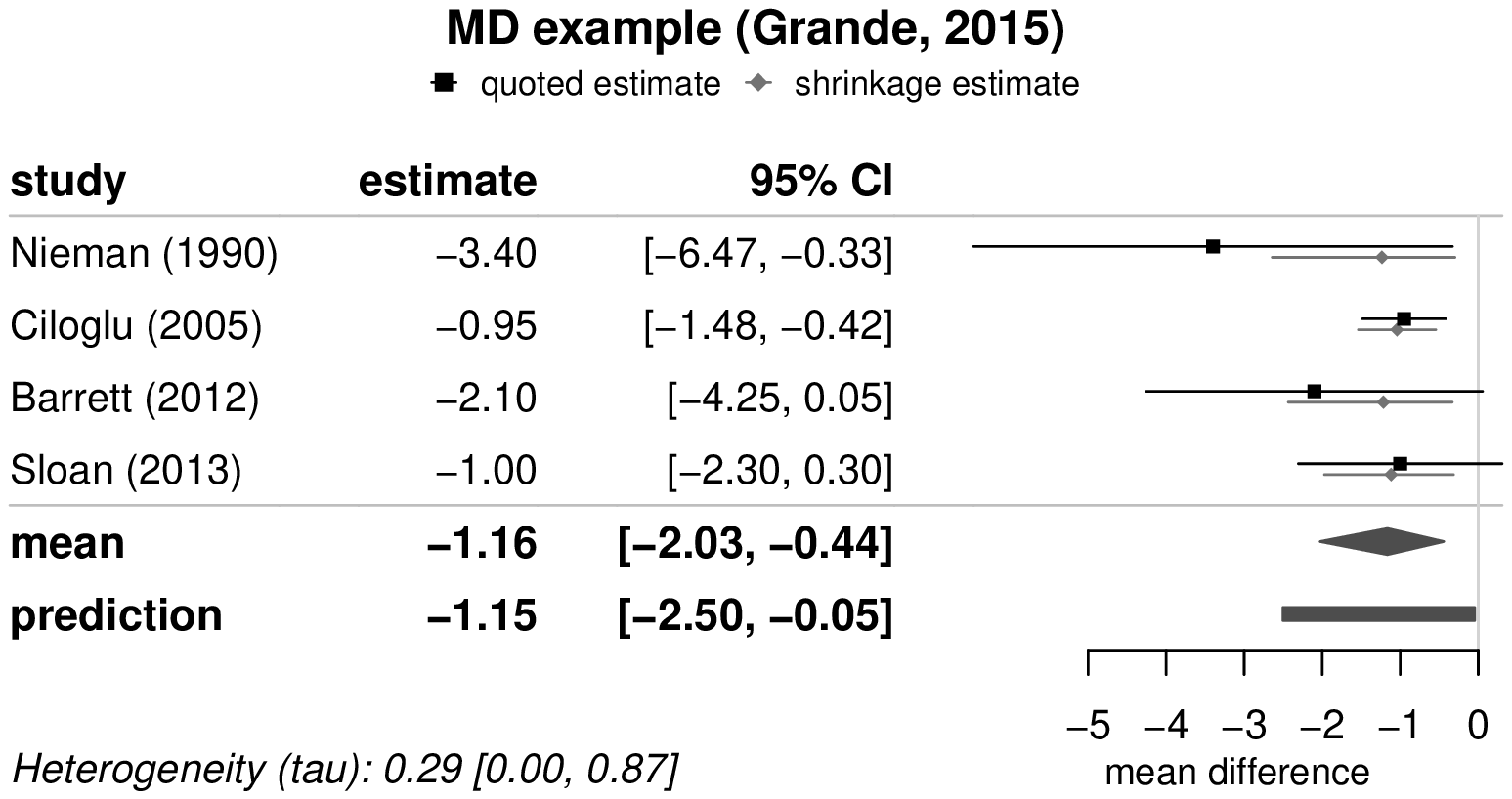}%
               \hspace{0.05\linewidth}%
               \includegraphics[width=0.47\linewidth]{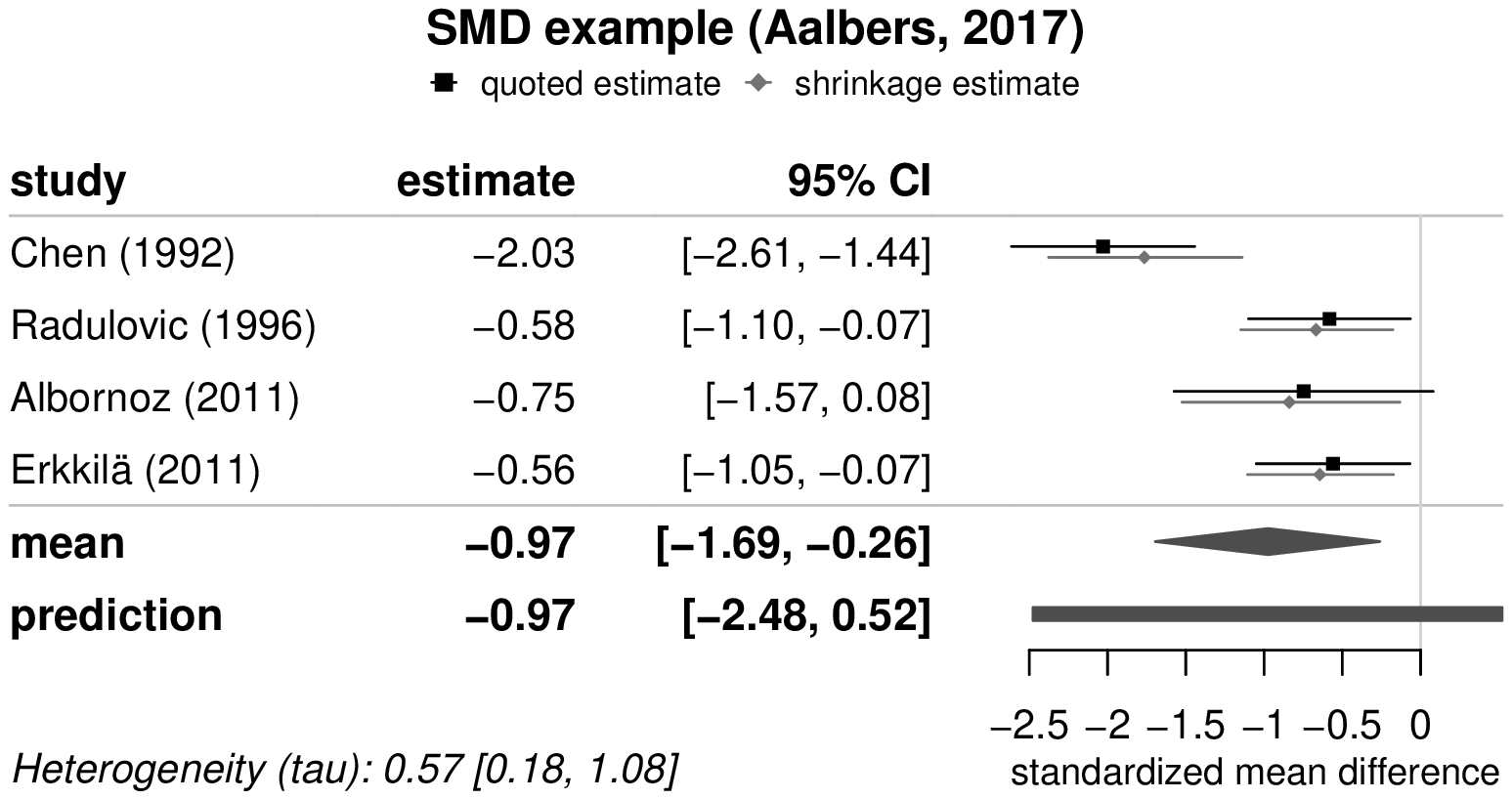}}
      \caption{\label{fig:forest01}Forest plots for the two examples
        discussed in Sections~\ref{sec:MDexample}
        and~\ref{sec:SMDexample}. In both cases, a
        half-normal(0.5)~prior for the heterogeneity~$\tau$ was
        used. Besides the intervals based on the quoted estimates, the
        shrinkage intervals are shown in grey. At the bottom, the
        credible interval for the overall mean ($\mu$) is shown along
        with the prediction interval for a ``new'' additional study
        effect~$\theta_{k+1}$. The estimated heterogeneity~($\tau$) is
        quoted in terms of the posterior median and shortest 95\%
        credible interval.}
    \end{figure}

    For the present example, we would then suggest a half-normal(0.5)
    prior as a slightly conservative choice, in order to reflect the
    potential heavy-tailedness suggested by Rhodes \emph{et~al.}
    (2015),\citep{RhodesEtAl2015} and to account for the fact that the
    empirical data might be of limited relevance for the present
    example data. A meta-analysis based on the half-normal(0.5)~prior
    is illustrated in Figure~\ref{fig:forest01}. Among the four
    studies, three consistently indicate estimates in the range $0.5$
    -- $0.8$, while the first one shows a huge effect estimate of the
    order of~$2.0$; a positive amount of heterogeneity appears to be
    present (the CI for~$\tau$ is in a strictly positive range; see
    also Figure~\ref{fig:tauposteriors}), and the eventual combined
    estimate indicates a ``small'' to ``very large'' average effect.
    Given the pronounced heterogeneity one might discuss whether the
    estimation of a pooled effect is meaningful. Nevertheless, we use
    this example to illustrate the use of Bayesian methods in
    heterogeneous situations, where heterogeneity cannot be explained
    and good reasons are available to perfom a quantitative
    meta-analysis despite of large heterogeneity.  The large estimated
    heterogeneity here results in a wide CI for the overall effect, a
    very wide prediction interval, and also very little shrinkage for
    the estimated study-specific effects~$\theta_i$.

  \subsection{Log-transformed effect scales}\label{sec:LOGexample}
    \subsubsection{Log odds ratio}\label{sec:logORexample}
      A systematic review was performed by Crins \emph{et~al.}
      (2014)\citep{CrinsEtAl2014} to investigate the effect of
      Interleukin-2 receptor antagonists (IL2-RA) on recovery of
      pediatric patients following liver transplantation. One aspect
      of interest was the occurrence of \emph{acute rejection (AR)}
      reactions as a common adverse event. Two randomized controlled
      trials reporting such data were found, the event counts along
      with the corresponding (logarithmic) odds ratios and standard
      errors are shown in Table~\ref{tab:example3}. Both studies
      indicated a reduction in the chances of an AR event for the
      treatment group.

      \begin{table}[b] 
        \caption{\label{tab:example3}Log-OR example
          data.\citep{CrinsEtAl2014} $a$ and $n_1$ as well as $c$ and
          $n_2$ denote the event counts and total numbers of patients
          in treatment and control groups, which together summarize
          the trial outcome in terms of a $2\!\times\!2$~table. The
          $y_i$~are the derived logarithmic odds ratios and $\sigma_i$
          are the associated standard errors that eventually go into
          the analysis (see Section~\ref{sec:nnhm}). Negative values
          here indicate a reduction of the event odds, i.e., a
          beneficial treatment effect.}  \centering
        \begin{tabular}{clcccccc}
          \toprule
            && \multicolumn{2}{c}{treatment group}
             & \multicolumn{2}{c}{control group}
             & \multicolumn{2}{c}{log-OR} \\
          \cmidrule(lr){3-4}
          \cmidrule(lr){5-6}
          \cmidrule(lr){7-8}
          $i$ & study & events ($a_i$) & total ($n_{1;i}$) & events ($c_i$)  & total ($n_{2;i}$)
          & $y_i$ & $\sigma_i$\\
          \midrule
          1 & Heffron (2003) & 14 & 61  &  15 & 20  &  -2.31 & 0.60 \\
          2 & Spada (2006)   & \phantom{0}4 & 36  &  11 & 36  &  -1.26 & 0.64 \\
          \bottomrule
        \end{tabular}
      \end{table}

      The treatment effect is expressed and analyzed on a logarithmic
      scale here.  A heterogeneity magnitude of $\tau=1.0$ would imply
      that any random pair of studies would be expected to exhibit
      effects differing by a factor of~2.6 (see
      Table~\ref{tab:conditional}), which seems quite extreme already;
      values like $\tau=0.5$ or below seem more plausible.  In a
      simular investigation involving 14~studies and based on adult
      patients (Goralczyk \emph{et~al.};
      2011),\citep{GoralczykEtAl2011} a mean treatment effect (log-OR)
      of $-0.26$, corresponding to an OR of $0.77$, was found. The
      UISD for a log-OR is at $\sigmau\approx 4$ per subject, while
      for the present data here we get an estimate of $\su = 5.4$.  An
      empirical study based on a large number of meta-analyses
      predicts a median (95\% quantile) of 0.28 (1.16) for the
      heterogeneity (Turner \emph{et~al.};
      2015),\citep{TurnerEtAl2015} and an investigation of
      heterogeneity \emph{estimates} found a median (95\% quantile) of
      0.00 (1.05) (G\"{u}nhan \emph{et~al.};
      2020).\citep{GunhanRoeverFriede2020} In the data from the
      closely related meta-analysis by Goralczyk \emph{et~al.}
      (2011),\citep{GoralczykEtAl2011} the heterogeneity is estimated
      at 0.12 (0.38).

      A half-normal(0.5) prior would mostly cover values $\tau<1.0$
      (up to ``fairly high'' heterogeneity according to
      Table~\ref{tab:SpiegelhalterCategories}) with an expectation and
      median below $0.5$ (see also Table~\ref{tab:predictive-1}). The
      resulting 95\% prior predictive interval would still include
      effects within a factor of~$3$ around the overall mean
      log-OR~$\mu$.  For the present investigation, we would then
      suggest a half-normal(0.5) prior as a reasonably conservative
      choice, which also agrees roughly with the empirical evidence
      (see Fig.~\ref{fig:TurnerPrior}).  A meta-analysis based on this
      prior is shown in Figure~\ref{fig:forest02}. In this example we
      have two studies only, demonstrating the somewhat speculative
      nature of infering heterogeneity based on sparse data, and
      higlighting the value of considering a-priori probabilities. In
      the present case, the two studies involved are not very large,
      and their resulting CIs are overlapping, which makes the data
      consistent with a wide range of heterogeneity values, from
      homogeneity ($\tau\!=\!0$) up to magnitudes of $\tau\!=\!10$ or
      $\tau\!=\!20$. Including the weakly informative heterogeneity
      prior, and effectively down-weighting unreasonably large
      heterogeneity values, then leads to an estimate of $-1.81$ for
      the log-OR, corresponding to a reduction in the odds of an
      AR~event down to $\exp(-1.81)=16\%$. While the uncertainty still
      is large (ranging roughly from $5\%$ up to $50\%$), the analysis
      clearly indicates a substantial reduction in AR~events here.
      The heterogeneity's posterior density is also shown in
      Figure~\ref{fig:tauposteriors}; here we can see that for the
      present example constellation, the posterior is very similar to
      the prior. With the very uncertain original estimates (due to
      the small sample sizes), the overall mean's CI is wide, but the
      additional width of the prediction interval is limited due to
      the (prior and empirical) information on the heterogeneity, and
      a noticeable shrinkage effect is also observable.

    \subsubsection{Log incidence rate ratio}\label{sec:logIRRexample}
      Four studies investigating the effect of ferric carboxymaltose
      vs.\ placebo in heart-failure patients with iron deficiency were
      jointly analyzed by Anker \emph{et~al.}
      (2018).\citep{AnkerEtAl2018} The main outcome was the
      \emph{incidence rate ratio (IRR)} with respect to the composite
      endpoint of recurrent cardiovascular (CV) hospitalisations or CV
      death. The relevant available data are shown in
      Table~\ref{tab:example4}.  The eventual analysis is based on the
      logarithmic ratio of the event rates (per 100~patient-years of
      follow-up) of treatment over placebo group.

      \begin{table}[b] 
        \caption{\label{tab:example4}Log-IRR example
          data.\citep{AnkerEtAl2018} The incidence rate ratios for the
          composite endpoint of recurrent cardiovascular (CV)
          hospitalisations and CV mortality are given for each
          study. For the analysis, the logarithmic rate ratio is
          considered.  Negative values here indicate a reduction of
          incidence rates, i.e., a beneficial treatment effect.}
        \centering
        \begin{tabular}{clcccccc}
          \toprule
                &&&&\multicolumn{2}{c}{log-IRR} \\
          \cmidrule(lr){5-6}
          $i$ & study & rate ratio [95\% CI] & $n_i$
          & $y_i$ & $\sigma_i$\\
          \midrule
          1 & \textsc{Fair-HF} (2009)     & 0.44 [0.22, 0.90] &           459 & -0.82 & 0.36\\
          2 & \textsc{Confirm-HF} (2015)  & 0.68 [0.38, 1.21] &           301 & -0.39 & 0.30\\
          3 & \textsc{Efficacy-HF} (2015) & 1.09 [0.21, 5.54] & \phantom{0}34 &  0.09 & 0.83\\
          4 & \textsc{Fer-Cars-01} (2018) & 0.87 [0.07, 10.4] & \phantom{0}45 & -0.14 & 1.28\\
          \bottomrule
        \end{tabular}
      \end{table}

      As in the previous example, the outcome is analyzed on the
      logarithmic scale, so that many arguments apply essentially
      analogously here. Regarding empirical evidence on previously
      encountered amounts of heterogeneity, there are no studies
      available that would be directly applicable for log-IRRs,
      however, odds ratios and rate ratios have quite some similarity,
      so that these findings also have some bearing here. The UISD
      here is at $\sigmau=2$~\emph{per event} (see
      Appendix~\ref{sec:UnitInfoIRR}); with a total of 114~events
      observed among a total of
      839~patients\citep[][Tab.~4]{AnkerEtAl2018} (a rate of $\approx
      0.14$ events per patient), this would correspond to $\sigmau
      \approx\frac{2}{\sqrt{0.14}}=5.4$ \emph{per patient}.  For the
      present data, we empirically get an average of $\su=6.6$.

      \begin{figure}
        \centering
        \makebox{\includegraphics[width=0.47\linewidth]{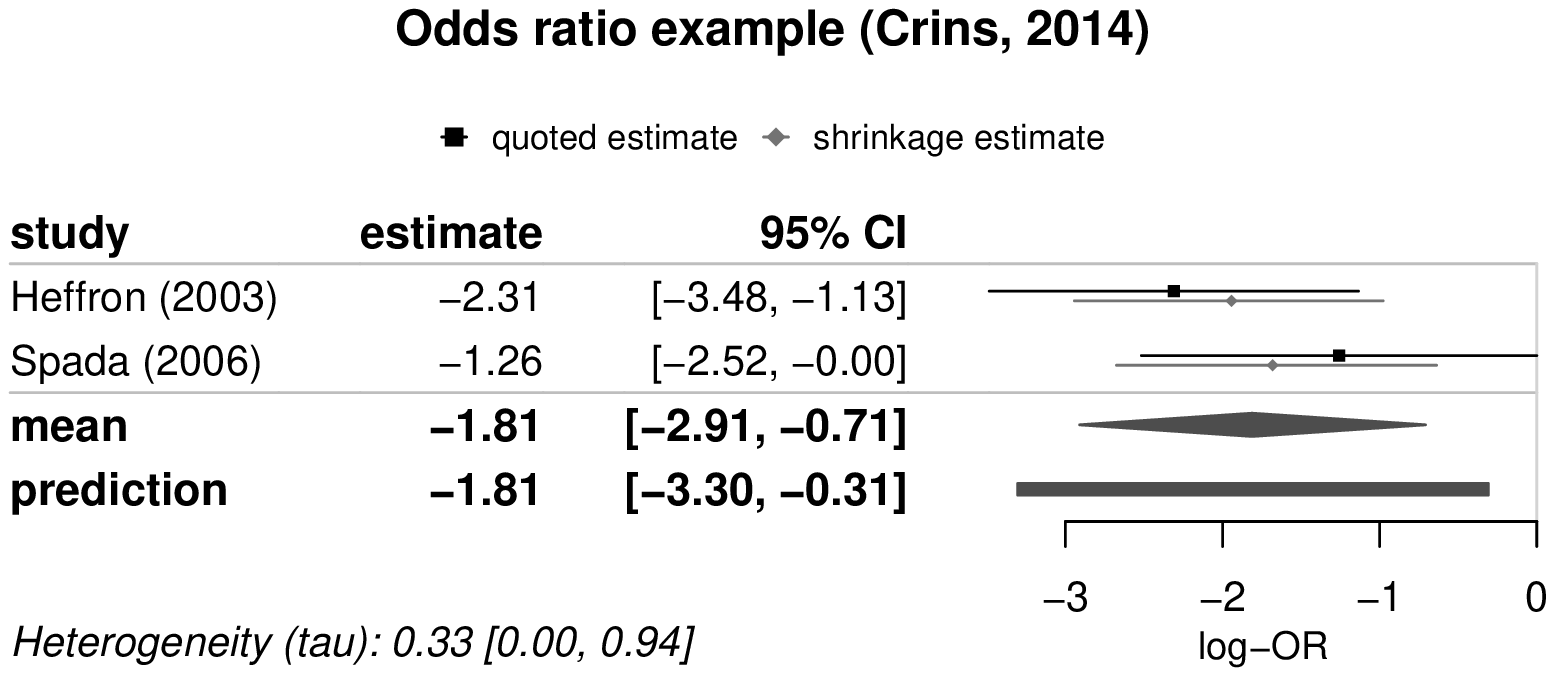}%
                 \hspace{0.05\linewidth}%
                 \includegraphics[width=0.47\linewidth]{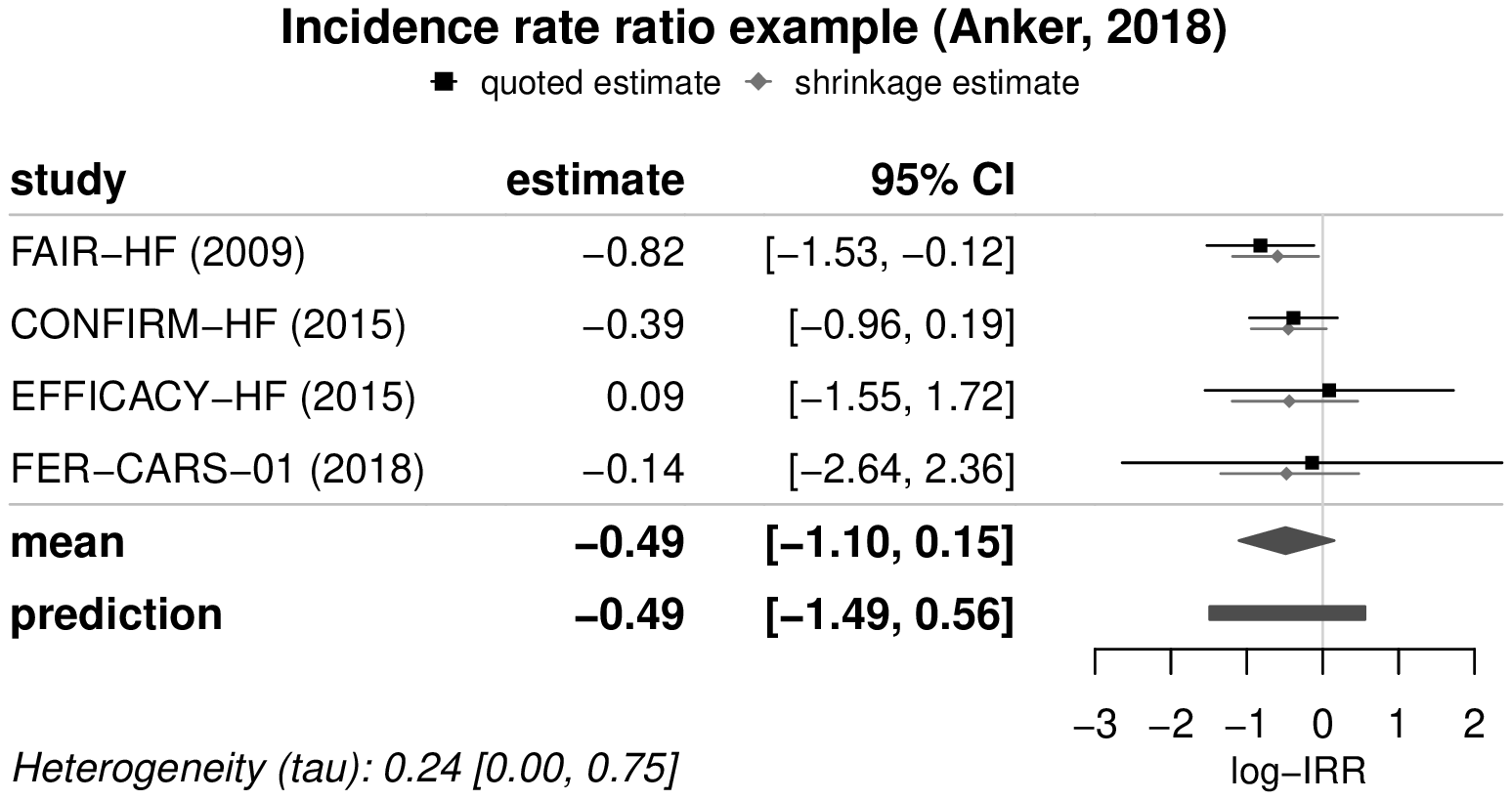}}
        \caption{\label{fig:forest02}Forest plots for the two examples
          discussed in Sections~\ref{sec:logORexample}
          and~\ref{sec:logIRRexample}. In both cases, a
          half-normal(0.5)~prior for the heterogeneity~$\tau$ was
          used.}
      \end{figure}

      For this example, we would again suggest a half-normal(0.5)
      prior. A meta-analysis based on this prior is shown in
      Figure~\ref{fig:forest02}. While the data \emph{look}
      homogeneous (all intervals have some overlap, also because some
      studies are very small and intervals are correspondingly wide),
      we would still anticipate the possibility of heterogeneity ---
      since from experience we know that heterogeneity is frequently
      present, and because we know that heterogeneous circumstances
      are still likely to produce data that may still ``look
      homogeneous''.\citep{FriedeRoeverWandelNeuenschwander2017a}
      Compared to our a-priori expectations of $\tau$~values up
      to~$0.98$ (see Table~\ref{tab:predictive-1}), the posterior then
      suggests a slightly lower heterogeneity range of up to~$0.75$,
      but the data do not provide very much evidence in this regard
      (see also the posterior in Figure~\ref{fig:tauposteriors}). The
      mean treatment effect eventually is at a log-IRR of~$-0.49$,
      corresponding to an IRR of~61\% (i.e., a reduction in the event
      rate), with a CI ranging from~33\% up to~116\%.  For these
      somewhat homogeneous estimates, one can see that the ones with
      very large associated standard errors eventually have shrinkage
      estimates close to the overall prediction interval.
      A sensitivity analysis investigating alternative prior choices
      for this example is also shown in
      Appendix~\ref{sec:SensitivityAppendix}.

    \subsubsection{Log odds}\label{sec:logOddsExample}
      Neuenschwander \emph{et~al.} investigated the use of historical
      data in order to inform the analysis of a new data
      set.\citep{NeuenschwanderEtAl2010} A meta-analysis of several
      trials in ulcerative colitis was performed in order to support
      the analysis of a subsequent phase~II trial. The figure of
      interest here was the probability for \emph{clinical remission
        at week~8} in placebo-treated patients, and the main interest
      was in a prediction for the new study's event probability, to
      then formally integrate this in a subsequent analysis using a
      meta-analytic-predictive (MAP) approach.\citep{SchmidliEtAl2014}
      Four previous randomized controlled trials reporting this
      endpoint were available, their data are shown in
      Table~\ref{tab:example5}.
      \begin{table}[b] 
        \caption{\label{tab:example5}Log-odds example data due to
          Neuenschwander \emph{et~al.}
          (2010).\citep{NeuenschwanderEtAl2010} The $n_i$ and $x_i$
          here denote total numbers and the numbers of remitting
          patients among these.  Analysis is done based on the derived
          log-odds~$y_i$ and their standard errors~$\sigma_i$.}
        \centering
        \begin{tabular}{clcccccc}
          \toprule
                && \multicolumn{2}{c}{remission}
                 & proportion & odds
                 & \multicolumn{2}{c}{log-odds}\\
          \cmidrule(lr){3-4}
          \cmidrule(lr){5-5}
          \cmidrule(lr){6-6}
          \cmidrule(lr){7-8}
          $i$ & study & events ($x_i$) & total ($n_i$) 
          & $p_i=\frac{x_i}{n_i}$
          & $\frac{x_i}{n_i-x_i}=\frac{p_i}{1-p_i}$
          & $y_i$ & $\sigma_i$\\
          \midrule
          1 & Feagan (2005)      & \phantom{0}9 & \phantom{0}63 & 0.143 & 0.167 & $-1.79$ & $0.36$ \\
          2 & Rutgeerts (2005a)  &           18 &           121 & 0.149 & 0.175 & $-1.74$ & $0.26$ \\
          3 & Rutgeerts (2005b)  & \phantom{0}7 &           123 & 0.057 & 0.060 & $-2.81$ & $0.39$ \\
          4 & Van~Assche (2006)  & \phantom{0}6 & \phantom{0}56 & 0.107 & 0.120 & $-2.12$ & $0.43$ \\
          \bottomrule
        \end{tabular}
      \end{table}
      Instead of working directly on the estimated probabilities~$p$,
      the analysis here is done based on the
      \emph{odds}~$\frac{p}{1-p}$, and a subsequent
      log-transformation.\citep{TrikalinosEtAl2013}

      Homogeneity of placebo rates is not expected --- differences
      between control rates are among the main reasons for requiring a
      control arm for each RCT, and for pursuing a contrast-based
      analysis.\citep{DiasAdes2016,WhiteEtAl2019} The studies were
      designed aiming for an estimate of the treatment effect, and the
      placebo rate originally way mostly a nuisance parameter here.
      However, \emph{some} amount of similarity still is anticipated,
      and the aim of this exercise is to carefully derive the
      predictive distribution, which of course depends on the amount
      of heterogeneity~$\tau$.

      The earliest of the four studies was planned anticipating a
      remission rate of~10\% for the placebo
      group,\citep{FeaganEtAl2005} and hence a UISD of $\sigmau
      \approx \sqrt{\frac{1}{0.1}+\frac{1}{0.9}} = 3.33$ may be
      expected.  Empirically, we get an estimate of $\su = 3.2$ from
      the present data set.

      As the endpoint are \emph{logarithmic} odds, we may again apply
      similar reasoning as in the previous subsections, regarding the
      anticipated ratios of odds. However, a major difference here is
      that while clinical trials are usually carefully designed to
      provide reliable estimates of treatment effects
      (treatment/control contrasts), this is not necessarily the case
      for the event rates that we are considering here; we may expect
      the log-odds to be more variable than the log-ORs. With this in
      mind, and considering conservatism and robustness particularly
      desirable in the present context, we would suggest a
      half-normal(1.0) prior here. From Table~\ref{tab:predictive-1},
      we can see that the implied 95\% prior predictive interval then
      spans a range of roughly a factor~9 around the
      median~$\mu$. Given the context, it may be of particular
      interest to consider the associated prior maximum sample
      size~$n^\star_\infty$ (see Section~\ref{sec:sigmau}); for the
      prior median of $\tau=0.67$, we have
      $\frac{\tau}{\sigmau}=\frac{0.67}{3.2}=0.21$, corresponding to a
      maximum size of $n^\star_\infty=23$ (compared to an original
      total of 363~subjects included in the analysis).  The prior's 95
      \% quantile is (approximately) at $\tau=2$, and larger values
      would effectively imply (with $n^\star_\infty<3$) an almost
      noninformative posterior predictive distribution.

      The eventual analysis is illustrated in
      Figure~\ref{fig:forest03}.  Looking at the heterogeneity's
      posterior (Figure~\ref{fig:tauposteriors}), one can see that
      heterogeneity here appeared to be less than anticipated. The
      prediction interval is relatively wide, and on the
      back-transformed scale is centered at a probability of 0.11 with
      its 95\% posterior predictive interval ranging from~0.03
      to~0.34.  The posterior predictive distribution's standard error
      is~0.70, and relative to the UISD, this roughly corresponds to
      an effective sample size of~$\frac{3.2^2}{0.70^2}=21$ subjects.

  \subsection{Regression slopes}\label{sec:RegExample}
    Bergau \emph{et~al.} (2017)\citep{BergauEtAl2017} investigated
    predictors of all-cause mortality among patients with an
    implantable cardioverter-defibrillator (ICD) device. Several
    potential covariables were considered, among these the \emph{left
      ventricular ejection fraction (LVEF)}, which is a measure of the
    efficiency of heart function that is usually determined via
    echocardiography. LVEF is commonly expressed in percent, where
    52\%--72\% are normally observed in healthy individuals, while
    values below 30\% are considered abnormal.\citep{LangEtAl2015}
    Criteria for an indicated ICD therapy include various conditions,
    including thresholds on the LVEF in the range
    30--40\%.\citep{TracyEtAl2013}.  Five studies were found that had
    reported on survival analyses including LVEF as a predictor, and a
    meta-analysis was performed based on the coefficients standardized
    to a \emph{5 percentage point decrease in LVEF}; the data are
    shown in Table~\ref{tab:example6}. The different studies also
    included different sets of additional covariates in their
    analyses.\citep{BergauEtAl2017}

      \begin{table}[b] 
        \caption{\label{tab:example6}Regression example
          data.\citep{BergauEtAl2017} Regression slopes result from
          survival analyses and are expressed in terms of hazard
          ratios (HRs) and with reference to a \emph{5~percentage
            point decrease in LVEF}. The baseline means and standard
          deviations of LVEF values are also shown.} \centering
        \begin{tabular}{clcccccccc}
          \toprule
                &&\multicolumn{2}{c}{LVEF (\%)}&&&\multicolumn{2}{c}{log-HR} \\
          \cmidrule(lr){3-4}
          \cmidrule(lr){7-8}
          $i$ & study 
          & mean & s.d.
          & HR [95\% CI] & $n_i$
          & $y_i$ & $\sigma_i$\\
          \midrule
          1 & Maci\k{a}g (2012)               & 28.0 & \phantom{0}4.0 & 1.16 [0.90, 1.51] & \phantom{0}121 & 0.148 & 0.132\\
          2 & Hage (2013)                     & 28.0 &           15.0 & 1.23 [1.08, 1.40] & \phantom{0}696 & 0.207 & 0.066\\
          3 & Demirel (2014)                  & 31.9 & \phantom{0}9.3 & 1.29 [1.02, 1.64] & \phantom{00}99 & 0.255 & 0.121\\
          4 & Konstantino (2016)              & 31.6 &           11.1 & 1.16 [1.02, 1.32] &           1125 & 0.148 & 0.066\\
          5 & Rodr\'{i}guez-Ma\~{n}ero (2016) & 26.2 & \phantom{0}7.6 & 1.23 [1.10, 1.36] &           1174 & 0.207 & 0.054\\
          \bottomrule
        \end{tabular}
      \end{table}

    The regressor, LVEF, here is expressed in percentages (between 0
    and 100), which might just as well have been expressed as a
    fraction (between 0 and 1), while for the analysis a unit of a
    \emph{5~percentage point decrease} was used --- this highlights
    the importance of clarifying the scale of the increment~$\Delta_x$
    that heterogeneity considerations are to be based
    on. Table~\ref{tab:example6} also shows the distributions of LVEF
    within studies; these are roughly similar and have standard
    deviations of the order of 10~percentage points.  For the
    ``reference'' increment~$\Delta_x$ for judging plausible
    heterogeneity magnitudes, we will then consider a difference of
    20~percentage points, which roughly spans the bulk of LVEF values
    encountered in each of the studies. This also coincides with the
    range of values considered ``normal'' (52\%--72\%) or the
    difference between ``normal'' and ``abnormal'' ranges ($\geq 52\%$
    vs.\ $<30\%$) here.  The (empirical) UISD for the present data is
    at $\su=1.9$ (for the 5\%~increments shown in
    Table~\ref{tab:example6}, corresponding to~$\su=7.5$ for a
    20\%~difference).

\begin{figure}
      \centering
      \makebox{\includegraphics[width=0.47\linewidth]{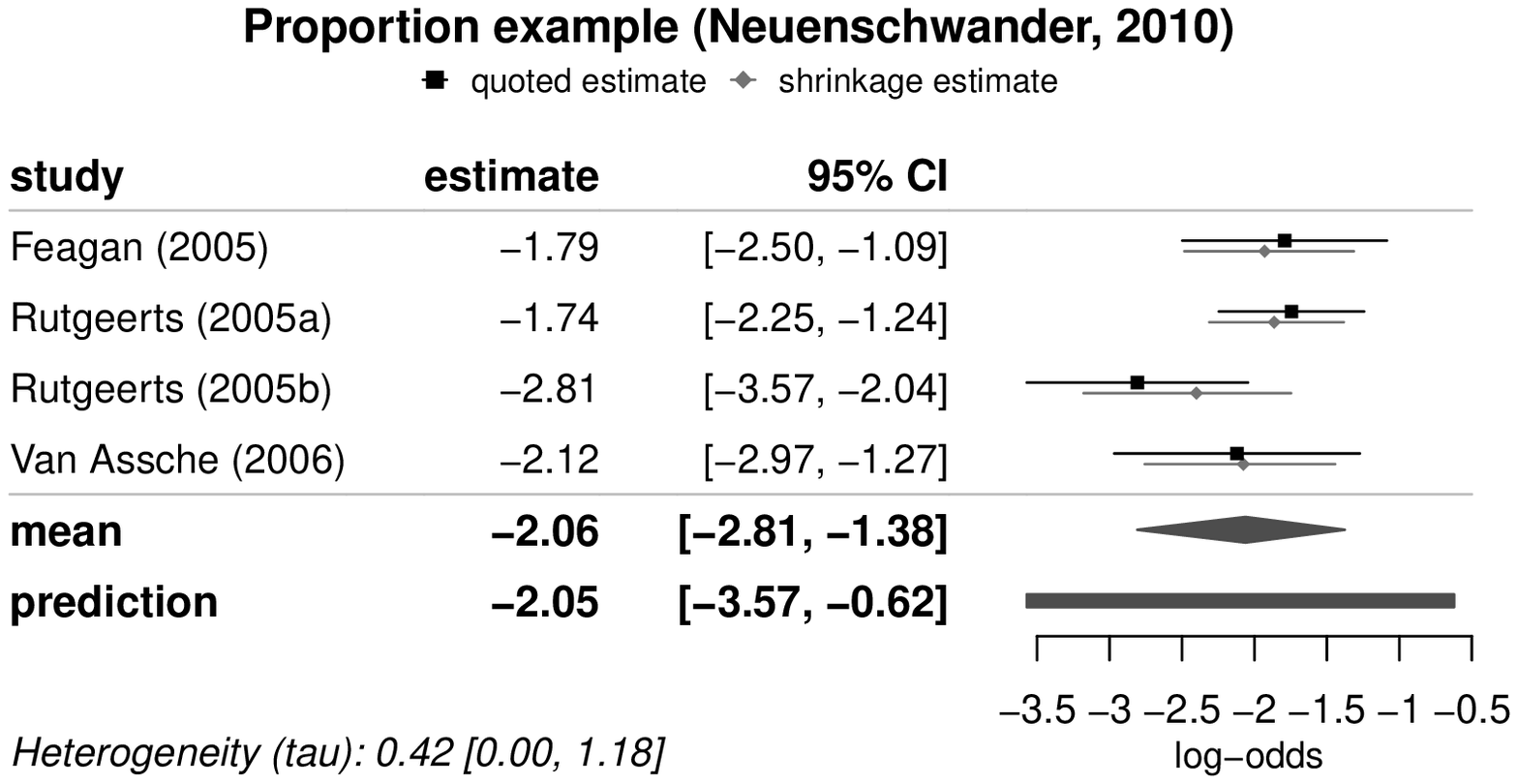}%
               \hspace{0.05\linewidth}%
               \includegraphics[width=0.47\linewidth]{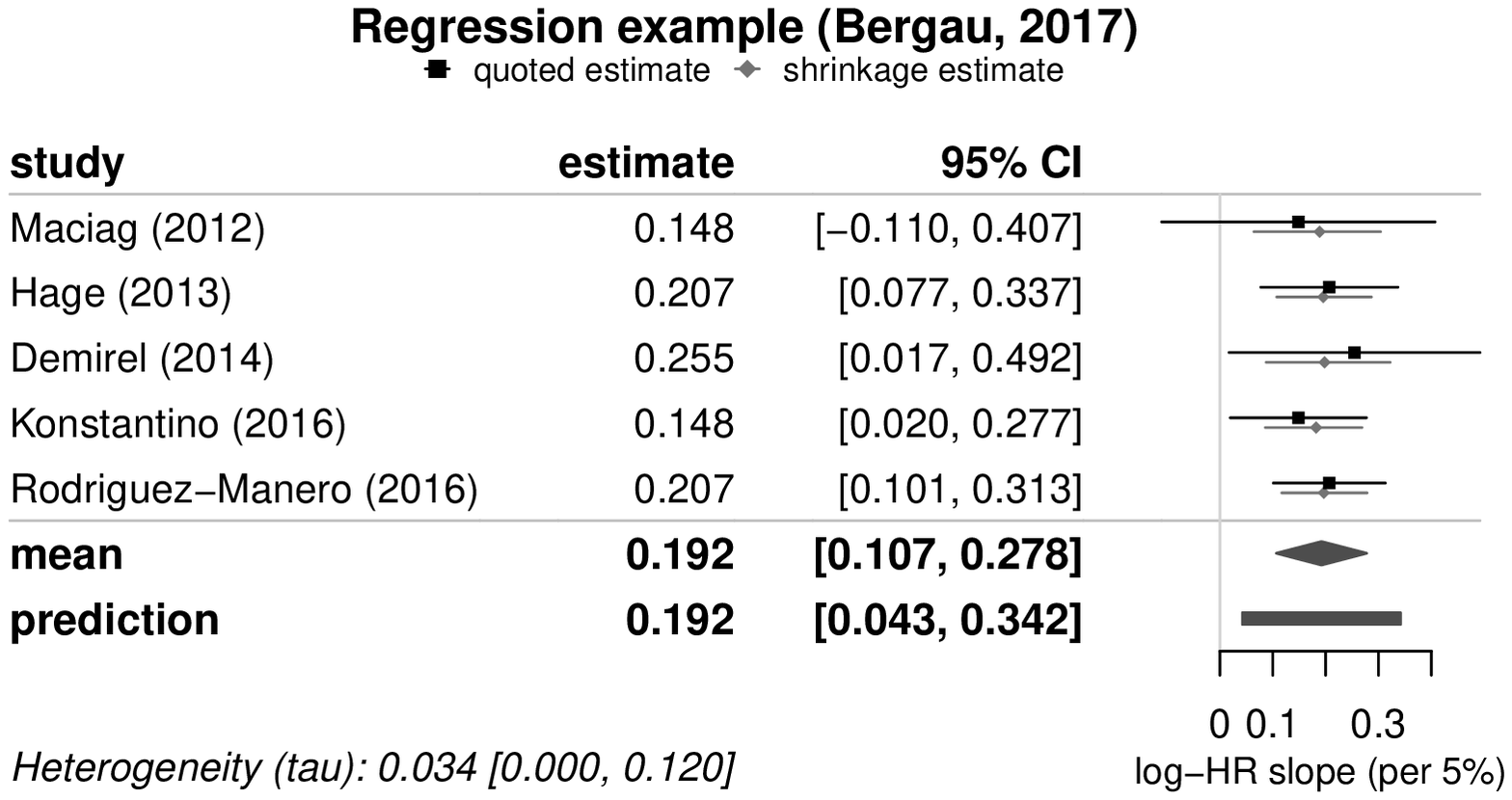}}
      \caption{\label{fig:forest03}Forest plots for the examples
        discussed in Sections~\ref{sec:logOddsExample}
        and~\ref{sec:RegExample}.  For the log-odds, a
        half-normal(1.0) prior was used, and for the log-HR regression slopes, a half-normal(0.125)~prior was used
        for the heterogeneity~$\tau$.}
    \end{figure}

    Since the regression coefficient is to be interpreted as a
    logarithmic HR, we will assume a half-normal(0.5) prior for the
    effect correponding to a $\Delta_x=20$~percentage point increment
    (analogously to the arguments made in
    Sections~\ref{sec:logORexample} and~\ref{sec:logIRRexample}).  For
    the 5~percentage point decreases considered in the analyses, this
    then implies a four-fold smaller heterogeneity, i.e., a
    half-normal(0.125) prior.  Analysis results for a
    half-normal(0.125) prior are illustrated in
    Figure~\ref{fig:forest03}.
    The estimates are very homogeneous, which is evident from the
    forest plot as well as from the estimated heterogeneity (see also
    Figure~\ref{fig:tauposteriors}). The overall log-HR estimate is
    at~0.19, corresponding to 1.21-fold increased mortality hazard for
    a 5~percentage point decrease (worsening) in LVEF\@.

  \subsection{Correlations}\label{sec:CORexample}
    Molloy \emph{et~al.} (2014)\citep{MolloyOCarrollFerguson2014}
    investigated the relationship between conscientiousness and
    medication adherence. A total of 16~relevant studies reporting
    correlation coefficients of the two factors were found, which were
    also graded according to their methodological quality. Three of
    the studies were rated with the highest quality score; their data
    are shown in Table~\ref{tab:example7}. The data are also available
    as part of the \texttt{metafor}
    \textsf{R}~package.\citep{Viechtbauer2010} In order to avoid
    problems due to the bounded parameter space of correlations~$r_i$
    (between $-1$ and $+1$), we will use the Fisher-$z$ transformed
    values instead.  Note that, since in the present example the
    reported correlations ($r_u$) are relatively close to zero, the
    corresponding Fisher-$z$ values ($y_i$) are almost identical here
    (see Table~\ref{tab:example7}; $r_i$~and $y_i$~values only differ
    in their third decimal place) and the transformation eventually
    makes little difference.

    \begin{table}[b] 
      \caption{\label{tab:example7}Correlation example
        data.\citep{Viechtbauer2010,MolloyOCarrollFerguson2014} $r_i$
        and $n_i$ here denote the empirical correlation coefficients
        and the underlying sample sizes. The $y_i$~are the Fisher-$z$
        transformed correlations and $\sigma_i$ the associated
        standard errors that eventually go into the analysis (see
        Section~\ref{sec:nnhm}). A positive effect size~$y_i$ here
        indicates a positive correlation.}  
      \centering
      \begin{tabular}{clcccc}
        \toprule
        & & & & \multicolumn{2}{c}{Fisher's~$z$} \\
        \cmidrule(lr){5-6}
        $i$ & study & Correlation $r_i$ & $n_i$
        & $y_i$ & $\sigma_i$\\
        \midrule
        1 & Stilley (2004) & 0.24 & 158 & 0.245 & 0.080 \\
        2 & Ediger (2007)  & 0.05 & 326 & 0.050 & 0.056 \\
        3 & Jerant (2011)  & 0.01 & 771 & 0.010 & 0.036 \\
        \bottomrule
      \end{tabular}
    \end{table}

    As elaborated in Section~\ref{sec:specialcase:correlation}, we
    expect smaller magnitudes of heterogeneity for correlation
    endpoints (say, mostly $\tau\leq 0.3$); the UISD is
    at~$\sigmau=1.0$, which also matches the figures we see
    empirically in the present data set ($\su=1.004$).  Van~Erp
    \emph{et~al.}  (2017)\citep{vanErpEtAl2017} report a median and
    95\% quantile of~0.12 and~0.29, respectively, for empirically
    observed heterogeneity estimates from published studies.
    Meta-analysing the remaining set of 13~studies from the present
    data set\citep{MolloyOCarrollFerguson2014} (using a uniform
    prior), in order to quantify the evidence ``external'' to the
    example data, yields a heterogeneity estimate of 0.07 with 95\% CI
    [0.00, 0.17].

    Heterogeneity values of $\tau=0.1$ or $\tau=0.2$ would imply
    differences between a random pair of studies of a similar order of
    magnitude (see Table~\ref{tab:conditional}).  A half-normal(0.2)
    prior for the heterogeneity would cover values mostly in the range
    below~$0.4$, with a prior median at $\tau=0.13$ (see
    Table~\ref{tab:predictive-1}).

    \begin{figure}[b]
      \centering
      \makebox{\includegraphics[width=0.47\linewidth]{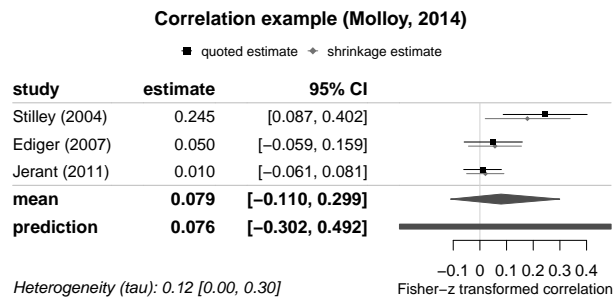}}
      \caption{\label{fig:forest04}Forest plot for the example
        discussed in Section~\ref{sec:CORexample}.  For the (Fisher-$z$
        transformed) correlations, a half-normal(0.2)~prior was used
        for the heterogeneity~$\tau$.}
    \end{figure}

\begin{figure}[t]
      \centering
      \makebox{\includegraphics[width=0.95\linewidth]{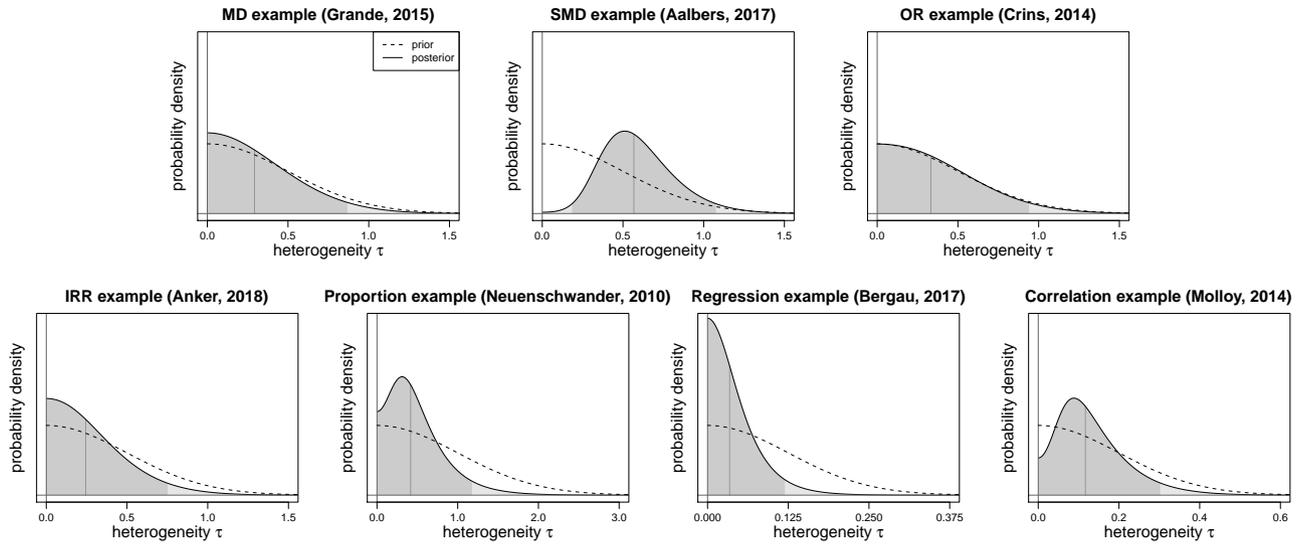}}
      \caption{\label{fig:tauposteriors} Marginal prior and posterior
        densities for the heterogeneity parameter~$\tau$ in the seven
        examples discussed in Section~\ref{sec:examples}. The dashed
        lines show the prior densities, the solid lines show the
        posteriors. The area shaded in dark grey indicates the 95\%
        credible interval, the vertical line is the posterior median.}
    \end{figure}

    For the present analysis, we would then suggest a half-normal(0.2)
    prior for the heterogeneity. A meta-analysis of the example data
    based on this prior is illustrated in Figure~\ref{fig:forest04}.
    The two traits were originally measured using differing scales, so
    that complete homogeneity might be considered especially unlikely.
    The heterogeneity's resulting posterior median is at $\tau=0.12$
    (with the 95\%~CI ranging up to 0.30), its posterior distribution
    is also illustrated in Figure~\ref{fig:tauposteriors}.  The three
    studies are of differing size and suggest neutral to slightly
    positive correlation between conscientiousness and medication
    adherence. The resulting mean estimate is positive at about~0.08,
    while the CI ranges from negative to positive ($-0.1$ to $+0.3$).

\section{Discussion}\label{sec:discussion}
  While executing a Bayesian meta-analysis is not technically
  difficult, specifying a widely acceptable prior remains a challenge,
  especially when it comes to the heterogeneity parameter~$\tau$.
  Although the problem may appear complex at first, it is usually
  possible to break down the specification into a number of more
  specific questions that are easier to approach one-by-one.  These
  steps are summarized in Table~\ref{tab:questions} and may be
  outlined as follows:
  (i)~what is the effect's scale?
  (ii)~what is the probable magnitude of other effects?  
  (iii)~how large is the unit information standard deviation (UISD)?
  (iv)~is relevant empirical information available?
  The information may then be related to more concrete prior
  specifications by constraining
  (v)~prior quantiles (of~$\tau$)
  (vi)~prior predictive quantiles (of~$\theta_i$), and
  (vii)~other prior properties.
  We have demonstrated the prior specification in seven applications
  involving few studies and covering a range of common effect scales
  and application areas, leading to sensible prior distributions and
  results in all examples.  Besides the case of few studies, another
  context in which (weakly) informative priors are useful is whenever
  marginal likelihoods (or Bayes factors) need to be
  computed.\citep{RoeverWandelFriede2018} Calculation of marginal
  likelihoods requires proper prior distributions, and special care
  must be taken in their selection in order to avoid (seemingly)
  paradoxical results.\citep{BDA3rd,Lindley1957}

  In many applications, the results will be robust to variations of
  the prior, which may also be checked in sensitivity analyses.  The
  prior specification will usually not be the most crucial or
  influential among the line of assumptions being made, which include
  normality,\citep{JacksonWhite2018} exchangeability, the selection of
  estimates to be pooled, or the choice between effect
  measures.\citep{DeeksAltman2001} Different prior specifications will
  of course leave their imprint on the posterior distribution, for
  example, results based on short- or heavy-tailed priors will reflect
  the differing assumptions, which may be based on emphasizing
  regularisation or robustness aspects.  There usually is no unique
  ``correct'' prior, and ``sceptical'' or ``enthusiastic'' results may
  be derived by implementing corresponding prior
  assumptions.\citep{SpiegelhalterEtAl1994} Even uncertainty in the
  prior distribution itself (or its scale) may be accommodated by
  using mixture priors.  Consideration of the stochastic ordering of
  heterogeneity priors may help assessing more or less conservative
  settings, which may be useful for the definition of sensitivity
  analyses. However, we would also like to warn against inflationary
  default specification and execution of multiple analyses here, as
  the resulting alternative estimates may lead to unnecessary
  ambiguity or inconsistent (flip-flopping) conclusions.
  In Appendix~\ref{sec:SensitivityAppendix}, sensitivity analyses are
  discussed in the context of the two examples from
  Sections~\ref{sec:MDexample} and~\ref{sec:logIRRexample}.
  Pre-specification of analyses (and their intended consequences) may
  help here.  
  In case there is genuine a-priori uncertainty about the
  heterogeneity's magnitude, this might better be reflected in a
  single prior (e.g., in terms of a mixture distribution).
  Either way, one needs to be prepared and willing to base the
  eventual analysis results on the posterior also when the data have
  little information on heterogeneity to add to the weakly informative
  prior, as was the case for some of the examples discussed here (see
  Figure~\ref{fig:tauposteriors}).  If it is not possible to specify a
  suitable (weakly) informative prior for the expected heterogeneity,
  then one might have to resort to a more conservative approach using
  uninformative priors.

  Another central assumption crucial to the validity of inference is
  the \emph{exchangeability} (see Section~\ref{sec:nnhm}). This might
  be compromised by selection effects, for example, publication
  bias\citep{RothsteinSuttonBorenstein} or reporting
  bias.\citep{WilliamsonEtAl2005} Especially in the case of only few
  studies, such effects might be hard to detect from the data, and
  information on the presence of selection effects may need to come
  from considerations of the context.

  Choice of heterogeneity priors has consequences for estimation of
  the overall mean parameter, but in particular also in prediction and
  shrinkage applications, as the inferred heterogeneity directly
  impacts on the amount of
  borrowing-of-strength;\citep{Roever2020,WandelNeuenschwanderRoeverFriede2017,SchmidliEtAl2014}
  smaller heterogeneity will lead to stronger pooling of estimates,
  and larger heterogeneity will imply that individual estimates are
  only loosely connected through the model.

  Especially in regulatory settings such as drug approval or health
  technology assessment (HTA) the definition of a standard prior
  distribution for the heterogeneity parameter is important to avoid
  post hoc discussions in case the use of different prior
  distributions leads to results suggesting conflicting
  interpretations. The Institute for Quality and Efficiency in Health
  Care (IQWiG) in Germany is currently looking into determining the
  empirical distribution for the between-study heterogeneity parameter
  from all published IQWiG reports with the goal to motivate a
  suitable prior distribution for HTA applications.

  While in the present manuscript we focused on the NNHM, some of the
  arguments laid out here are analogously transferable to other models
  for pairwise meta-analysis, for example, a Binomial-Normal model.
  Additional parameters and their priors may need to be specified in
  regard to baselines (which are often nuisance parameters and
  assigned vague
  priors).\citep{GunhanRoeverFriede2020,DiasAdes2016,WhiteEtAl2019,WangEtAl2020}
  More complex applications in evidence synthesis such as
  meta-regression or network-meta-analysis would again require similar
  prior specifications regarding between-study heterogeneity in the
  effects, but would then entail additional model components, e.g., in
  order to accommodate individual-patient data
  (IPD).\citep{DebrayEtAl2013a,BurkeEnsorRiley2017,Kontopantelis2018}
  Analogous arguments also extend more generally to hierarchical or
  multilevel models, such as generalized linear mixed models
  (GLMMs).\citep{GelmanHill,BrownPrescott}
  The sensitivity analyses shown in
  Appendix~\ref{sec:SensitivityAppendix} suggest that (for a given
  prior median) the prior distribution's shape has little impact on
  the results, as compared to the scaling of the pior.  As it might
  simplify prior specification further, it will be interesting to
  investigate whether or to what extent this feature holds more
  generally.
  In summary, the application of Bayesian methods with weakly
  informative prior distribution for the heterogeneity parameter can
  be recommended for meta-analyses with random effects especially in
  the common case of only few studies. This paper provides guidance on
  the choice of useful prior distributions for various effect measures
  and data situations.

\section*{Highlights}
  \begin{itemize}
    \item \emph{What is already known:}
      \begin{itemize}
        \item A Bayesian approach to meta-analysis may often be
          useful, in particular in cases of only few studies, and in
          order to derive predictions and shrinkage estimates.
        \item Careful specification (and justification) of prior distributions is
          required, especially for the heterogeneity parameter.
      \end{itemize}
    \item \emph{What is new:}
      \begin{itemize}
        \item Prior selection may usually be narrowed down
          considerably using a structured approach.
        \item A series of questions to guide choice and justification
          of the prior distribution was devised.
        \item Unit information standard deviations (UISDs) were
          derived for some commonly used effect measures.
      \end{itemize}
    \item \emph{Potential impact for \emph{Research Synthesis Methods}
      readers outside the authors' field:}
      \begin{itemize}
        \item Similar approaches may be useful also in related
          fields where hierarchical models or generalized linear mixed
          models (GLMMs) are used.
      \end{itemize}
  \end{itemize}

\ack
\section*{Acknowledgment}
  Support from the \emph{Deutsche Forschungsgemeinschaft (DFG)} is
  gratefully acknowledged (grant number \mbox{FR~3070/3-1}).

\section*{Conflicts of interest}
  The authors have declared no conflict of interest.

\section*{Data availability}
  The data that supports the findings of this study are available in the supplementary material of this article.

  \bibliography{../../literature/literature}

\appendix

\section{Unit information standard deviations}\label{sec:UnitInfoAppendix}
  \subsection{Standardized mean differences (SMDs)}\label{sec:UnitInfoSMD}
    Defining an SMD simply as $\delta_i =
    \frac{\mu_{2;i}-\mu_{1;i}}{\varsigma_i}$ (see
    Section~\ref{sec:specialcase:SMD}), this figure is in practice
    estimated based on empirical group-averages~$\bar{x}_{1;i}$
    and~$\bar{x}_{2;i}$. Neglecting uncertainty in variance estimation
    and assuming a known common standard deviation~$\varsigma_i$ for
    both treatment groups then leads to
    $\var(\delta_i)=\var\Bigl(\frac{\bar{x}_{2;i}-\bar{x}_{1;i}}{\varsigma_i}\Bigr) = \frac{\var(\bar{x}_{2;i}) + \var(\bar{x}_{1;i})}{\varsigma_i^2} = \frac{\frac{\varsigma_i^2}{n_{2;i}} + \frac{\varsigma_i^2}{n_{1;i}}}{\varsigma_i^2} = \frac{1}{n_{1;i}} + \frac{1}{n_{2;i}}$. 
    Furthermore assuming equal group sizes
    ($n_{1;i}=n_{2;i}=\frac{n_i}{2}$) then leads to an approximate
    standard error of $\frac{2}{\sqrt{n_i}}$ and hence a UISD of
    $\sigmau=2$.

  \subsection{Logarithmic odds (logits)}\label{sec:UnitInfoLogit}
    The variance of a logarithmic odds (or logit-proportion) estimate
    is $\frac{1}{n}\bigl(\frac{1}{p}+\frac{1}{1-p}\bigr)$, where
    $n$~is the sample size and $p$~is the true proportion. The
    variance (squared standard error) is in practice commonly
    estimated by $\bigl(\frac{1}{x}+\frac{1}{n-x}\bigr)$, where $x$~is
    the observed event count.\citep{TrikalinosEtAl2013} The UISD then
    is given by $\sigmau=\sqrt{\frac{1}{p}+\frac{1}{1-p}} \geq 2$.

    Note the similarity to the standard error of an \emph{logarithmic
      odds ratio},\citep{Roever2020} which may be expressed as the
    difference of two log-odds. For $p=\frac{1}{2}$, the resulting
    UISD~$\sigmau$ is twice as large (i.e. the variance~$\sigmau^2$ is
    four times as large), since (i) the two logits' variances add,
    while (ii) each of the two logits has twice the variance since it
    is only based on ``half as many'' subjects (per total number of
    subjects~$n$).

  \subsection{Logarithmic incidence rate ratios (log-IRRs)}\label{sec:UnitInfoIRR}
    An (approximate) standard error for an incidence rate ratio is
    given by $\sqrt{\frac{1}{a}+\frac{1}{c}}$, where $a$ and $c$ are
    the event counts in treatment- and control-groups,
    respectively.\citep[][Sec.~6.7.1]{CochraneHandbook} Assuming a
    total number~$m$ of events, and, for simplicity, $a=c=\frac{m}{2}$
    then yields a standard error of $\frac{2}{\sqrt{m}}$, implying a
    \emph{per-event} UISD of~$2$. For a given event rate~$\lambda$
    (per subject), the per-subject standard deviation then is at
    $\sigmau=\frac{2}{\sqrt{\lambda}}$.

\section{Prior distribution families}\label{sec:PriorTableAppendix}

  Table~\ref{tab:priorFamilies} characterizes some of the probability
  distribution families that are discussed in
  Section~\ref{sec:heteroPriors} in more detail (see also
  Figure~\ref{fig:densities} and Table~\ref{tab:predictive-2}). The
  distribution families considered are half-normal,
  half-Student\mbox{-}$t$, half-Cauchy, half-logistic, exponential,
  Lomax, log-normal and (proper)
  uniform.\citep{JohnsonKotzBalakrishnan,TurnerEtAl2015}.  The
  distributions' parameters, probability density functions, medians,
  95\% quantiles, means, variances, and coefficients of variation
  ($\varcoef=\frac{\sqrt{\var(X)}}{\expect[X]}$) are listed.

  In particular for families including several parameters, some of the
  expressions may get somewhat complex (e.g., the moments of a general
  half-Student\mbox{-}$t$ distribution, which are omitted in the
  table).\citep{PsarakisPanaretos1990}  However, if only a scale
  parameter is present, then quantiles, expectation and standard
  deviation are simply proportional to the scale, and the
  coefficient of variation is a constant. Examples are the half-normal
  distribution, or half-Student\mbox{-}$t$ or Lomax distributions with
  fixed shape parameters.

  Note that for the exponential distribution, which is most commonly
  parameterized using a \emph{rate} (or \emph{inverse scale})
  parameter, the inverse of the rate is a scale parameter.  Similarly,
  for the log-normal distribution, $\exp(\mu)$~would be a scale
  parameter, and the corresponding expressions then
  factor as multiples of~$\exp(\mu)$.
  Some of the expressions given below are not always defined, e.g.,
  expectation and variance of the half-$t$ distribution are only
  defined for $\nu>1$ and $\nu>2$,
  respectively,\citep{PsarakisPanaretos1990} and the first two moments
  of the Lomax distribution are only finite for $\alpha>1$ and
  $\alpha>2$, respectively.\citep{JohnsonKotzBalakrishnan}

  \begin{sidewaystable}
    \caption{\label{tab:priorFamilies}Some properties of potential
      prior distribution families that were discussed in
      Section~\ref{sec:heteroPriors}. An asterisk~($\ast$) means that
      the corresponding expression is somewhat complex and hence
      omitted here, and a dash~(\mbox{---}) means the figure is not
      defined. $\varcoef$~denotes the coefficient of variation (the
      ratio of standard deviation over expectation).}  \centering
    \begin{tabular}{lllccccc}
      \toprule
      distribution & parameter(s) & density function $p(x)$ & median & 95\% quantile & expectation & variance & $\varcoef$\\ 
      \midrule
      half-normal & scale~$\sigma$ & $\frac{\sqrt{2}}{\sqrt{\pi}\,\sigma}\exp\Bigl(-\frac{1}{2}\bigl(\frac{x}{\sigma}\bigr)^2\Bigr)$ & $0.674\sigma$ & $1.96\sigma$ & $0.798\sigma$ & $(0.603\sigma)^2$ & $0.756$ \\
      half-Student-$t$ & shape~$\nu$, scale~$\sigma$ & $\frac{2\,\Gamma\bigl(\frac{\nu+1}{2}\bigr)}{\Gamma\bigl(\frac{\nu}{2}\bigr) \, \sqrt{\nu\pi}\, \sigma} \, \Bigl(1+\frac{1}{\nu}\bigl(\frac{x}{\sigma}\bigr)^2\Bigr)^{-\frac{\nu+1}{2}}$ & $\ast$ & $\ast$ & $\ast$ & $\ast$ & $\ast$ \\
      $\;\;$ half-Student-$t_{\nu=3}$ & scale~$\sigma$ & $\frac{4}{\pi \, \sqrt{\nu}\, \sigma} \, \Bigl(1+\frac{1}{3}\bigl(\frac{x}{\sigma}\bigr)^2\Bigr)^{-2}$ & $0.765\sigma$ & $3.18\sigma$ & $1.10\sigma$ & $(1.34\sigma)^2$  & $1.21$ \\
      half-Cauchy & scale~$\sigma$ & $\frac{2}{\pi\,\sigma}\,\Bigl(1+\bigl(\frac{x}{\sigma}\bigr)^2\Bigr)^{-1}$ & $\sigma$ & $12.7\sigma$ & --- & --- & --- \\
      half-logistic & scale~$\sigma$ & $\frac{2\,\exp\bigl(-\frac{x}{\sigma}\bigr)}{\sigma\,\bigl(1+\exp\bigl(-\frac{x}{\sigma}\bigr)\bigr)^2}$ & $1.10\sigma$ & $3.66\sigma$ & $1.39\sigma$ & $(1.17\sigma)^2$ & $0.844$ \\
      exponential & rate~$\lambda$ & $\lambda\,\exp(-\lambda x)$ & $0.693\frac{1}{\lambda}$ & $3.00\frac{1}{\lambda}$ & $\frac{1}{\lambda}$ & $\bigl(\frac{1}{\lambda}\bigr)^2$ & $1$\\
      Lomax & shape~$\alpha$, scale~$\lambda$ & $\frac{\alpha}{\lambda}\bigl(1+\frac{x}{\lambda}\bigr)^{-(\alpha+1)}$ & $(2^\frac{1}{\alpha}\!-\!1)\lambda$ & $(20^\frac{1}{\alpha}\!-\!1)\lambda$ & $\frac{1}{\alpha-1}\lambda$ & $\Bigl(\sqrt{\frac{\alpha}{(\alpha-1)^2 (\alpha-2)}}\,\lambda\Bigr)^2$ & $\sqrt{\frac{\alpha}{\alpha-2}}$\\
      $\;\;$ Lomax($\alpha\!=\!6$) & scale~$\lambda$ & $\frac{6}{\lambda}\bigl(1+\frac{x}{\lambda}\bigr)^{-7}$ & $0.122\lambda$ & $0.648\lambda$ & $\frac{1}{5}\lambda$ & $(0.245\lambda)^2$ & $1.22$\\
      $\;\;$ Lomax($\alpha\!=\!1$) & scale~$\lambda$ & $\frac{\lambda}{(x+\lambda)^2}$ & $\lambda$ & $19\lambda$ & --- & --- & --- \\
      log-normal & shape~$\mu$, shape~$\sigma$ & $\frac{1}{x\sqrt{2\pi}\,\sigma}\,\exp\Bigl(-\frac{(\log(x)-\mu)^2}{2\sigma^2}\Bigr)$ & $\exp(\mu)$ & $\scriptstyle \exp(1.64\sigma)\exp(\mu)$ & $\scriptstyle \exp\bigl(\frac{\sigma^2}{2}\bigr) \exp(\mu)$ & $\scriptstyle \bigl(\sqrt{\exp(2\sigma^2)-\exp(\sigma^2)}\exp(\mu)\bigr)^2$ & $\scriptstyle \sqrt{\exp(\sigma^2)-1}$ \\
      (proper) uniform & scale~$a$ & $\left\{\begin{array}{ll} 1/a\;\; & \mbox{if $0 \leq x \leq a$} \\ 0 & \mbox{otherwise} \end{array}\right.$ & $\frac{1}{2}a$ & $0.95a$ & $\frac{1}{2}a$ & $(0.289a)^2$ & $0.577$ \\
      \bottomrule
    \end{tabular}
  \end{sidewaystable}

\section{Scale mixture parametrisations}\label{sec:MixtureAppendix}
  \subsection[Motivating Lomax and Student-t distributions as scale mixtures]{Motivating Lomax and Student-$t$ distributions as scale mixtures}
    Heavy-tailed priors may be constructed as \emph{scale mixtures} of
    shorter-tailed distributions. For example, a distribution
    $p(\theta|\scale)$ that has a scale parameter~$\scale>0$ may be
    generalized by specifying a \emph{mixing distribution} $p(\scale)$
    and subsequently marginalizing over it, yielding the mixture
    $p(\theta)=\int_0^\infty
    p(\theta|\scale)\,p(\scale)\,\differential\scale$.\citep{OHaganPericchi2012,Lindsay}
    In order to make the connection to the original (conditional)
    distribution $p(\theta|\scale)$, it is instructive to consider the
    mixing distribution's location and spread, e.g. in terms of
    expectation~$\mu=\expect[\scale]$ and coefficient of
    variation~$\varcoef=\frac{\sqrt{\var(\scale)}}{\expect[\scale]}$.
    For small~$\varcoef$, the mixture will closely resemble the
    original distribution ($p(\theta|\scale)$), for larger~$\varcoef$,
    it will be heavier-tailed. Note that, since the scale parameter's
    domain is the positive real line, an increasing coefficient of
    variation also implies an increasingly skewed mixing distribution.
    In the following, we show how Lomax and Student\mbox{-}$t$
    distributions result as scale mixtures of exponential and normal
    distributions, respectively, and how these may be parameterised in
    terms of pre-specified expectation and coefficient of variation of
    their scale parameters.  Specification of a prior in terms of a
    scale mixture may be seen as a case of a ``contaminated'' prior
    also considering variations of a prior that are ``close to an
    elicited one''.\citep{BergerBerliner1986}
    \citep[][Sec.~3.5.3]{Berger}

  \subsection{The Lomax distribution as an exponential scale mixture}\label{sec:LomaxMixtureAppendix}
    The exponential distribution may be parameterized in terms of
    \emph{rate (inverse scale)}~$\lambda$, or
    \emph{scale}~$\scale=\frac{1}{\lambda}$, where the expected value
    is given by $\expect[X]=\scale=\frac{1}{\lambda}$.  Suppose that
    the scale~$\scale$ is uncertain with expectation
    $\expect[\scale]=\mu$ and coefficient of variation
    $\frac{\sqrt{\var(\scale)}}{\expect[\scale]} = \varcoef$. Then
    $\scale$ may be modelled using an inverse-gamma distribution with
    matched moments, using shape~$\alpha=2+\frac{1}{\varcoef^2}$ and
    scale $\beta=\mu\bigl(1+\frac{1}{\varcoef^2}\bigr)$ (implying a
    gamma-distribution for the rate~$\lambda$ with shape~$\alpha$ and
    scale~$\frac{1}{\beta}$).  A mixture of exponential distributions
    with inverse-gamma-distributed scale (or gamma-distributed rate)
    then results as a Lomax distribution parameterized by
    shape~$\alpha=\alpha$ and scale~$\lambda=\beta$, with
    expectation~$\frac{\lambda}{\alpha-1}$ and
    variance~$\frac{\lambda^2\alpha}{(\alpha-1)^2(\alpha-2)}$.  By
    pre-sprecifying the exponential scale's expectation and
    uncertainty (in terms of the coefficient of variation), we can
    then derive the corresponding Lomax distribution.
    For example, if we are aiming for an exponential scale mixture in
    which the scale has expectation~$\mu=0.5$ and coefficient of
    variation~$\varcoef = 0.5$, this implies Lomax parameters of
    shape~$\alpha=2+\frac{1}{\varcoef^2}=2+\frac{1}{0.5^2}=6$ and
    scale~$\lambda=\mu\bigl(1+\frac{1}{\varcoef^2}\bigr)=0.5\bigl(1+\frac{1}{0.5^2}\bigr)=2.5$.

  \subsection[The (half-) Student-t distribution as a normal scale mixture]{The (half-) Student-$t$ distribution as a normal scale mixture}\label{sec:NormalMixtureAppendix}
    The Student\mbox{-}$t$ distribution (with $\nu$~degrees of
    freedom) is classically defined as the distribution of a variable
    $X=\frac{Y}{\sqrt{Z/\nu}}$, where $Y$~follows a standard normal
    distribution, and $Z$~is independent and follows a $\chi^2_\nu$
    distribution (with $\nu$~degrees of freedom).  The
    Student\mbox{-}$t$ family includes the Cauchy distribution as a
    special case (for~$\nu\!=\!1$) and the normal distribution as a
    limiting case (for~$\nu\!\rightarrow\!\infty$).  Alternatively,
    the distribution of $X$~may be expressed via
    $X|\sigma\sim\normaldistn(0,\sigma^2)$ and
    $\sigma\sim\mbox{Inv-}\chi\bigl(\nu,\sqrt{\nu}\bigr)$, where the
    distribution of the normal scale~$\sigma$ is a \emph{scaled
      inverse $\chi$~distribution} with $\nu$ degrees of freedom and
    scale $s=\sqrt{\nu}$.  The latter formulation then makes the scale
    mixture connection more obvious. The arguments in the following
    then equally apply for Student\mbox{-}$t$ and
    half-Student\mbox{-}$t$ distributions.  The inverse
    $\chi$~distribution is simply defined as the distribution of the
    inverse of the square root of a $\chi^2_\nu$-distributed deviate;
    it is a special case of a square-root inverted-gamma
    distribution\citep{BernardoSmith} (with $\alpha=\frac{\nu}{2}$ and
    $\beta=\frac{1}{2}$).  The \emph{scaled} inverse
    $\chi$~distribution then results by introducing an additional
    scale parameter~$s$.\citep[][Sec.~VII.6.2]{MGB}
    Its probability density function is given by
    \begin{equation}\label{eqn:scaledInvChiDensity}
      p(\theta|\nu,s) 
      \; = \; 
      \frac{2^{(1 - \nu/2)}}{s\,\Gamma(\nu/2)}\;\Bigl(\frac{s}{\theta}\Bigr)^{(\nu+1)}\;\exp\Bigl(-\frac{s^2}{2\,\theta^2}\Bigr)\mbox{.}
    \end{equation} 
    Cumulative distribution function, quantiles, etc. may be computed
    via the $\chi^2_\nu$ distribution.  Its moments are given by
    \begin{equation}\label{eqn:scaledInvChiMoments}
      \expect[\theta|\nu,\scale] \;=\; \scale \, \frac{\Gamma\bigl(\frac{\nu-1}{2}\bigr)}{\sqrt{2}\,\Gamma\bigl(\frac{\nu}{2}\bigr)}
      \qquad \mbox{and} \qquad
      \var(\theta|\nu,\scale) \;=\; \scale^2\,\biggl(\frac{1}{\nu-2}-\frac{1}{2}\Bigl(\frac{\Gamma((\nu-1)/2)}{\Gamma(\nu/2)}\Bigr)^2\biggr)
    \end{equation}
    (for $\nu>1$ and $\nu>2$, respectively). Its coefficient of
    variation depends only on the degrees of freedom parameter~$\nu$.
    This means that if, analogously to the previous section, we want
    to define a Student\mbox{-}$t$ distribution corresponding to a
    normal scale mixture where the normal scale has a pre-specified
    expectation and coefficient of variation, we can first determine
    the associated degrees of freedom~$\nu$ and subsequently the
    scale~$s$.  Table~\ref{tab:invChiCV} lists corresponding
    coefficients of variation for a selected set of degrees of freedom
    values (according to
    equation~\eqref{eqn:scaledInvChiMoments}). Inversion of the
    relationship may be done numerically; degrees of freedom~$\nu$
    settings corresponding to certain coefficients of
    variation~$\varcoef$ are shown in Table~\ref{tab:invChiDf}.

    \begin{table}[t]
      \caption{\label{tab:invChiCV} Coefficients of variation
        ($\varcoef$) corresponding to certain settings of the degrees
        of freedom~($\nu$) of an inverse $\chi$ distribution.}
      \centering
      \begin{tabular}{cc}
        \toprule
        $\nu$ & $\varcoef$ \\
        \midrule
        $\phantom{0}2.5$ & $1.09$ \\
        $\phantom{0}3\phantom{.0}$ & $0.76$ \\
        $\phantom{0}4\phantom{.0}$ & $0.52$ \\
        $\phantom{0}5\phantom{.0}$ & $0.42$ \\
        $10\phantom{.0}$ & $0.24$ \\
        $20\phantom{.0}$ & $0.17$ \\
        $50\phantom{.0}$ & $0.10$ \\
        \bottomrule
      \end{tabular}
    \end{table}

    \begin{table}
      \caption{\label{tab:invChiDf} Degrees of freedom ($\nu$)
        settings corresponding to certain pre-specified coefficient of
        variation~($\varcoef$) of an inverse $\chi$ distribution.}
      \centering
      \begin{tabular}{cc}
        \toprule
        $\varcoef$ & $\nu$ \\
        \midrule
        $2$ & $\phantom{0}2.2$ \\
        $1$ & $\phantom{0}2.6$ \\
        $1/2$ & $\phantom{0}4.2$ \\
        $1/3$ & $\phantom{0}6.7$ \\
        $1/4$ & $10.2$ \\
        $1/5$ & $14.7$ \\
        $1/10$ & $52.2$ \\
        \bottomrule
      \end{tabular}
    \end{table}

    For example, if one was aiming for a normal scale mixture with
    expectation~$\mu=0.5$ and coefficient of variation~$\varcoef =
    0.5$, this first of all implies $\nu=4.2$ degrees of freedom
    (Table~\ref{tab:invChiDf}).  Using an ``plain'' Student\mbox{-}$t$
    distribution now would correspond to a scaled inverse $\chi$
    mixing distribution of the normal scale~$\sigma$ with degrees of
    freedom~$\nu=4.2$ and scale $\scale=\sqrt{4.2}=2.05$, and,
    according to equation~\eqref{eqn:scaledInvChiMoments}, with
    $\expect[\sigma]=1.24$. In order to set the expectation to the
    intended $\mu=0.5$ instead, the (half-) Student\mbox{-}$t$
    distribution needs be scaled by a factor of
    $\frac{0.5}{1.24}=0.40$.
    So a half-Student\mbox{-}$t$ distribution with $\nu=4.2$ degrees
    of freedom and a scale of $0.40$ may be motivated as a half-normal
    scale mixture with $\expect[\sigma]=0.5$ and
    $\frac{\sqrt{\var(\sigma)}}{\expect[\sigma]}=0.5$.
    On the other hand, a setting of $\nu=4$ and Student\mbox{-}$t$
    scale~0.5 would imply $\varcoef=0.52$ and $\mu=0.63$.

  \subsection{Scale mixture examples}\label{sec:MixtureExampleAppendix}
    Tables~\ref{tab:mixtures-1} and~\ref{tab:mixtures-2} below show a
    number of Lomax and Student\mbox{-}$t$ distributions that result
    as scale mixtures for pre-specified mean and variance for the
    exponential or half-normal scale parameter.  Note that, due to
    linearity, simple re-scaling of the (exponential or half-normal)
    scale's distribution implies proportional re-scaling of
    heterogeneity and predictive distribution. For example, the
    Lomax$_{\alpha=6}$(8.17)-distribution from
    Table~\ref{tab:predictive-2} results from re-scaling of the
    Lomax$_{\alpha=6}$(5)-distribution from Table~\ref{tab:mixtures-1}
    by a factor of~$\frac{1}{0.612}$ so that the prior median is
    at~$1.0$.
    Note also that by fixing the expectation and increasing the
    coefficient of variation, one get an increasingly skewed mixing
    distribution with a decreasing median.

    \begin{sidewaystable} 
      \caption{\label{tab:mixtures-1}Lomax prior distributions
        resulting as scale mixtures of exponential distributions. An
        inverse-gamma distributed scale (or \emph{inverse rate})
        parameter for the exponential distribution marginally yields a
        Lomax distribution. Pre-specifying expectation and coefficient
        of variation~($\varcoef$) for the scale (shown in bold)
        implies a unique inverse-gamma and resulting Lomax
        distribution. The table illustrates distributions of
        exponential scale~($s$), heterogeneity~($\tau$) and
        predictions~($\theta_i$). The first line corresponds to a
        ``plain'' exponential distribution with fixed scale.}
      \centering
      \begin{tabular}{lccccccccccccc}
        \toprule
        & \multicolumn{6}{c}{\textbf{exponential scale}}
        & \multicolumn{6}{c}{\textbf{heterogeneity}}
        & \multicolumn{1}{c}{\textbf{prediction}}\\
        & \multicolumn{6}{c}{}
        & \multicolumn{6}{c}{$\tau | s \;\sim\; \mathrm{exponential}(s)$,}
        & \multicolumn{1}{c}{$(\theta_i-\mu) | \tau \;\sim\; \normaldistn(0,\tau^2)$,}\\
        & \multicolumn{6}{c}{$s|\alpha,\beta\;\sim\; \invgamma(\alpha,\,\beta)$}
        & \multicolumn{6}{c}{$\tau|\alpha,\beta \;\sim\; \mbox{Lomax}_{\alpha}(\beta)$}
        & \multicolumn{1}{c}{$(\theta_i-\mu)|\alpha,\beta \;\sim\; \mbox{normal mixture}$}\\
        \cmidrule(lr){2-7}
        \cmidrule(lr){8-13}
        \cmidrule(lr){14-14}
        \textbf{$\tau$~prior} & shape $\alpha$ & scale $\beta$   & mean & $\varcoef$ & median & $q_{95\%}$ 
                              & shape $\alpha$ & scale $\lambda$ & mean & $\varcoef$ & median & $q_{95\%}$ 
                              & $q_{2.5\%} / q_{97.5\%}$\\
        \midrule
        exponential(1.0) &                            &                            & \textbf{1.0} & \textbf{0.0} & 1.00 & 1.00 &                            &                            & 1.0 & 0.00 & 0.69 & 3.00 & $\pm$2.052 \\[0.75ex]
        Lomax$_{\alpha=102}$(101)  &           102\phantom{.00} &           101\phantom{.00} & \textbf{1.0} & \textbf{0.1} & 0.99 & 1.17 &           102\phantom{.00} &           101\phantom{.00} & 1.0 & 1.01 & 0.69 & 3.01 & $\pm$2.052 \\
        Lomax$_{\alpha=27}$(26)     & \phantom{0}27\phantom{.00} & \phantom{0}26\phantom{.00} & \textbf{1.0} & \textbf{0.2} & 0.97 & 1.36 & \phantom{0}27\phantom{.00} & \phantom{0}26\phantom{.00} & 1.0 & 1.04 & 0.68 & 3.05 & $\pm$2.049 \\
        Lomax$_{\alpha=6}$(5)       & \phantom{00}6\phantom{.00} & \phantom{00}5\phantom{.00} & \textbf{1.0} & \textbf{0.5} & 0.88 & 1.91 & \phantom{00}6\phantom{.00} & \phantom{00}5\phantom{.00} & 1.0 & 1.22 & 0.61 & 3.24 & $\pm$2.022 \\
        Lomax$_{\alpha=3}$(2)       & \phantom{00}3\phantom{.00} & \phantom{00}2\phantom{.00} & \textbf{1.0} & \textbf{1.0} & 0.75 & 2.45 & \phantom{00}3\phantom{.00} & \phantom{00}2\phantom{.00} & 1.0 & 1.73 & 0.52 & 3.43 & $\pm$1.937 \\
        Lomax$_{\alpha=2.25}$(1.25) & \phantom{00}2.25           & \phantom{00}1.25           & \textbf{1.0} & \textbf{2.0} & 0.65 & 2.72 & \phantom{00}2.25           & \phantom{00}1.25           & 1.0 & 3.00 & 0.45 & 3.48 & $\pm$1.834 \\
        \bottomrule
      \end{tabular}
    \end{sidewaystable}

    \begin{sidewaystable} 
      \caption{\label{tab:mixtures-2} Half-Student\mbox{-}$t$ prior
        distributions resulting as scale mixtures of half-normal
        distributions. An inverse-Chi distributed scale parameter for
        the half-normal distribution marginally yields a
        half-Student\mbox{-}$t$ distribution. Pre-specifying
        expectation and coefficient of variation~($\varcoef$) for the
        scale (shown in bold) implies a unique inverse-Chi and
        resulting half-Student\mbox{-}$t$ distribution. The table
        illustrates distributions of half-normal scale~($\sigma$),
        heterogeneity~($\tau$) and predictions~($\theta_i$). The first
        line corresponds to a ``plain'' half-normal distribution with
        fixed scale.}  
      \centering
      \begin{tabular}{lccccccccccccc}
        \toprule
        & \multicolumn{6}{c}{\textbf{half-normal scale}}
        & \multicolumn{6}{c}{\textbf{heterogeneity}}
        & \multicolumn{1}{c}{\textbf{prediction}}\\
        & \multicolumn{6}{c}{}
        & \multicolumn{6}{c}{$\tau | \sigma \;\sim\; \halfnormaldistn(\sigma)$,}
        & \multicolumn{1}{c}{$(\theta_i-\mu) | \tau \;\sim\; \normaldistn(0,\,\tau^2)$,}\\
        & \multicolumn{6}{c}{$\sigma|\nu,s \;\sim\; \invchi(\nu, \, s)$}
        & \multicolumn{6}{c}{$\tau | \nu,s \;\sim\; \halftdistn\bigl(\nu,\, s / \sqrt{\nu}\bigr)$}
        & \multicolumn{1}{c}{$(\theta_i-\mu) | \nu,s \;\sim\; \mbox{normal mixture}$}\\
        \cmidrule(lr){2-7}
        \cmidrule(lr){8-13}
        \cmidrule(lr){14-14}
        \textbf{$\tau$~prior} & d.f. $\nu$ & scale $s$   & mean & $\varcoef$ & median & $q_{95\%}$ 
                              & d.f. $\nu$ & scale & mean & $\varcoef$ & median & $q_{95\%}$ 
                              & $q_{2.5\%} / q_{97.5\%}$\\
        \midrule
        half-normal(1.0)                &                &      & \textbf{1.0} & \textbf{0.0} & 1.00 & 1.00 &       $\infty$ & 1.00 & 0.80 & 0.76 & 0.68 & 1.96 & $\pm$2.18 \\[0.75ex]
        half-Student-$t_{\nu=52.2}$(0.99) &           52.2 & 7.12 & \textbf{1.0} & \textbf{0.1} & 0.99 & 1.18 &           52.2 & 0.99 & 0.80 & 1.02 & 0.67 & 1.98 & $\pm$2.19 \\
        half-Student-$t_{\nu=14.7}$(0.95) &           14.7 & 3.64 & \textbf{1.0} & \textbf{0.2} & 0.97 & 1.37 &           14.7 & 0.95 & 0.80 & 1.08 & 0.66 & 2.02 & $\pm$2.21 \\
        half-Student-$t_{\nu=4.2}$(0.81)  & \phantom{0}4.2 & 1.65 & \textbf{1.0} & \textbf{0.5} & 0.88 & 1.86 & \phantom{0}4.2 & 0.81 & 0.80 & 1.39 & 0.60 & 2.20 & $\pm$2.27 \\
        half-Student-$t_{\nu=2.6}$(0.67)  & \phantom{0}2.6 & 1.09 & \textbf{1.0} & \textbf{1.0} & 0.78 & 2.25 & \phantom{0}2.6 & 0.67 & 0.80 & 2.10 & 0.53 & 2.35 & $\pm$2.30 \\
        half-Student-$t_{\nu=2.2}$(0.60)  & \phantom{0}2.2 & 0.88 & \textbf{1.0} & \textbf{2.0} & 0.71 & 2.42 & \phantom{0}2.2 & 0.60 & 0.80 & 3.73 & 0.48 & 2.41 & $\pm$2.28 \\
        \bottomrule
      \end{tabular}
    \end{sidewaystable}

\clearpage
\section{Example applications}
  \subsection[R code for conditional prior predictive distribution]{\textsf{R} code to illustrate the conditional prior predictive distribution}\label{sec:PrevostRCode}
    The following \textsf{R} code illustrates the \emph{conditional}
    prior predictive distribution $p(\theta_i|\mu,\,\tau)$ (see
    Section~\ref{sec:fixedTau}) for the example case discussed by
    Prevost \emph{et~al.} (2000).\citep{PrevostEtAl2000} A fixed value
    of $\tau=0.35$ implies a conditional distribution of effects (RRs,
    $\exp(\theta_i)$) within factors of 0.5 and 2.0 with 95\%
    probability.
\begin{verbatim}
# generate log-RRs based on (fixed) tau=0.35:
N <- 1000
theta <- rnorm(N, mean=0, sd=0.35)
# show quantiles:
quantile(theta, prob=c(0.025, 0.975))
log(c(0.5, 2.0))
# (approximately 95% are within log(0.5) and log(2.0))

# derive RRs:
rr <- exp(theta)
# show histogram:
hist(rr)
# show quantiles:
quantile(rr, prob=c(0.025, 0.975))
# (approximately 95% are within 0.5 and 2.0)

# conditional quantiles may also be computed numerically:
qnorm(c(0.025, 0.975), mean=0, sd=0.35)
exp(qnorm(c(0.025, 0.975), mean=0, sd=0.35))
\end{verbatim}

  \subsection[R code for marginal prior predictive distribution]{\textsf{R} code to illustrate the marginal prior predictive distribution}\label{sec:DiasRCode}
    The following \textsf{R} code illustrates the \emph{marginal}
    prior predictive distribution $p(\theta_i|\mu)$ (see
    Section~\ref{sec:randomTau}) for the example case discussed by
    Dias \emph{et~al.} (2013).\citep{DiasEtAl2013} A
    half-normal($0.32$)-distribution for $\tau$ implies a marginal
    distribution of effects (ORs, $\exp(\theta_i)$) within factors of
    0.5 and 2.0 with 95\% probability.
\begin{verbatim}
# generate tau values from half-normal(0.32) distribution:
N <- 1000
tau <- abs(rnorm(N, mean=0, sd=0.32))
# generate log-ORs based on above tau values:
theta <- rnorm(N, mean=0, sd=tau)
# show quantiles:
quantile(theta, prob=c(0.025, 0.975))
log(c(0.5, 2.0))
# (approximately 95% are within log(0.5) and log(2.0))

# derive ORs:
or <- exp(theta)
# show histogram:
hist(or)
# show quantiles:
quantile(or, prob=c(0.025, 0.975))
# (approximately 95% are within 0.5 and 2.0)

# marginal quantiles may also be computed numerically:
library("bayesmeta")
nm032 <- normalmixture(cdf=function(t){phalfnormal(t, scale=0.32)})
nm032$quantile(c(0.025, 0.975))
exp(nm032$quantile(c(0.025, 0.975)))
\end{verbatim}

  \subsection[R code to reproduce examples]{\textsf{R} code to reproduce examples}\label{sec:RCode}
    The following \textsf{R} code shows how to use the
    \texttt{bayesmeta} library\citep{Roever2020} to perform a
    meta-analysis of the example data from
    Section~\ref{sec:logORexample}\citep{CrinsEtAl2014} using a
    half-normal(0.5) prior.
\begin{verbatim}
# load library:
require("bayesmeta")

# load data:
data("CrinsEtAl2014")

# calculate effect measures (ORs) for 2 randomized AR studies:
effsize <- escalc(measure="OR",
                  ai=exp.AR.events,  n1i=exp.total,
                  ci=cont.AR.events, n2i=cont.total,
                  slab=publication, data=CrinsEtAl2014,
                  subset=(CrinsEtAl2014$randomized=="yes"))

# perform meta-analysis:
bma <- bayesmeta(effsize, tau.prior = function(t){dhalfnormal(t, scale=0.5)})

# show results:
print(bma)

# show forest plot:
forestplot(bma)
\end{verbatim}

\clearpage
\subsection{Sensitivity analyses}\label{sec:SensitivityAppendix}
  \subsubsection{General remarks}
    In the following we illustrate some sensitivity analyses for
    the prior choice based on the MD example from
    Section~\ref{sec:MDexample} (Grande \emph{et~al.},
    2015),\citep{GrandeEtAl2015} and on the IRR example from
    Section~\ref{sec:logIRRexample} (Anker \emph{et~al.},
    2018),\citep{AnkerEtAl2018} both involving $k=4$~studies.
    Sensitivity analyses are commonly suggested and often easy to do,
    however, sensitivity to the choice of prior alone should only be a
    reason for concern if the prior is not convincingly
    motivated. Also, prior sensitivity must not be confused with the
    (weak/strong) informativeness of a prior; these are two quite
    separate aspects. A sensitivity analysis will also contribute
    little to the question of whether a particular prior is
    ``appropriate'' or not.
    Besides investigating variations of a given prior, analysis
    results may also be contrasted with those obtained when using a
    \emph{noninformative} prior (presuming that this is possible;
    e.g., the improper uniform prior requires $k\geq 3$~studies in
    order to yield a proper posterior). The aim here may then be to
    investigate to what extent results are determined by prior or data
    (likelihood).
    It should also be noted that the prior is only one among several
    crucial aspects of the model that might be challenged; additional
    aspects include normality,\citep{JacksonWhite2018} exchangeability
    (selection
    effects),\citep{RothsteinSuttonBorenstein,WilliamsonEtAl2005} the
    choice of effect measure,\citep{DeeksAltman2001} the model
    parametrisation,\citep{JacksonEtAl2018} or the use (by deliberate
    choice or due to a zero heterogeneity estimate) of a common-effect
    model.\citep{ChungEtAl2013a,FriedeRoeverWandelNeuenschwander2017a}
    The default statement of several (seemingly) alternative results
    might also encourage inconsistent (flip-flopping) conclusions from
    the data; if in fact there is uncertainty about the shape or scale
    of the prior, this might more appropriately be addressed via
    specification of a mixture prior reflecting this uncertainty.

    In Sections~\ref{sec:MDexample} and~\ref{sec:logIRRexample}
    half-normal priors with scale~0.5 were suggested for both
    examples. In order to investigate sensitivity of the analysis to
    details of the heterogeneity prior specification, we will vary the
    prior scale (within the half-normal family) as well as the
    distribution family (while keeping the prior median fixed).  For
    the sensitivity check, we will then consider scales half or twice
    as large.  For the investigation of sensitivity with respect to
    the choice of the prior distribution's shape, we consider the
    distribution families shown in Figure~\ref{fig:densities} and
    Table~\ref{tab:predictive-2}, which are half-Student\mbox{-}$t$
    (with $\nu=3$~d.f.), half-Cauchy, half-logistic, exponential and
    Lomax (with shape parameters~6 or~1).  The prior median for the
    original half-normal(0.5) prior was at~$\tau=0.34$, the different
    distributions then were scaled to have a matching median. Some of
    the reasons why one might choose one of these distribution
    families were discussed in Section~\ref{sec:priorFamilies};
    differences between these in particular relate to their behaviour
    near zero or towards their upper tail, or to their motivation as
    mixture distributions (see Appendix~\ref{sec:MixtureAppendix}).
    In order to contrast results with those obtained by using a
    noninformative prior, we selected the improper uniform prior
    in~$\tau$, as well as the Jeffreys prior for~$\tau$ for
    comparison. Both of these are improper and ``noninformative'' in a
    particular sense, and, since $k\geq 3$ in both examples, both
    yield proper posteriors here.\citep{Roever2020}

  \subsubsection{Mean difference example (Grande \emph{et al.}; 2015)}
\begin{figure}[t]
      \centering
      \makebox{\includegraphics[width=0.47\linewidth]{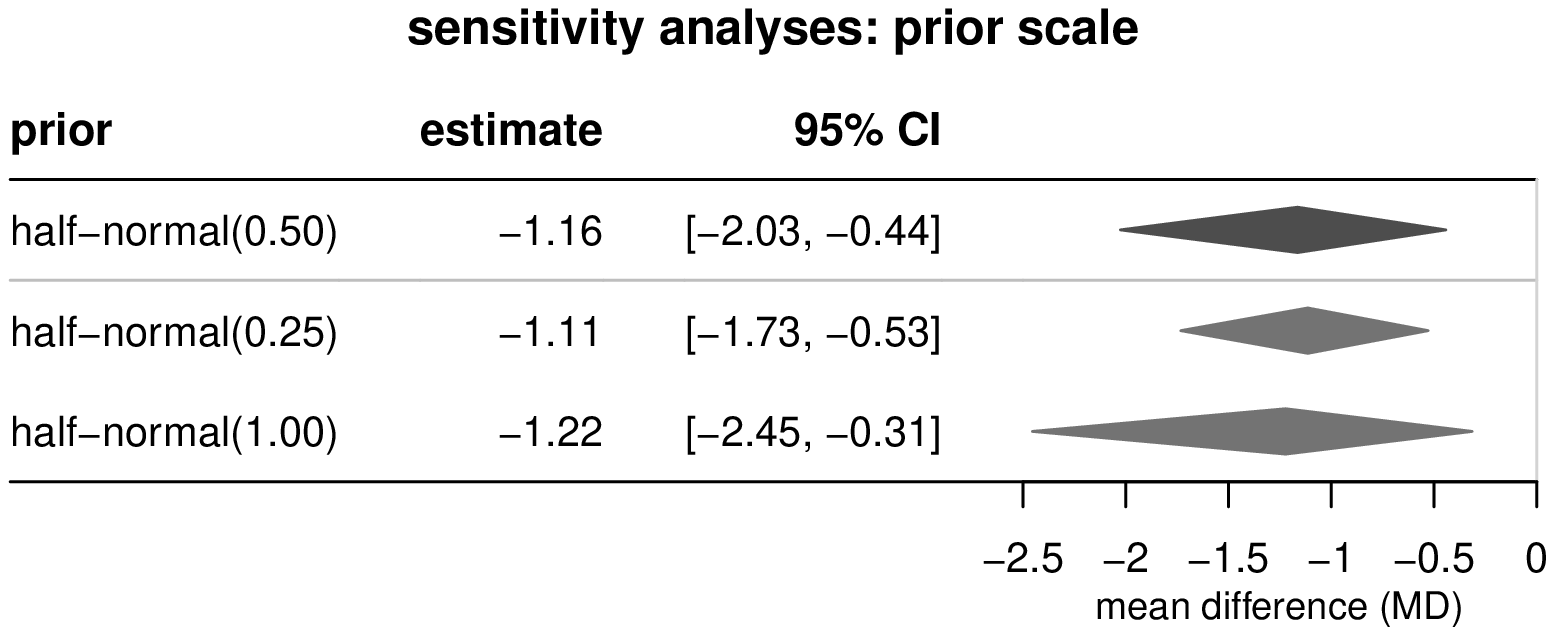}%
               \hspace{0.05\linewidth}%
               \includegraphics[width=0.47\linewidth]{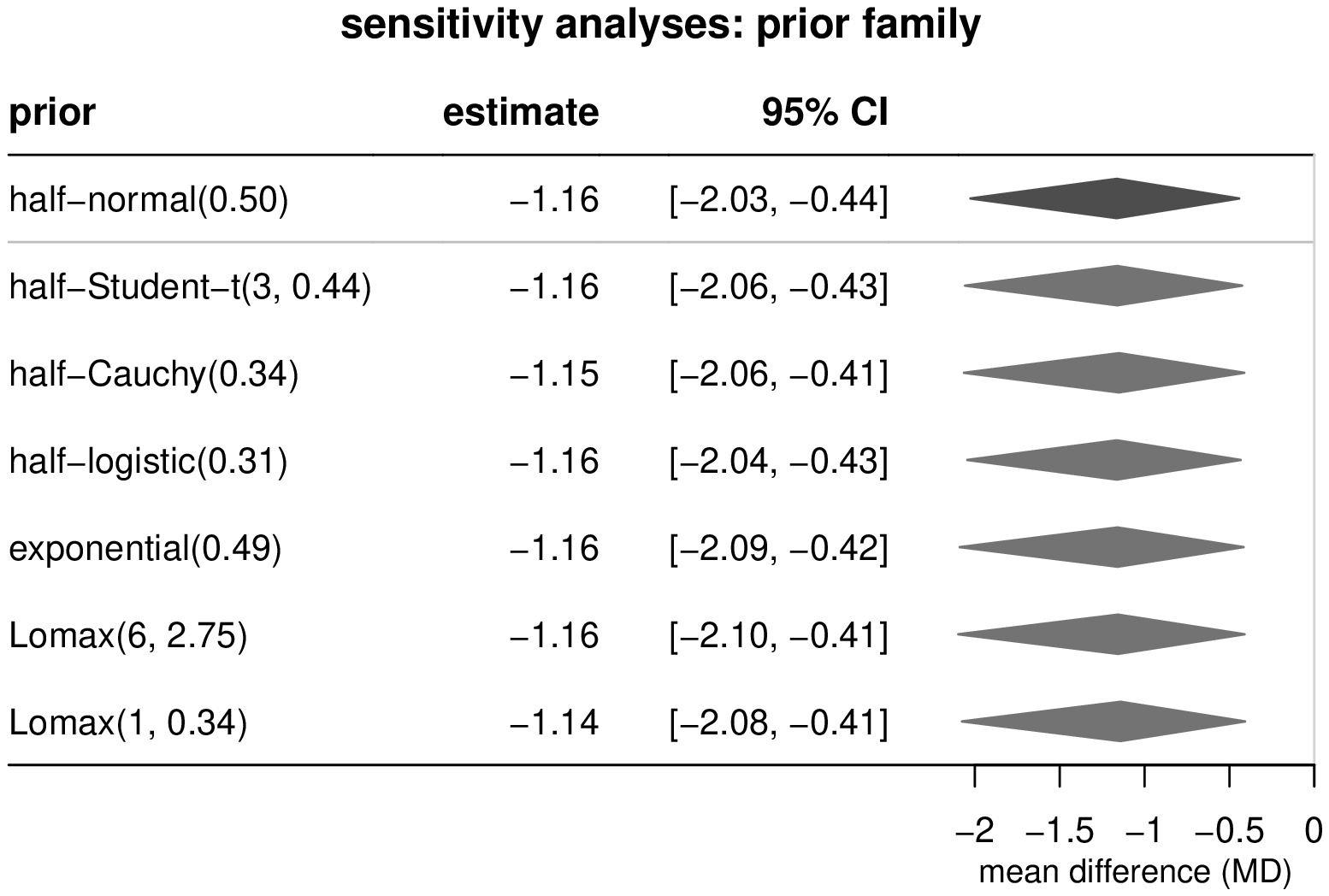}}
      \caption{\label{fig:sensitivityGrande} Sensitivity analyses for
        the MD example (Grande \emph{et~al.};
        2015)\citep{GrandeEtAl2015} from Sec.~\ref{sec:MDexample}
        investigating alternative prior distributions for the
        heterogeneity parameter~$\tau$.  The left panel shows
        variations of the scale parameter within the half-normal
        family, while the right panel corresponds to different prior
        distribution families with a fixed prior median (here:
        at~0.34). The ``original'' result from
        Sec.~\ref{sec:MDexample} (half-normal prior with scale~0.5) is
        shown at the top of each panel.}
    \end{figure}

    Varying the prior scale by a factor of two here implies a
    re-scaling of the prior predictive distribution by the same factor
    (see also the discussion in Section~\ref{sec:randomTau} and
    especially Table~\ref{tab:predictive-1}). Instead of a-priori
    considering between-study variations of $\pm 1$~day around the
    overall mean most plausible, this would mean focusing on a range
    of half a day of two days instead, respectively.
    Figure~\ref{fig:sensitivityGrande} (left panel) shows the overall
    effect estimates corresponding to the three prior settings. Most
    notably, with larger heterogeneity deemed plausible, the effect
    CI's lower bound includes more extreme values, while median and
    upper bound are less affected. This is consistent with the
    empirical data here (see Figure~\ref{fig:forest01}), as larger
    heterogeneity implies greater weight for the most extreme, yet
    also most uncertain first estimate, while lower heterogeneity
    implies that weighting is closer to the inverse-variance
    weights.\citep{Roever2020}

    When varying the prior distribution's shape (and keeping the prior
    median fixed), the effect on the resulting overall estimate is
    remarkably small, despite the different priors' different
    properties and appearances (see Table~\ref{tab:predictive-2} and
    Figure~\ref{fig:densities}).  Figure~\ref{fig:sensitivityGrande}
    (right panel) illustrates the corresponding effect estimates,
    where differences are barely discernible visually.

    Parameter estimates for the above analyses (also for
    heterogeneity~$\tau$ and prediction~$\theta_{k+1}$) are shown in
    Table~\ref{tab:sensitivityGrande}. When contrasting results with
    those obtained based on the noninformative uniform or Jeffreys
    priors, the estimates differ more substantially.  One may argue
    that, without the use of a weakly informative prior, the empirical
    data alone are not sufficient to rule out implausible ranges of
    heterogeneity here. For instance, in case of the improper uniform prior,
    heterogeneity values beyond $\tau=4.0$ would be considered
    a-posteriori plausible. This is more than the estimated UISD
    ($\su=3.9$), and, looking at Table~\ref{tab:predictive-2}, this
    would imply variability of $\theta_i$~values (differences in mean
    numbers of symptom days) within ranges of more than $\mu \pm
    1\,\mbox{week}$, while the numbers of symptom days themselves were
    only of the order of one week (see Table~\ref{tab:example1}). Use of
    a weakly informative prior then allows to rule out such
    implausible parameter ranges.

    \begin{table}[t]
      \caption{\label{tab:sensitivityGrande}Estimates and 95\%~CIs for
        the heterogeneity~($\tau$), the overall mean effect~($\mu$)
        and the prediction~($\theta_{k+1}$) corresponding to the
        discussed sensitivity analyses for the MD example (Grande
        \emph{et~al.}; 2015)\citep{GrandeEtAl2015}. The first line
        shows the estimates resulting from the half-normal(0.5) prior
        that was originally proposed in Section~\ref{sec:MDexample}.}
      \centering
      \begin{tabular}{lcccccc}
        \toprule
          & \multicolumn{2}{c}{heterogeneity~$\tau$} 
          & \multicolumn{2}{c}{effect~$\mu$} 
          & \multicolumn{2}{c}{prediction~$\theta_{k+1}$}\\
        \cmidrule(lr){2-3} \cmidrule(lr){4-5} \cmidrule(lr){6-7}
        prior                     & median & 95\% CI    & median & 95\% CI      & median & 95\% CI \\
        \midrule
        half-normal(0.50)         & 0.29 & [0.00, 0.87] & -1.16 & [-2.03, -0.44] & -1.15 & [-2.50, -0.05] \\ 
        \cmidrule(lr){1-7}
        half-normal(0.25)         & 0.16 & [0.00, 0.47] & -1.11 & [-1.73, -0.53] & -1.11 & [-1.92, -0.36] \\ 
        half-normal(1.00)         & 0.47 & [0.00, 1.47] & -1.22 & [-2.45, -0.31] & -1.19 & [-3.34, 0.48] \\ 
        \cmidrule(lr){1-7}
        half-Student-$t$(3, 0.44) & 0.28 & [0.00, 0.98] & -1.16 & [-2.06, -0.43] & -1.14 & [-2.58, 0.00] \\ 
        half-Cauchy(0.34)         & 0.23 & [0.00, 1.08] & -1.15 & [-2.06, -0.41] & -1.13 & [-2.61, 0.02] \\ 
        half-logistic(0.31)       & 0.29 & [0.00, 0.92] & -1.16 & [-2.04, -0.43] & -1.15 & [-2.55, -0.02] \\ 
        exponential(0.49)         & 0.26 & [0.00, 1.05] & -1.16 & [-2.09, -0.42] & -1.14 & [-2.65, 0.04] \\ 
        Lomax(6, 2.75)            & 0.25 & [0.00, 1.09] & -1.16 & [-2.10, -0.41] & -1.14 & [-2.67, 0.05] \\ 
        Lomax(1, 0.34)            & 0.20 & [0.00, 1.16] & -1.14 & [-2.08, -0.41] & -1.13 & [-2.66, 0.04] \\ 
        \cmidrule(lr){1-7}
        uniform                   & 0.86 & [0.00, 4.60] & -1.30 & [-3.98, 0.58] & -1.25 & [-6.57, 3.13] \\ 
        Jeffreys                  & 0.62 & [0.01, 2.61] & -1.27 & [-3.03, -0.05] & -1.24 & [-4.55, 1.38] \\ 
        \bottomrule
      \end{tabular}
    \end{table}

  \subsubsection{Log incidence rate ratio example (Anker \emph{et al.}; 2018)}
\begin{figure}[t]
      \centering
      \makebox{\includegraphics[width=0.47\linewidth]{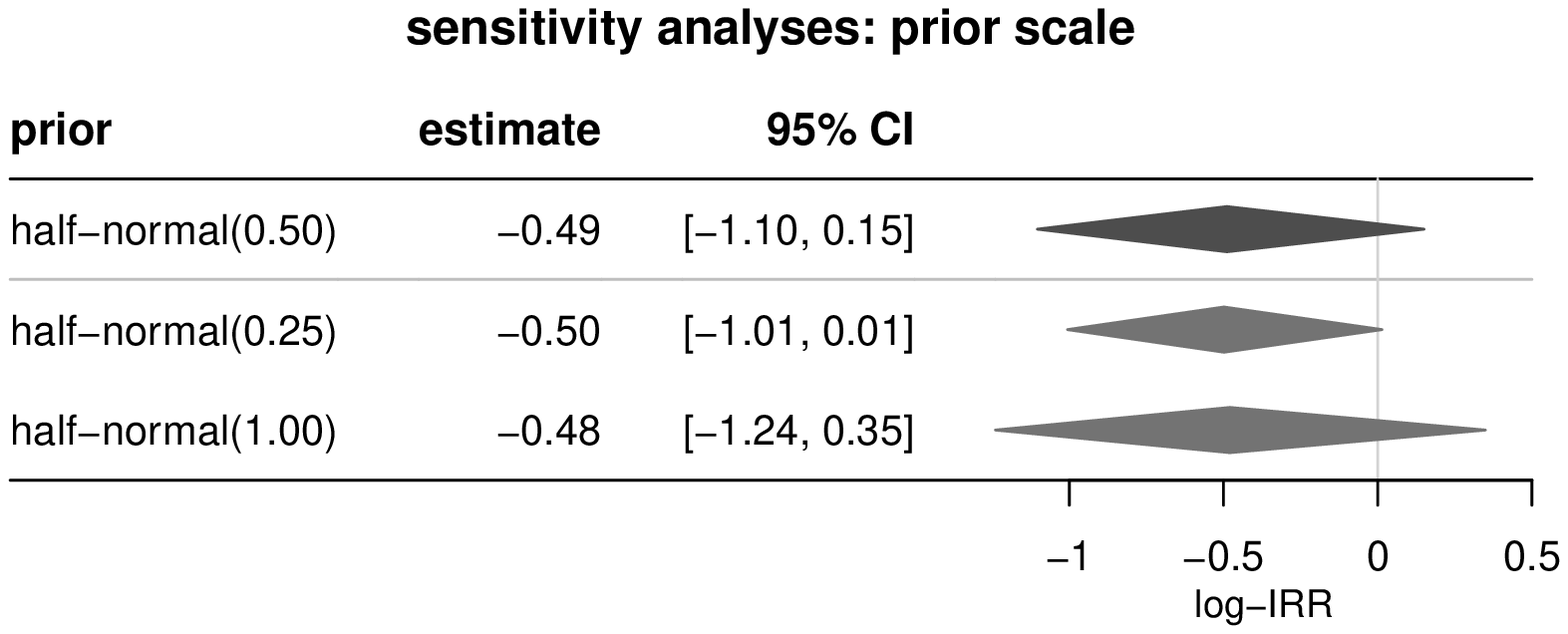}%
               \hspace{0.05\linewidth}%
               \includegraphics[width=0.47\linewidth]{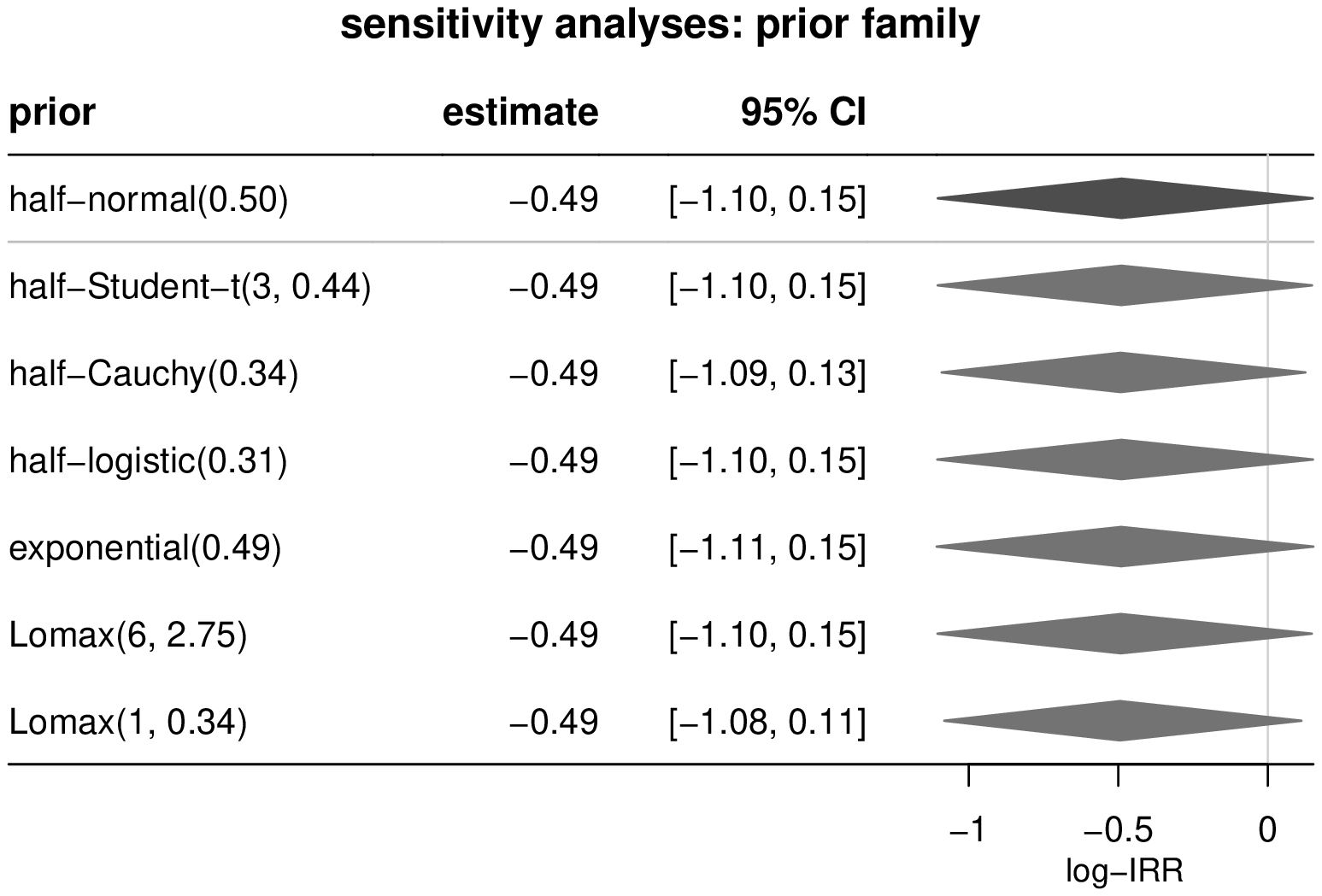}}
      \caption{\label{fig:sensitivityAnker} Sensitivity analyses for
        the IRR example (Anker \emph{et~al.};
        2018)\citep{AnkerEtAl2018} from Sec.~\ref{sec:logIRRexample}
        investigating alternative prior distributions for the
        heterogeneity parameter~$\tau$.  The left panel shows
        variations of the scale parameter within the half-normal
        family, while the right panel corresponds to different prior
        distribution families with a fixed prior median (here:
        at~0.34). The ``original'' result from
        Sec.~\ref{sec:logIRRexample} (half-normal prior with
        scale~0.5) is shown at the top of each panel.}
    \end{figure}

    Regarding the implications of varying the prior scale, the
    half-normal(1.0) prior was also discussed in
    Sections~\ref{sec:randomTau} and~\ref{sec:specialcase:logtrafo}
    (see especially Table~\ref{tab:predictive-1} and
    Figure~\ref{fig:TurnerPrior}).  In the present context, the
    half-normal(0.25) prior would assign 31\% probability to ``small''
    values, 64\% and 4.5\% to ``reasonable'' and ``fairly high''
    heterogeneity, respectively, and allow only 0.0063\% probability
    for ``fairly extreme'' values. The 95\% predictive interval (for
    $\theta_i-\mu$) ranges across $\pm0.55$, corresponding to
    multiplicative effects of factors of $0.58$ or $1.73$ on the
    exponentiated scale.
    With that, the half-normal(0.25) prior confines heterogeneity to a
    rather optimistic range of values, which would need to be
    discussed in the exact context of the example. One might need to
    argue why mostly up to ``reasonable'' (but virtually no ``fairly
    high'' or ``fairly extreme'') heterogeneity would be expected; in
    the present case, such an argument might be made based on the fact
    that the studies were performed according to similar protocols and
    executed by the same sponsor.\citep{AnkerEtAl2018}
    Figure~\ref{fig:sensitivityAnker} (left panel) illustrates the
    overall effect estimates corresponding to the three prior
    settings. Although especially the CI~width changes slightly, and
    for the more ``optimistic'' half-normal(0.25) prior almost
    excludes zero, the conclusions do not change drastically.

    Figure~\ref{fig:sensitivityAnker} (right panel) illustrates the
    overall effect estimates corresponding to the different prior
    distribution families.  Despite their different appearances and
    properties (see also Figure~\ref{fig:densities}), the resulting
    overall effect estimates and CIs again are remarkably similar.

    The noninformative priors yield CIs that are wider by factors of
    roughly~1.5 or~2.1 than for the analysis based on the proposed
    half-normal(0.5) prior, so the precision gain is quite substantial
    here.
    In addition to investigating the priors' influence on the overall
    effect~($\mu$), it may also be of interest to consider its effect
    on prediction intervals, shrinkage estimates, or the
    heterogeneity's posterior.  Table~\ref{tab:sensitivityAnker} lists
    some estimates corresponding to all of the sensitivity analyses
    discussed above.
    \begin{table}[b]
      \caption{\label{tab:sensitivityAnker}Estimates and 95\%~CIs for
        the heterogeneity~($\tau$), the overall mean effect~($\mu$)
        and the prediction~($\theta_{k+1}$) corresponding to the
        discussed sensitivity analyses for the IRR example (Anker
        \emph{et~al.}; 2018)\citep{AnkerEtAl2018}. The first line
        shows the estimates resulting from the half-normal(0.5) prior
        that was originally proposed in
        Section~\ref{sec:logIRRexample}.}  \centering
      \begin{tabular}{lcccccc}
        \toprule
          & \multicolumn{2}{c}{heterogeneity~$\tau$} 
          & \multicolumn{2}{c}{effect~$\mu$} 
          & \multicolumn{2}{c}{prediction~$\theta_{k+1}$}\\
        \cmidrule(lr){2-3} \cmidrule(lr){4-5} \cmidrule(lr){6-7}
        prior                     & median & 95\% CI    & median & 95\% CI      & median & 95\% CI \\
        \midrule
        half-normal(0.50)         & 0.24 & [0.00, 0.75] & -0.49 & [-1.10, 0.15] & -0.49 & [-1.49, 0.56] \\
        \cmidrule(lr){1-7}
        half-normal(0.25)         & 0.15 & [0.00, 0.44] & -0.50 & [-1.01, 0.01] & -0.50 & [-1.18, 0.20] \\ 
        half-normal(1.00)         & 0.34 & [0.00, 1.17] & -0.48 & [-1.24, 0.35] & -0.49 & [-1.89, 1.02] \\
        \cmidrule(lr){1-7}
        half-Student-$t$(3, 0.44) & 0.23 & [0.00, 0.78] & -0.49 & [-1.10, 0.15] & -0.49 & [-1.49, 0.55] \\ 
        half-Cauchy(0.34)         & 0.19 & [0.00, 0.77] & -0.49 & [-1.09, 0.13] & -0.50 & [-1.44, 0.50] \\ 
        half-logistic(0.31)       & 0.24 & [0.00, 0.77] & -0.49 & [-1.10, 0.15] & -0.49 & [-1.49, 0.56] \\ 
        exponential(0.49)         & 0.21 & [0.00, 0.82] & -0.49 & [-1.11, 0.15] & -0.49 & [-1.50, 0.57] \\ 
        Lomax(6, 2.75)            & 0.20 & [0.00, 0.82] & -0.49 & [-1.10, 0.15] & -0.49 & [-1.49, 0.56] \\ 
        Lomax(1, 0.34)            & 0.16 & [0.00, 0.79] & -0.49 & [-1.08, 0.11] & -0.50 & [-1.42, 0.48] \\
        \cmidrule(lr){1-7}
        uniform                   & 0.46 & [0.00, 2.52] & -0.47 & [-1.68, 0.91] & -0.48 & [-3.05, 2.30] \\ 
        Jeffreys                  & 0.43 & [0.01, 1.61] & -0.47 & [-1.39, 0.55] & -0.48 & [-2.29, 1.47] \\ 
        \bottomrule
      \end{tabular}
    \end{table}

\end{document}